\documentstyle[a4,12pt,epsfig,eqnumber]{article}
\def\be{\begin{equation}}
\def\ee{\end{equation}}
\def\bq{\begin{eqnarray}}
\def\eq{\end{eqnarray}}
\def\g{\gamma}
\def\ra{\rightarrow}
\def\vp{\varphi}
\begin{document}
\thispagestyle{empty}
\setcounter{page}{0}
\begin{flushright}
WUE-ITP-98-011\\
CERN-TH/98-124\\
\end{flushright}
\vspace{2em}
\begin{center}
{\Large \bf Exclusive Decays of Charm and Beauty}\\
\vspace{2em}
\large
\vspace{2em}
Reinhold R\"uckl\\
\vspace{2em}
{\small Institut f\"ur Theoretische Physik,
Universit\"at W\"urzburg, D-97074 W\"urzburg, Germany}\\
{\small and}\\
{\small Theory Division, CERN, CH-1211 Gen\`eve, Switzerland}\\
\end{center}
\bigskip
%
%
\begin{abstract}
Exclusive decays of $B$ and $D$ mesons are discussed with the focus
on the theoretical problem how to calculate the relevant hadronic
matrix elements of weak operators in the framework of QCD sum rules.  
The three lectures are devoted to\\
\hspace*{3cm} 1. leptonic decays and decay constants\\
\hspace*{3cm} 2. semileptonic decays and form factors\\
\hspace*{3cm} 3. nonleptonic decays and nonfactorizable amplitudes.\\
I shall introduce some of the basic concepts, describe various calculational 
techniques, and illustrate the numerical results. 
The latter are compared with lattice and quark model predictions
and, where possible, with experimental data. Applications include
the determination of $V_{ub}$ from 
$B \rightarrow \pi \bar{l} \nu_l$ and 
$B \rightarrow \rho \bar{l} \nu_l$, and an estimate of the
phenomenological coefficient $a_2$ for $B \rightarrow J/\psi K$.
\end{abstract}
\bigskip
%
%
\section{Introduction}

Inclusive and exclusive decays of heavy flavours play a complementary role 
in the determination of fundamental parameters of the 
electroweak standard model and in
the development of a deeper understanding of QCD.
While the theory of inclusive decays
\footnote{See, e.g., the lectures 
by N.G. Uraltsev in this volume \cite{uraltsev}.} is well
advanced, inclusive measurements are generally quite difficult.
Conversely, exclusive decays into few-body final
states are often easier to measure,
but the theory of exclusive processes is more demanding and hence           
still underdeveloped.
In view of the exciting
experimental prospects at future bottom and charm factories,
where many exclusive channels are expected to be measured accurately, it 
is pressing to make further progress on the theoretical
side.

The difference in complexity of inclusive and exclusive observables
is illuminated
by the fact that the simple parton picture of charm and beauty
decays illustrated in Tab. 1 provides surprisingly reasonable 
estimates for the lifetimes and inclusive branching ratios,
whereas no such description exists for exclusive decays. The theory
of both inclusive and exclusive processes is based on 
operator product expansion (OPE) which
allows to separate the dynamics at short and long distances.
At very large scale, $\mu = O(m_W)$, charm and beauty decays are
described by second order weak interaction
involving $W$ exchange. Since the momentum transfer
$p^2 \ll m_W^2$, one effectively has four-fermion interactions described
by a hamiltonian of the form
\be
H_W = \frac{G_F}{\sqrt{2}} j^{\mu} j'_{\mu} ~,
\label{HF}
\ee
where $j_\mu$ and $j'_\mu$ are $V-A$ quark or lepton currents. 
In the nonleptonic case, evolution of the hamiltonian to the physical scale 
of the order of the heavy quark mass, $\mu = O(m_Q)$,
leads to considerable modifications of (\ref{HF}) due to 
strong interactions. These effects can be taken 
into account by OPE and renormalization group methods. The result
is an effective hamiltonian 
\be
H_{NL} = \frac{G_F}{\sqrt{2}} \sum_i  c_i(\alpha_s, \mu) O_i(\mu) 
\label{HW}
\ee
involving a sum of local operators $O_i$ with coefficients
$c_i$ which can be calculated in perturbation theory as long
as $\mu \gg \Lambda_{QCD}$ so that one is only dealing with 
interactions of quarks and gluons at short distances.
The decay amplitudes are then given by  
\be
A(P \to X_h) = \frac{G_F}{\sqrt{2}} \sum_i  c_i(\alpha_s, \mu) 
 \langle X_h \mid O_i(\mu) \mid P\rangle 
\label{amplinc}
\ee
showing the main problem, that is the calculation of the matrix elements 
of the weak operators $O_i$ which incorporate the long-distance effects. 
For semileptonic decays
one instead has to compute matrix elements of weak currents:    
\be
A(P \to l \bar{\nu}_l X_h) = \frac{G_F}{\sqrt{2}}
\langle X_h \mid j_{\mu} \mid P\rangle 
\langle l \bar{\nu_l} \mid j'_{\mu} \mid 0\rangle ~.
\label{amplincl}
\ee
This simplifies the problem.
 
The usual way to proceed in inclusive decays is to assume 
duality of the sum over 
all possible hadronic states $X_h$ and the quark (and gluon) final states.
The theory is further improved by including radiative 
corrections to the decay widths, quark mass
effects, and initial bound state corrections \cite{uraltsev}. 
In comparison, 
the calculation of the hadronic matrix elements 
in (\ref{amplinc}) and (\ref{amplincl})
for exclusive final states $X_h$ constitutes
an ab initio nonperturbative problem.
Tab. 2 illustrates typical exclusive decays considered in these
lectures, and the corresponding matrix elements 
such as decay constants, form factors and nonleptonic two-body amplitudes. 
Here, the interplay between weak and strong interactions
is even more intricate than in inclusive decays. Obviously,
in order to disentangle the properties of weak interactions such as the 
charged current structure, flavour mixing and CP violation
from exclusive processes, one must understand the impact of QCD
beyond perturbation theory.  

\begin{figure}
\mbox{
\epsfig{file=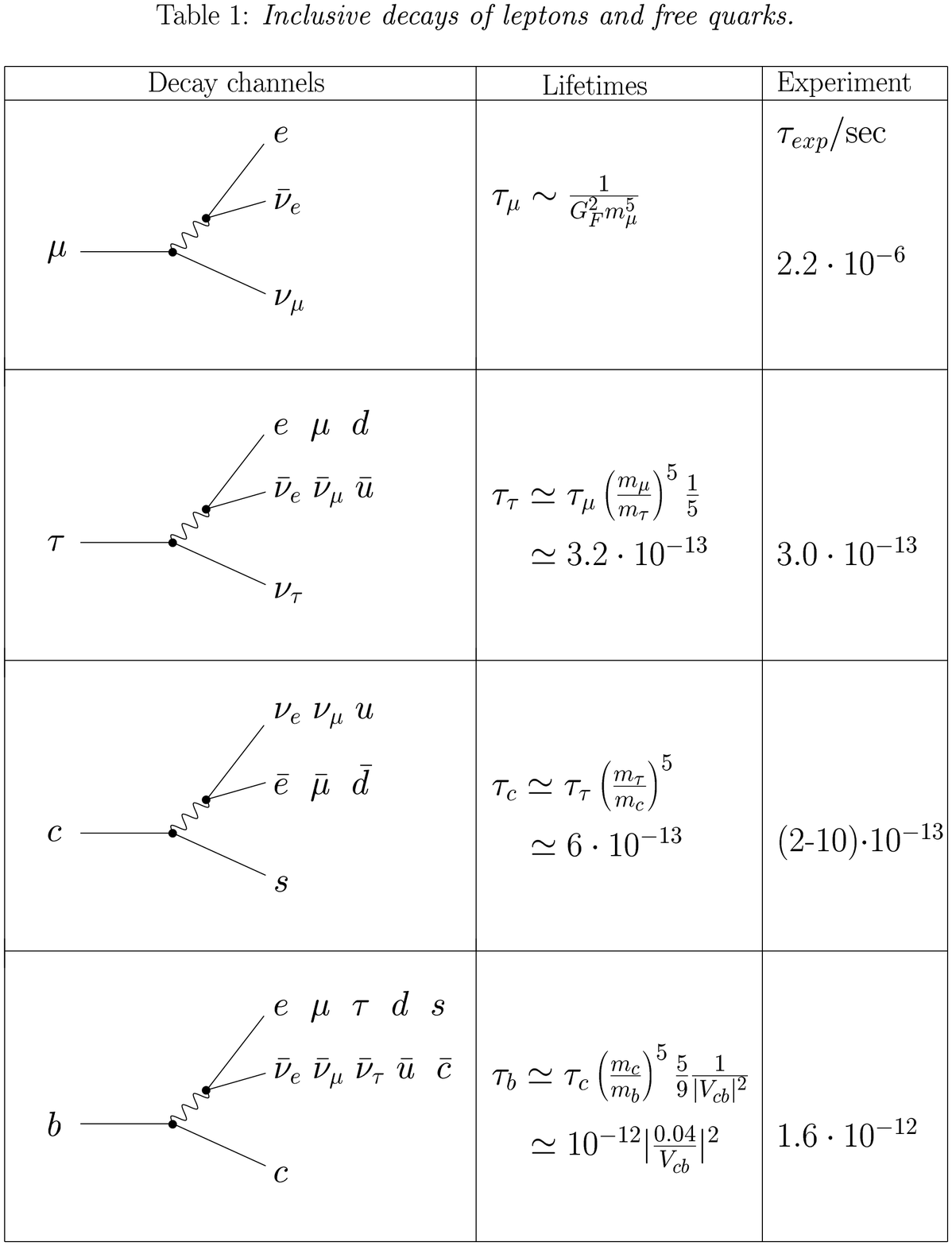,width=\textwidth,bbllx=0pt,bblly=60pt,bburx=600pt,%
bbury=850pt,clip=}
}
\end{figure}

\begin{figure}
\mbox{
\epsfig{file=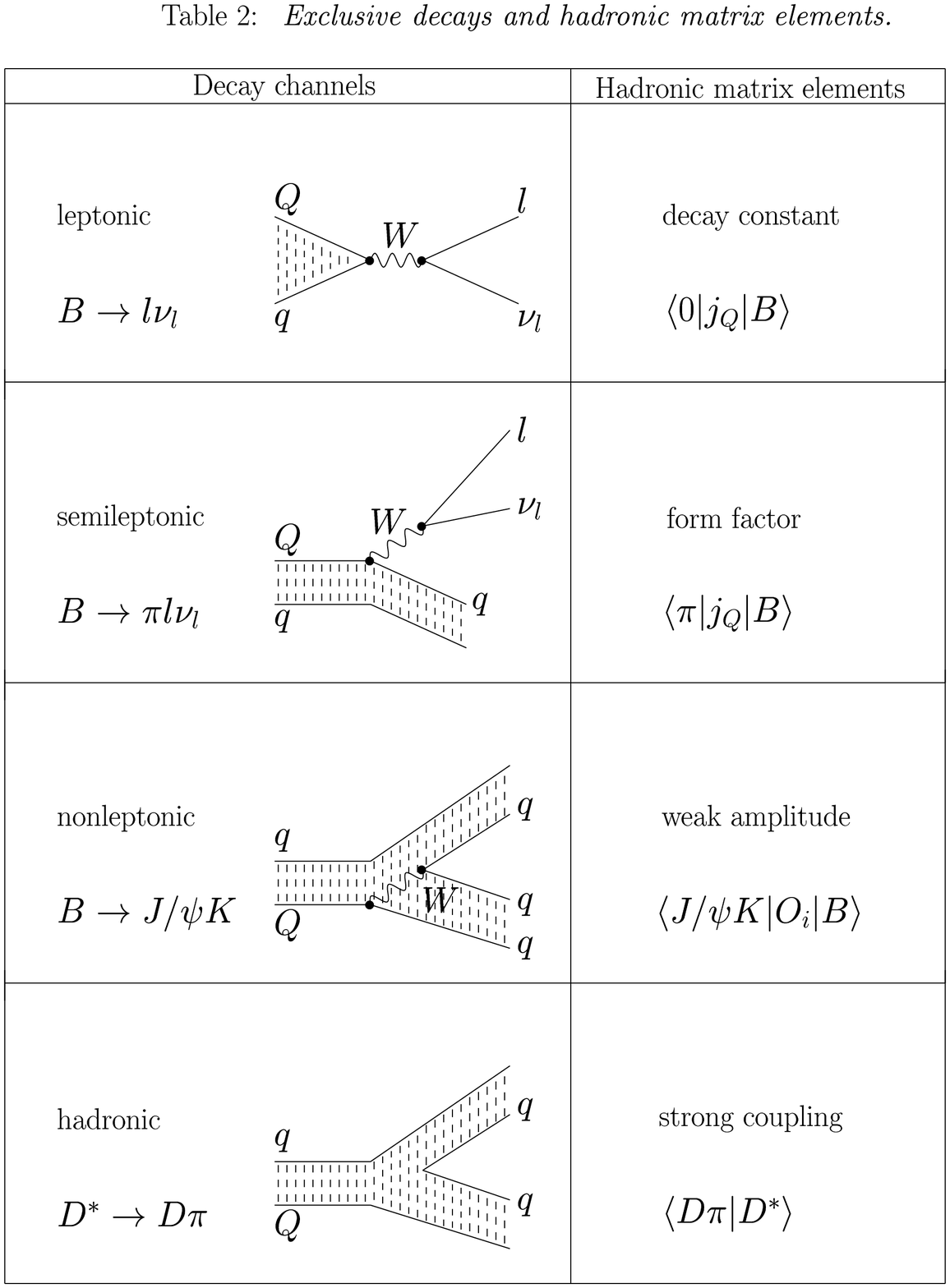,width=\textwidth,bbllx=6pt,bblly=78pt,bburx=600pt,%
bbury=840pt,clip=}
}
\end{figure}

Current approaches include lattice calculations, QCD sum rules, heavy
quark effective theory (HQET), chiral perturbation theory (CHPT),
and phenomenological quark models.
Each of these approaches has advantages and disadvantages.
For example, quark models are easy to use and good
for intuition. However, their relation to QCD is unclear.
On the other hand, lattice calculations are rigorous from the
point of view of QCD, but they suffer from lattice artifacts
and uncertain extrapolations.
Furthermore, effective theories are usually applicable only to
a restricted class of problems, and sometimes require substantial
corrections which cannot be calculated within the same framework.
For example, HQET is very powerful in treating  $b \ra c $ transitions,
but a priori less suitable for $b\to u$ transitions, while CHPT is 
designed for processes involving soft pions and kaons.

These lectures focus on applications of
QCD sum rules \cite{SVZ,RRY,Volume}
to exclusive $B$ and $D$ decays.
\footnote{Further details can be found in a recent review by 
A. Khodjamirian and the present author \cite{review}.} 
Proceeding from the firm basis
of QCD perturbation theory this approach
systematically incorporates  
nonperturbative elements of QCD. Schematically, 
hadronic matrix elements of the kind indicated in Tab. 2
are extracted from suitable correlation functions of quark currents at 
unphysical external momenta,
rather than estimated directly. Thereby, 
one makes use of OPE, the analyticity principle
(dispersion relations), and $S$-matrix unitarity. In addition, one assumes
the validity of quark-hadron duality. 
The long-distance dynamics is parameterized in terms of vacuum
condensates or light-cone wave functions. At present, these 
nonperturbative quantities cannot
be calculated directly from QCD. Therefore, they are determined from
experimental data. Nevertheless,
because of the universal nature of the nonperturbative input,
the sum rule approach retains its predictive power. 
Moreover, the method is rather flexible and can therefore be
applied to a large variety of problems in hadron physics. Experience
has shown that sum rules are particularly suited for heavy quark physics.
The more recent development 
of sum rule techniques for exclusive $B$ and $D$ decays 
appears similarly promising.

The lectures are organized as follows.
The first lecture is devoted to
leptonic decays and decay constants,
the second one to semileptonic decays and form factors,
and the third to nonleptonic decays and the problem
of nonfactorizable amplitudes.
I shall introduce the basic concepts, 
describe various calculational techniques, 
and show selected numerical results,
aiming at a compromise between the demands by theorists
and the needs of experimenters.
The results are compared with lattice and quark model predictions
and, where possible, with experimental data. 
Phenomenological applications include
the determination of $V_{ub}$ from 
$B \rightarrow \pi \bar{l} \nu_l$ and 
$B \rightarrow \rho \bar{l} \nu_l$, and an estimate of the
effective coefficient $a_2$ for $B \rightarrow J/\psi K$.
In addition, I shall present a few illustrative predictions for $D$ mesons.

Although the charm and beauty flavours could be treated in parallel,  
for definiteness, I will usually refer to $B$ mesons.
In most cases, it is obvious how to obtain the corresponding results 
for $D$ mesons. However, when convenient I will also use the
generic notation $P$ ($V$) for a heavy pseudoscalar (vector) meson 
and $Q$ for a heavy quark.  

%
\section{Leptonic decays and decay constants}

\subsection{Decay width}

The leptonic $B$ decays are induced by the weak annihilation
process $b\bar{u}\to l \bar{\nu}_l$ where $l = e, \mu, \tau$. 
The relevant weak hamiltonian is given by 
\be
H_L= \frac{G_F}{\sqrt{2}}V_{ub}
     (\bar{u}\Gamma^\mu b)(\bar{l}\Gamma_\mu \nu_l) + h.c. 
\label{HL}
\ee
with $\Gamma_\mu = \gamma_\mu(1-\gamma_5)$. Generically,
using the decay constant $f_P$ as defined by the matrix element
of the axial-vector current,
\be
\langle 0 \mid\bar{q}\gamma_\mu \gamma_5 Q\mid P\rangle =if_P q_\mu~,
\label{fB}
\ee
$q_\mu$ being the $P$ four-momentum,
one obtains the following expression for the decay width: 
\be
\Gamma(P \to l \bar{\nu}_l) = \frac{G_F^2}{8\pi} |V_{qQ}|^2
m_Pm_l^2 \left(1-\frac{m_l^2}{m_P^2}\right)^2 f_P^2 ~.
\label{width}
\ee
The decay constant $f_P$
characterizes the size of the $P$-meson wave function at the origin,
and therefore the annihilation probability. The proportionality
to the square of the lepton mass $m_l$ is enforced by helicity 
conservation leading to a strong suppression of decays into light
leptons. 

From the measurement of a given leptonic
decay width one can directly determine
the product of a specific CKM matrix element and
decay constant. Unfortunately, 
because of multiple suppression it is difficult to measure 
leptonic decays. So far, only 
the mode $D_s \to \bar{\mu} \nu_{\mu}$ has been observed
\cite{fBDexp}, while a bound \cite{m3} exists
on the Cabibbo-suppressed decay $D \to \bar{\mu} \nu_{\mu}$.
Whether or not it is possible to study $B \to \bar{\mu} \nu_{\mu}$
and $B \to \bar{\tau} \nu_{\tau}$  
at future $B$ factories is not completely clear \cite{bfact}.

Theoretically, the decay constants $f_P$ are extremely interesting quantities.
They represent the simplest hadronic matrix elements
and therefore provide crucial tests of nonperturbative methods in QCD.
Moreover, the decay constants are needed in sum rule analyses
of other hadronic
properties of $B$ and $D$ mesons, e.g., form factors.
Finally, $f_B$ and $f_{B_s}$ are important parameters of 
mixing and CP-violation in the $B$ system.
For these and other reasons which will be pointed out in the course 
of the lectures, accurate theoretical
predictions on the various decay constants are very desirable.

\subsection{QCD sum rule for decay constants}

The QCD sum rule estimate of $f_B$ is based
on an analysis of the two-point correlation function
\cite{RRY,6auth,BG,AE}
\be
\Pi(q^2)= i\int d^4x e^{iqx}\langle 0\mid
T\{\bar{q}(x)i\gamma_5 b(x), \bar{b}(0)i\gamma_5 q(0)\}
\mid 0\rangle ~,
\label{corr2}
\ee
$q$ being the external momentum. 
By inserting a complete set of states 
with $B$-meson quantum numbers between the currents in (\ref{corr2}) 
one obtains a formal hadronic representation of $\Pi (q^2)$.
The term arising  from the ground state $B$-meson is proportional to
$\langle 0\mid \bar{q}i\gamma_5 b\mid B \rangle
\langle B\mid \bar{b}i\gamma_5q\mid 0 \rangle$, i.e., to 
$f_B^2$ as can be seen from the relation
\be
m_b\langle 0 \mid\bar{u}i\gamma_5 b\mid B\rangle =m_B^2f_B ~
\label{fB2}
\ee
which is equivalent to (\ref{fB}).
The hadronic expression for (\ref{corr2}) 
can be written in the form of a dispersion relation:
\be
\Pi(q^2)=
\int_{m_B^2}^\infty
\frac{\rho (s)ds}{s-q^2}
\label{disp2}
\ee
with the spectral density 
\be
\rho (s)=\delta (s-m_B^2)\frac{m_B^4f_B^2}{m_b^2}
+ \rho^{h}(s)\Theta(s-s^h_0)~.
\label{spect2}
\ee
Obviously, the $\delta$-function term on the r.h.s. of (\ref{spect2})
represents the $B$ meson, while $\rho^h(s)$
and $s^h_0$ are the spectral density and threshold energy squared
of the excited resonances and continuum states, respectively.
In order to make the dispersion integral (\ref{disp2}) 
ultraviolet-finite, one actually has to subtract the first two
terms of the Taylor expansion of $\Pi(q^2)$ at $q^2=0$.
These subtraction terms are removed by Borel
transformation with respect to $q^2$:
\be
{\cal B}_{M^2}\Pi(q^2)=\lim_{\stackrel{-q^2,n \to \infty}{-q^2/n=M^2}}
\frac{(-q^2)^{(n+1)}}{n!}\left( \frac{d}{dq^2}\right)^n \Pi(q^2)
\equiv \Pi(M^2)~.
\label{Borel2}
\ee
With
\be
{\cal B}_{M^2}
\left(\frac1{s-q^2}\right)^k
=\frac1{(k-1)!}\left(\frac1{M^2} \right)^{k-1}e^{-s/M^2}
\label{Borel1}
\ee
and 
\be
{\cal B}_{M^2}(-q^2)^k=0 
\label{Borel22}
\ee
for $k \ge 0$ one readily finds
\be
\Pi(M^2)=\frac{m_B^4f_B^2}{m_b^2}e^{-m_B^2/M^2}+
\int_{s^h_0}^\infty \rho^h(s)e^{-s/M^2}ds~.
\label{Borel12}
\ee
We see that Borel transformation removes arbitrary
polynomials in $q^2$ and suppresses the
contributions from excited and continuum states 
exponentially relative to the ground-state contribution.
The second point is actually the main motivation for this 
transformation.

In the space-like momentum region, $q^2 < 0$, where one has no poles
and cuts associated with physical states
it is possible to calculate the
correlation function $\Pi(q^2)$ in QCD in terms of quark and gluon degrees
of freedom. To this end,
the $T$--product of currents in (\ref{corr2}) 
is expanded in a series of regular
local operators $\Omega_d$: 
\be
\Pi(q^2) = \sum_d C_d(q^2,\mu)\langle 0 \mid \Omega_d(\mu) \mid 0 \rangle~,
\label{ope112}
\ee
$d$ being the dimension of a given operator and 
$\mu$ being the renormalization scale.
The lowest-dimensional operators ($d\leq6$) are
as follows:
\be
\Omega_d = 1,~ \bar{q}q ,~ G^a_{\mu\nu}G^{a\mu\nu}, ~
\bar q\sigma_{\mu\nu} \frac{\lambda ^a}2 G^{a\mu\nu}q,~
(\bar{q}\Gamma_r q)( \bar{q}\Gamma_s q)~.
\label{oper2}
\ee
Here, $\lambda^a$ denotes the usual $SU(3)$-colour matrices, 
$G^a_{\mu\nu}$ is the gluon field strength tensor, and 
$\Gamma_t$ stands for a given
combination of Lorentz and colour matrices.
While the strong interaction effects at momenta larger than $\mu$ are included 
in the Wilson coefficients $C_d(q^2,\mu)$, the effects at momenta
smaller than $\mu$
are absorbed into the matrix elements of the operators 
$\Omega_d(\mu)$. Thus, if $\mu \gg \Lambda_{QCD}$ 
the coefficients depend only on 
short-distance dynamics, and can therefore be
calculated in perturbation
theory, while the long-distance effects are taken into 
account by the vacuum averages
$\langle 0 \mid \Omega_d \mid 0 \rangle \equiv \langle \Omega_d \rangle $.
These so-called condensates describe properties
of the full QCD vacuum and are process-independent.
At present, they can only be estimated in
some crude approximations. For this reason, the vacuum condensates 
are determined
empirically by fitting selected sum rules to experimental data.

It is essential for the whole approach that at $q^2 \ll m_b^2$
the expansion (\ref{ope112}) can be cut off after a few
terms. The reason is that the higher the dimension 
of $\Omega_d$, the more suppressed by inverse powers of 
$m_b^2-q^2$ is the corresponding Wilson coefficient 
$C_d$. In leading order and up to $d=6$, the coefficients can be calculated
from the diagrams depicted in Fig. 1 as explained 
in \cite{RRY} (see also \cite{review}). In order to improve
the accuracy of the OPE one can further 
include perturbative QCD corrections to the Wilson coefficients
originating from hard gluon exchanges in the diagrams of Fig. 1.
Most important are the $O(\alpha_s)$ effects on the
coefficient $C_0$ \cite{RRY,BG} given by
the two-loop diagrams of Fig. 2.

\begin{figure}[ht]
\mbox{
\epsfig{file=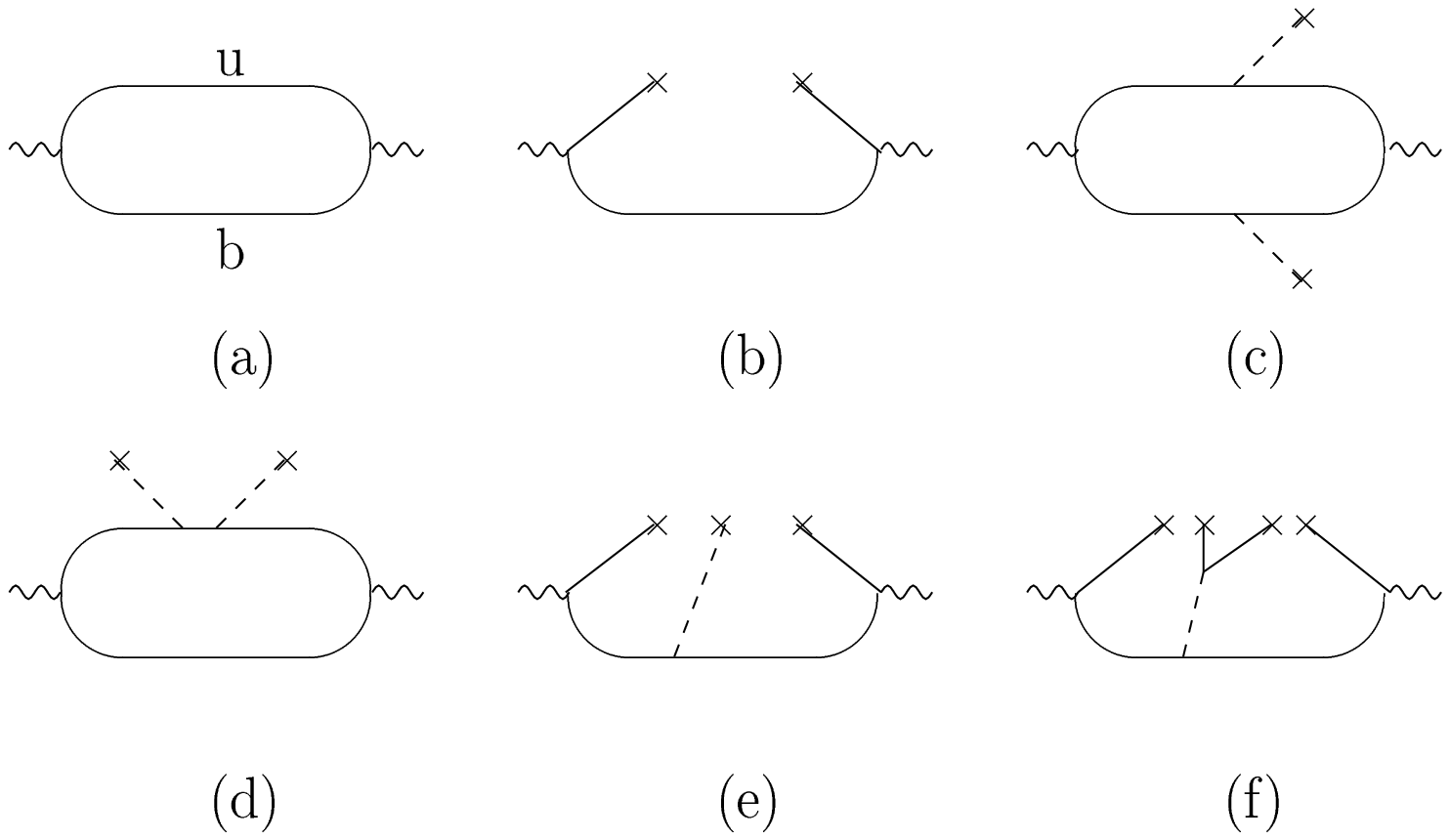,width=\textwidth,bbllx=0pt,bblly=260pt,bburx=600pt,%
bbury=570pt,clip=}
}
\caption{\it Diagrams determining the Wilson coefficients
in the OPE of the two-point correlation function (\ref{corr2}):
$C_0$ (a), $C_3$ (b), $C_4$ (c,d), $C_5$ (b,e), $C_6$ (b,f).
Solid lines denote quarks, dashed lines gluons, wavy lines
external currents. Crosses indicate vacuum fields.}
\end{figure}

\begin{figure}[ht]
\mbox{
\epsfig{file=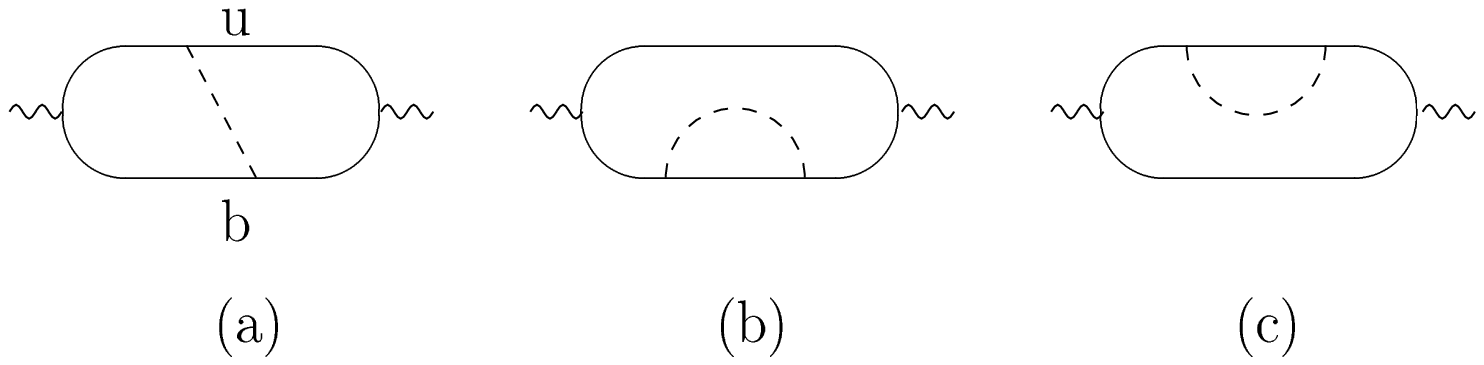,width=\textwidth,bbllx=0pt,bblly=340pt,bburx=600pt,%
bbury=480pt,clip=}
}
\caption{\it Feynman diagrams of the
$O(\alpha_s)$ correction to the Wilson coefficient $C_0(q^2)$.}
\end{figure}

The complete result for the Borel-transformed 
correlation function (\ref{corr2}) obtained in \cite{AE} 
is given by
$$
\Pi(M^2)=
\frac{3}{8\pi^2}\int_{m_b^2} ^{\infty}ds
\frac{(s-m_b^2)^2}{s}
\left(1+ \frac{4\alpha_s}{3\pi}f(s,m_b^2)\right)
\exp(-\frac{s}{M^2})
$$
$$
+\Bigg(-m_b\langle \bar q q \rangle +
\frac1{12}\langle  \frac{\alpha_s}{\pi}
G^a_{\mu\nu}G^{a\mu\nu} \rangle
-\frac{m_b}{2M^2}(1-\frac{m_b^2}{2M^2})
m_0^2\langle \bar q q \rangle
$$
\be
-\frac{16\pi\alpha_s}{27M^2}
(1-\frac{m_b^2}{4M^2}-\frac{m_b^4}{12M^4}) 
\langle \bar q q \rangle^2
\Bigg) \exp(-\frac{m_b^2}{M^2})~,
\label{ope12}
\ee
where
$$
f(s,m_b^2) = \frac94 - 2\int^{\frac{m_b^2}s}_0\frac{dt}tln(1-t)
+
\ln\frac{s}{m_b^2}\ln\frac{s}{s-m_b^2}
$$
\be
+ \frac32 \ln\frac{m_b^2}{s-m_b^2}+\ln\frac{s}{s-m_b^2}
+\frac{m_b^2}s \ln\frac{s-m_b^2}{m_b^2} +
\frac{m_b^2}{s-m_b^2}\ln\frac{s}{m_b^2}~.
\label{oalphas}
\ee
The $b$-quark mass appearing in the perturbative contribution 
(first term in (\ref{ope12})) is defined to be the pole mass, 
while the choice of $m_b$ in the leading-order coefficients of the 
contributions from the higher-dimensional operators is arbitrary. 
Usually, the pole mass is used everywhere. 
In order to keep the formula readable,
the scale dependence of the running coupling 
and the vacuum condensates is not made explicit
in the above. Furthermore, use has been made of the conventional
parametrization for the quark-gluon condensate density,
\be
\langle  \bar q \sigma_{\mu\nu}
\frac{\lambda^a}{2}G^{a\mu\nu}q \rangle =
m_0^2 \langle  \bar q q  \rangle~,
\label{qqbarG2}
\ee
and of vacuum saturation reducing four-quark condensates
to squares of quark condensates with known
coefficients \cite{SVZ}:
\be
\langle
 \bar{q}\Gamma_rq \bar{q}\Gamma_sq\rangle=
\frac{1}{(12)^2} \{ (Tr\Gamma_r)(Tr\Gamma_s)-Tr(\Gamma_r\Gamma_s) \}
\langle  \bar{q}q\rangle^2~.
\label{vac4q2}
\ee

The two representations of $\Pi(M^2)$, (\ref{Borel12}) and (\ref{ope12}),
yield an interesting equation which connects hadronic
properties with QCD parameters.
However, in order to determine the value of $f_B$
from it, one has to subtract the unknown contribution 
from the excited and continuum states to (\ref{Borel12}).
In a reasonable approximation based on quark-hadron duality, one may
estimate the integral over the hadronic spectral density $\rho^h$ 
by the corresponding integral over the perturbative density. 
This is achieved by substituting
\be
\rho^h(s)\Theta(s-s_0^h) = \frac1{\pi} \mbox{Im} C_0(s) \Theta (s-s_0^B)
\label{high2}
\ee
in (\ref{Borel12}), whereby $\sqrt{s_0^B}$ is treated as an effective 
threshold energy of the
order of the mass of the first excited $B$ resonance. In this
approximation, continuum
subtraction amounts to a simple 
change of the upper limit of integration in (\ref{ope12})
from $\infty$ to $s_0^B$.
The uncertainty from this rough procedure 
should not be too harmful because of the suppression 
of excited and continuum states after Borel transformation. 

Solving then the equation provided by (\ref{Borel12}) and (\ref{ope12})
for $f_B$, one obtains the following sum rule:
$$
f_B^2m_B^4=
\frac{3m_b^2}{8\pi^2}
\int_{m_b^2} ^{s_0^B}ds
\frac{(s-m_b^2)^2}s\left( 1+ \frac{4\alpha_s}{3\pi}f(s,m_b^2)\right )
\exp\left(\frac{m_B^2-s}{M^2}\right)
$$
$$
+m_b^2\Bigg \{-m_b \langle \bar q q \rangle \left(1+
\frac{m_0^2}{2M^2}\left(1-\frac{m_b^2}{2M^2}\right)\right)
+\frac1{12}\langle  \frac{\alpha_s}{\pi}
G^a_{\mu\nu}G^{a\mu\nu} \rangle
$$
\be
-\frac{16\pi}{27}\frac{\alpha_s\langle \bar q q \rangle^2}{M^2}
\left(1-\frac{m_b^2}{4M^2}-\frac{m_b^4}{12M^4}\right)
\Bigg\}
\exp\left(\frac{m_B^2-m_b^2}{M^2}\right)~.
\label{fB1}
\ee
Since the l.h.s. of this relation is a measurable quantity, 
any scale dependence on the r.h.s. must cancel, at least in principle. 
In practice, this can of course only be achieved approximately.

\subsection{Numerical results}

The numerical input in the sum rule (\ref{fB1}) for $f_B$ is as follows.
The pole mass of the $b$ quark is taken to be
\be
m_b=4.7 \pm 0.1 ~\mbox{GeV}~,
\label{bmass}
\ee
covering the range of 
estimates obtained from bottomonium sum rules \cite{mb},
while the continuum threshold in the $B$ channel
is fixed to the range
\be
s_0^B = 35 \mp 2 ~\mbox{GeV}^2 ~.
\label{s0B}
\ee
The scale $\mu$ at which 
$\alpha_s$ and the condensates are to be evaluated 
is still ambiguous in the approximation considered.  
A reasonable choice is 
\be
\mu^2 = O(M^2)
\label{scale}
\ee
characterizing the average virtuality of the quarks in the 
correlator (\ref{corr2}). Moreover, 
the products $m_b\langle \bar{q}q\rangle $ 
and $\langle \frac{\alpha_s}{\pi} G^a_{\mu\nu}G^{a\mu\nu} \rangle$
are renormalization-group invariant.
Using the numerical value of the
quark condensate density at $\mu = O(1~GeV)$ as derived
from the PCAC relation \cite{SVZ,Leutw}:
\be
\langle \bar q q  \rangle(1 ~\mbox{GeV}) = 
-\frac{f_\pi^2m_\pi^2}{2(m_u +m_d)}
\simeq-(240 ~\mbox{MeV})^3 ~,
\label{condens}
\ee 
and the running mass $m_b(1 GeV)$ corresponding to the central value 
(\ref{bmass}) of the pole mass, 
one gets
\be
m_b\langle \bar{q}q\rangle \simeq -0.084 ~\mbox{GeV}^4.
\ee
This is the estimate adopted for the present analysis 
together with the gluon condensate density 
determined from charmonium sum rules \cite{SVZ}:
\be
\langle \frac{\alpha_s}{\pi}
G^a_{\mu\nu}G^{a\mu\nu} \rangle \simeq 0.012 ~\mbox{GeV}^4~.
\label{glue}
\ee
The remaining parameters are assumed to be 
\be
m_0^2(\mbox{1 GeV}) \simeq 0.8  ~\mbox{GeV}^2
\label{m02}
\ee
as extracted from sum rules for light baryons \cite{BI},
and 
\be
\alpha_s\langle \bar{q}q\rangle^2 = 8\cdot 10^{-5} ~\mbox{GeV}^6
\label{4qc}
\ee
as given in \cite{AE}.
The scale dependence 
of $m_0^2 \langle \bar{q}q\rangle$ and  $\alpha_s\langle \bar{q}q\rangle^2$
is negligible. The numerical uncertainties in the condensate estimates
vary from 10 \% to about 50 \% or more. They nevertheless
play only a minor role for the total theoretical uncertainty on $f_B$.

%
\begin{table}[htb]
\setcounter{table}{2}
\caption{\label{tab2}
{\it Decay constants of
$B$ and $D$ mesons in MeV.}}
\begin{center}
\begin{tabular}{|c||c|c|c|c|c|}
\hline
&&&&&\\
Method & Ref.& $f_B$ & $f_{B_s}$ & $f_D$& $f_{D_s}$ \\
&&&&&\\
\hline
&&&&&\\
&$\stackrel{\mbox{this}}{\mbox{review}}$& 180 $\pm$ 30 & ~~~--&
190 $\pm$ 20&~~~-- \\
QCD sum rules &&&&&\\
&\cite{Dom}~$^{a)}$ & 175& 210 & $180 \pm 10$ & $220 \pm 10$ \\
&&&&&\\
\hline
&&&&&\\
& \cite{Flynn}& 175$\pm$ 25 & 200$\pm$ 25 & 205$\pm$ 15 &235 $\pm$ 15\\
&&&&&\\
Lattice &\cite{APE}&180$\pm$ 32 & 205$\pm$ 35 &221$\pm$ 17&237$\pm$ 16\\
&&&&&\\
&\cite{Wittig}&$172 ^{+27}_{-31}$ &$196^{+30}_{-35}$ &$191^{+19}_{-28} $
&$206^{+18}_{-28}$\\
&&&&&\\
\hline
&&&&&\\
&\cite{m3}&~~~--&~~~-- &$< 310$
&~~~--\\
Experiment&&&&&\\
&\cite{fBDexp}~$^{b)}$ &~~~-- &~~~-- &~~~-- &241 $\pm$ 21$\pm$ 30\\
&&&&&\\
\hline
\end{tabular}
\\
\end{center}
\hspace*{1.5cm}
$^{a)}$ update of results in \cite{RRY,AE,fB} taking 
        $m_b=4.67$ GeV, $s_0^B=35$ GeV$^2$, \\
\hspace*{1.9cm}
        and $m_c=1.3$ GeV, $s_0^D=5.5$ GeV$^2$.\\
\hspace*{1.5cm}
$^{b)}$ world average  \\
\end{table}

Next, one has to determine the range of values of the 
Borel parameter $M^2$ for which 
the sum rule (\ref{fB1}) can be trusted. 
On the one hand, $M^2$ has to be small enough such that 
the contributions from excited and continuum states are 
exponentially damped 
and the approximation (\ref{high2}) is acceptable. 
On the other hand, the scale $M^2$ must be large enough 
such that higher-dimensional operators are suppressed and 
the OPE converges sufficiently fast. Furthermore,
the sum rule should be stable under variations 
of $M^2$ in the allowed window. 
Whether or not these requirements can be met 
has to be investigated for each sum rule separately. 
In the case at hand, one finds 
that at
\be
3 ~\mbox{GeV}^2\leq M^2 \leq 5.5 ~\mbox{GeV}^2
\label{Borelint}
\ee
the necessary conditions are fulfilled:
the quark-gluon condensate contributes
less than 15 \%, and the four--quark operators a negligible amount,
while the excited and continuum states contribute less than 30 \%.
Moreover, $f_B$ changes insignificantly when $M^2$ varies
in the range (\ref{Borelint}). 

The predictions on $f_B$ and $f_{B_s}$ derived from the sum rule 
(\ref{fB1}) with the input specified above are summarized in Tab. 3 
together with the corresponding results for $D$ mesons
obtained with $m_c=1.3\pm 0.1 $ GeV, $s_0^D=6 \mp 1 $ GeV$^2$, and
1 GeV$^2 < M^2 < 2$ GeV$^2$.
The theoretical uncertainties are dominated by the uncertainties 
in the quark masses and continuum thresholds. For instance,
the ranges of values of $f_B$ and $f_D$ expressed by
the errors in the first row of Tab. 3
have been determined by varying $m_Q$ and $s_0^P$ simultaneously such that
the stability of the sum rule is preserved. The correlation 
is indicated by the reversed $\pm$ sign in (\ref{s0B}).
It should also be emphasized that the $O(\alpha_s)$ corrections
have a sizeable effect. Without these corrections the decay constants 
would be considerably smaller:
\be
\overline{f}_B \equiv f_B(\alpha_s=0)
=140 \pm 30 ~\mbox{MeV}~,
\label{fB0}
\ee
\be
\overline{f}_D \equiv f_D(\alpha_s=0)
=170 \pm 20 ~\mbox{MeV}~.
\label{fDres}
\ee
For comparison, Tab. 3 also shows recent lattice results and 
the available experimental data.
The mutual agreement within the uncertainties of the two theoretical 
approaches is satisfactory. On the other hand, the data are just beginning
to challenge theory.

%
\section{Semileptonic decays and form factors}

\subsection{Distribution in momentum transfer}

The weak hamiltonian (\ref{HL}) also applies to the semileptonic
decay $B \to \pi \bar{l} \nu_l$. 
The appropriate hadronic matrix element 
is parametrized by two independent form factors: 
\be
\langle\pi(q)|\bar{u} \gamma_\mu b |B(p+q)\rangle=
2f^+(p^2)q_\mu +
\left(f^+(p^2)+f^-(p^2)\right)p_\mu
\label{form3}
\ee
with $p+q$, $q$ and $p$ being the $B$ and $\pi$ four-momenta, and the
momentum transfer, respectively.
Frequently, in particular in nonleptonic decays, one also uses
the scalar form factor $f^0$ defined by  
\be
f^0(p^2) = f^+(p^2) + \frac{p^2}{m_B^2-m_\pi^2} f^-(p^2) ~.
\label{f0}
\ee
The distribution of the momentum transfer squared 
in $B \to \pi \bar{l} \nu_l$
is given by
$$
\frac{d\Gamma}{dp^2} =
\frac{G^2|V_{ub}|^2}{24\pi^3}
\frac{(p^2-m_l^2)^2 \sqrt{E_\pi^2-m_\pi^2}}{p^4m_B^2}
\Bigg\{ \left(1+\frac{m_l^2}{2p^2}\right)
m_B^2(E_\pi^2-m_\pi^2)\left[f^+(p^2)\right]^2
$$
\be
+ \frac{3m_l^2}{8p^2}(m_B^2-m_\pi^2)^2
\left[f^0(p^2)\right]^2\Bigg\} ~,
\label{dG}
\ee
where $E_\pi= (m_B^2+m_\pi^2 -p^2)/2m_B$ is
the pion energy in the $B$ rest frame. 

For $l = e, \mu$,
the form factor $f^0$ plays a negligible role 
because of the smallness of the lepton masses.
These channels can thus be used to 
directly determine the product $|V_{ub} f^+(p^2)|$,
while the heavy mode $B \to \pi \bar{\tau} \nu_{\tau}$ 
offers a possibility to test the scalar form factor $f^0(p^2)$. 
Recently, the decays $B\to \pi \bar{l} \nu_l$
and $B\to \rho \bar{l} \nu_l$ have been observed by CLEO \cite{CLEO}.
If the relevant form factors can be calculated with sufficient
accuracy these measurements eventually provide interesting 
alternatives to the determination of $V_{ub}$ from inclusive $b\to u$
transitions. In these lectures, the sum rule method is advocated.
It is one of the virtues of the sum rule calculations, 
in comparison to other approaches,  that they can be
tested and consistently improved by investigating the analogous $D$-meson
form factors. The already existing experimental data on 
$D\to h \bar{l} \nu_l$, $h=K, K^*,\pi,\rho,\eta$ and $l=e,\mu$ 
\cite{charm} have not yet been fully exploited in this respect.

\subsection{Light-cone vs. short-distance expansion}

For processes involving a light meson, e.g., a $\pi$, $K$, or $\rho$
there is an interesting alternative  
to the short-distance OPE of vacuum-vacuum correlation 
functions in terms of condensates,
namely the expansion 
of vacuum-meson correlators near the light-cone 
in terms of meson wave functions \cite{exclusive,BL,CZ}.
The latter are defined by matrix elements of composite operators
and classified by twist, 
similarly as the deep-inelastic structure functions.
The light-cone variant of QCD sum rules has been 
suggested in \cite{BBK,BF,CZ1}. In this approach,
the light mesons are described by means of light-cone
wave functions, whereas
the heavy meson channels are treated in the 
usual way: choice of a generating current,
dispersion relation, Borel transformation and continuum subtraction.

The general idea is explained in the following using 
the $B \rightarrow \pi$ matrix element (\ref{form3}) as an example.
The detailed derivation of the light-cone sum rules
for $f^+$ and $f^-$ is reported in \cite{BKR,BBKR,KRW} and
reviewed in \cite{review}.
The central object is the vacuum-pion correlation function 
\be
F_\mu (p,q)=
i \int d^4xe^{ipx}\langle \pi(q)\mid T\{\bar{u}(x)\gamma_\mu b(x),
\bar{b}(0)i\gamma_5 d(0)\}\mid 0\rangle
\label{1a4}
\ee
$$
= F(p^2,(p+q)^2) q_\mu + \tilde{F}(p^2,(p+q)^2) p_\mu~.
$$
From the comparison with (\ref{form3}) it is clear that
the two invariant functions $F$ and $\tilde{F}$ have some kind
of relation with the form factors $f^+$ and $f^+ + f^-$, respectively.
Since the pion is on-shell, $q^2 = m_\pi^2$ vanishes in the
chiral limit adopted throughout this discussion.
For the momenta in the $\bar{b}d$ and $\bar{u}b$ channels 
one requires  $(p+q)^2 \ll m_b^2$ and 
$p^2 \leq m_b^2 -2 m_b \chi$, 
$\chi$ being a $m_b$-independent scale of order $\Lambda_{QCD}$.
This assures that the $b$ quark is far off mass-shell.
Contracting the $b$-quark fields in (\ref{1a4}) 
and inserting the free $b$-quark propagator, 
one gets
\bq
F_\mu(p,q)&=&i\int \frac{
d^4x\,d^4k}{(2\pi )^4(m_b^2-k^2)}
e^{i(p-k)x}\left(m_b
\langle \pi (q)|\bar{u}(x)\gamma_\mu\gamma_5d(0)|0\rangle \right.
\nonumber
\\
&&{}+\left.
k^\nu \langle\pi(q) |\bar{u}(x)\gamma_\mu\gamma_\nu\gamma_5
d(0)|0\rangle\right)~.
\label{234}
\eq
This contribution is diagrammaticly depicted in Fig.~3a.
It is important to note that
the path-ordered gluon operator
\be
Pexp \left\{ig_s \int ^1_0d\alpha~ x_\mu A^\mu (\alpha x)\right \}
\label{factor}
\ee
necessary for gauge invariance of the matrix elements in (\ref{234}),
is unity in the light-cone gauge, $x_\mu A^\mu=0$,
assumed here. 

\begin{figure}[ht]
\mbox{
\epsfig{file=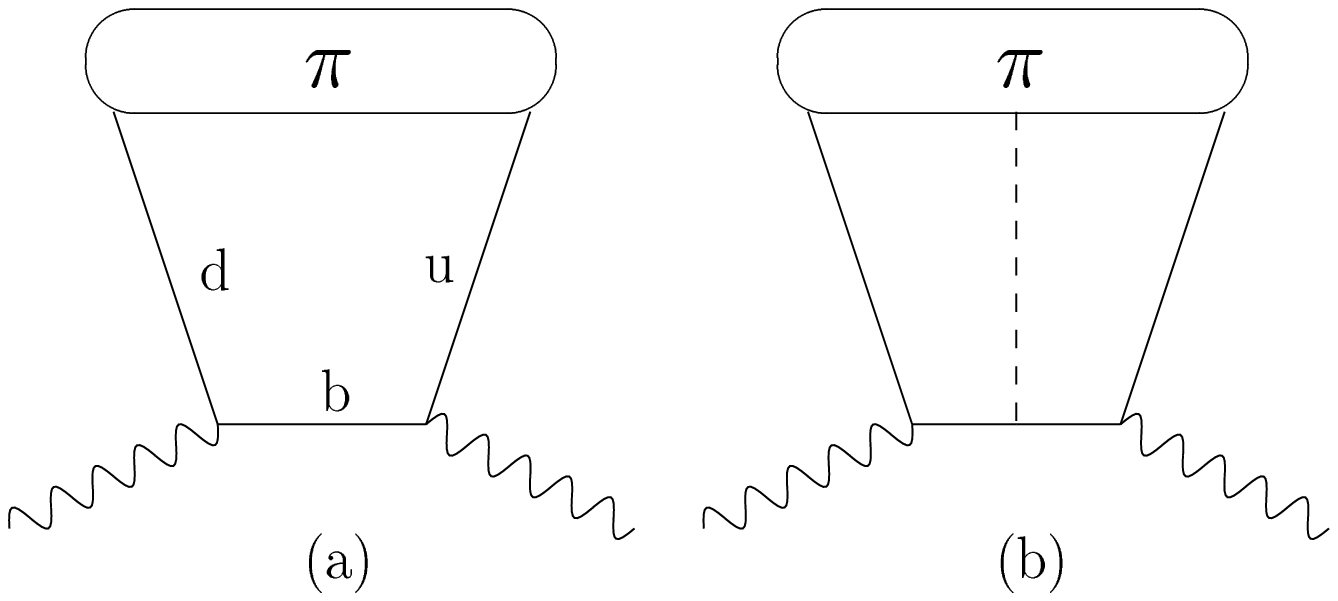,width=\textwidth,bbllx=0pt,bblly=290pt,bburx=600pt,%
bbury=520pt,clip=}
}
\caption{\it Diagrammatic representation of the correlation
function (\ref{1a4}): terms involving two-particle (a) 
and three-particle (b) wave functions.
Solid lines represent quarks, dashed lines gluons,
wavy lines external currents, and
ovals light-cone wave functions of the pion.}                       
\end{figure}

Let us first consider the matrix element
\be
\langle \pi (q)|\bar{u}(x)\gamma_\mu\gamma_5 d(0)|0\rangle ,
\label{matrix4}
\ee
and expand the bilocal quark-antiquark
operator around $x=0$: 
\be
\bar{u}(x)\gamma_\mu\gamma_5d(0)=\sum_r\frac{1}{r!}
\bar{u}(0)(\stackrel{\leftarrow}{D}\cdot x)^r\gamma_\mu\gamma_5 d(0)~.
\label{expan4}
\ee
The matrix elements of the local operators
can be written in the form  
\be
\langle\pi(q) |\bar{u}\stackrel{\leftarrow}{D}_{\alpha_1}
\stackrel{\leftarrow}{D}_{\alpha_2}...
\stackrel{\leftarrow}{D}_{\alpha_r}\g_\mu\g_5 d |0\rangle        
 =(i)^r q_\mu q_{\alpha_1} q_{\alpha_2}...q_{\alpha_r}M_r + ...~,
\label{local4}
\ee
where the ellipses stand for 
additional terms containing the  
metric tensor $g_{\alpha_i\alpha_k}$ in all possible combinations.
Note that the lowest twist \footnote{
Twist is defined as the difference between
the canonical dimension and the spin of a traceless
and totally symmetric local operator.} in (\ref{local4})
is equal to 2.
Substituting (\ref{expan4}) and (\ref{local4}) in the first
term of (\ref{234}), integrating over $x$ and $k$,
and comparing the result with (\ref{1a4}), one obtains
\bq
F(p^2,(p+q)^2)=i\frac{m_b}{m_b^2-p^2}
\sum_{r=0}^\infty \xi ^rM_r~
\label{expans4}
\eq
with 
\bq
\xi =\frac{2(p \cdot q)}{m_b^2-p^2}= \frac{(p+q)^2-p^2}{m_b^2-p^2}~.
\label{ksi4}
\eq
Obviously, if the ratio $\xi$
is finite one must keep an infinite series of matrix elements
in (\ref{expans4}). All of them give
contributions of the same order in the heavy quark propagator
$1/(m_b^2-p^2)$, differing only by powers
of the dimensionless parameter $ \xi$.
In other words,
the expansion (\ref{expan4})
is useful only in the case  $ \xi \rightarrow 0$,
i.e., $p^2 \ra (p+q)^2$
or equivalently, when the pion momentum vanishes.
In this case, the series in
(\ref{expans4}) can be truncated after a few terms 
involving only a manageable
number of unknown matrix elements $M_r$. However, generally, when 
$p^2 \neq (p+q)^2$ one has to sum up the infinite series of matrix elements
of local operators in some way.

One possible solution of this problem is provided by the light-cone
expansion of the bilocal operator (\ref{expan4}). 
For the matrix element (\ref{matrix4}) the
first term of this expansion is given by 
\bq
\langle\pi(q)|\bar{u}(x)\g_\mu\g_5d(0)|0\rangle_{x^2=0}=
-iq_\mu f_\pi\int_0^1du\,e^{iuqx}\varphi_\pi (u)  ~.
\label{pionwf4}
\eq
The function $\varphi_\pi(u)$ is known as the twist 2
light-cone wave function of the pion.
It represents the distribution in the fraction
of the light-cone momentum $q_0 +q_3 $ of the pion carried by
a constituent quark. As can be seen by putting
$x=0$ in the above, $\varphi_\pi(u)$ 
is normalized to unity. Substitution of (\ref{pionwf4}) in (\ref{234})
and integration over $x$ and $k$ yields 
\be
F(p^2,(p+q)^2)=m_bf_\pi\int_0^1\frac{du ~\varphi_\pi(u) }{m_b^2-(p+uq)^2} ~.
\label{Fzeroth4}
\ee
We see that the infinite series of matrix elements of local operators
encountered before in (\ref{expans4})
is effectively replaced by a wave function.
Formal expansion of (\ref{Fzeroth4}) in $q$,
\be
F(p^2,(p+q)^2)=m_bf_\pi\sum_{r=0}^{\infty}
\frac{(2p\cdot q)^r}{(m_b^2-p^2)^{r+1}}
\int_0^1du ~u^r\varphi_\pi(u)~,
\label{Fexp}
\ee
and comparison with (\ref{expans4}) yields the
relation between the matrix elements $M_r$
defined in (\ref{local4})
and the moments of $\varphi_\pi(u)$:
\be
M_r= -if_\pi
\int_0^1du ~u^r\varphi_\pi(u).
\label{local}
\ee

Including the next-to-leading terms in $x^2$, the light-cone
expansion  of the matrix element (\ref{matrix4})
is given by
\begin{eqnarray}
\langle\pi(q)|\bar{u}(x)\gamma_\mu\gamma_5d(0)|0\rangle =
-iq_\mu f_\pi\int_0^1du\,e^{iuqx}
\left(\varphi_\pi (u)+x^2g_1(u)\right)
\nonumber \\
+
f_\pi\left( x_\mu -\frac{x^2q_\mu}{qx}\right)\int_0^1
du\,e^{iuqx}g_2(u),
\label{phi}
\end{eqnarray}
where  $g_1$
and $g_2$ are twist~4 wave functions.
Proceeding to the second term
in (\ref{234}) and using the relation  
\be
\gamma_\mu\gamma_\nu\gamma_5= g_{\mu\nu}\gamma_5-
i\sigma_{\mu\nu}\gamma_5~,
\label{gamma1}
\ee
one encounters two further matrix elements:
\be
\langle\pi(q)\mid \bar{u}(x)i\gamma_5d(0)\mid 0\rangle=
f_\pi \mu_\pi \int_0^1du~e^{iuqx}\varphi_{p}(u) 
\label{phip}
\ee
and
\be
\langle\pi(q)\mid\bar{u}(x)\sigma_{\mu\nu}\gamma_5d(0)\mid 0\rangle=
i(q_\mu x_\nu -q_\nu x_\mu )\frac{f_\pi \mu_\pi}{6}
\int_0^1 du ~e^{iuqx}\varphi_{\sigma }(u) \, 
\label{phisigma}
\ee
with $\mu_\pi=m_\pi^2/(m_u + m_d)$. 
Only the leading terms of the expansion are considered here. 
They have twist 3 and are parameterized by the
wave functions $\varphi_p$ and $\varphi_\sigma$.

Beyond twist 2, one should also take into account
the higher-order term illustrated graphically in Fig. 3b,
where a gluon is emitted from the heavy quark line. This correction
is determined by quark-antiquark-gluon wave functions \cite{BB}.
Explicit expressions for this contribution can be found in 
\cite{BKR,BBKR} and also in \cite{review}. Below, this term is
denoted by $F_G$. 
Gluon radiation from the light quark lines in Fig. 3a  is effectively
included in the twist 3 and 4 wave functions as was shown in
\cite{BF,Gorsky,BF2}. 
Components of the pion wave function with two extra gluons,
or with an additional $\bar q q$ pair
are neglected.
Including all operators up to twist 4 one obtains the following results  
for the two invariant functions defined in (\ref{1a4}):
\bq
F(p^2,(p+q)^2)&=&f_\pi\int\limits_{0}^{1}\frac{du}{m_b^2-(p+uq)^2}
\Bigg\{m_b\varphi_\pi (u)
\nonumber\\
&&{}+\mu_\pi\Bigg[ u\varphi_p(u)+\frac16 \left(
2+\frac{p^2+m_b^2}{m_b^2-(p+uq)^2}\right)\varphi_\sigma (u)\Bigg]
\nonumber\\
&&{}+m_b\left[\frac{2ug_2(u)}{m_b^2-(p+uq)^2}-
\frac{8m_b^2}{(m_b^2-(p+uq)^2)^2}\left(g_1(u) -\int^u_0dv 
g_2(v)\right)\right]\Bigg\}
\nonumber\\
&&{} + F_G(p^2,(p+q)^2)
\label{F}
\\
\widetilde{F}(p^2,(p+q)^2)&=&f_\pi\int\limits_{0}^{1} \frac{du}
{m_b^2-(p+uq)^2}\Bigg\{\mu_\pi\varphi_{p}(u)
+\frac{\mu_\pi \varphi_{\sigma }(u) }{6u}
\nonumber\\
&&{}\times\Bigg[ 1-\frac{m_b^2-p^2}{m_b^2-(p+uq)^2}\Bigg]
+ \frac{2m_b g_2(u)}{m_b^2-(p+uq)^2} \Bigg\} ~.
\label{Ftilde}
\eq
We see that every power of $x^2$
in the light-cone expansion of the integrand in (\ref{1a4})
leads to an additional power of
$m_b^2-(p+uq)^2$ in the denominators of the coefficients.
This justifies the neglect of higher-twist operators,
provided $(p+q)^2$ and 
$p^2$ are sufficiently smaller than $m_b^2$. 
It is also interesting to note that there are no contributions 
from twist 2 and from three-particle wave functions
to $\tilde{F}$.

\subsection{Light-cone wave functions} 

Before proceeding with the derivation of the light-cone sum rules
for the form factors $f^{\pm}$, it may be useful to make a few 
more remarks on the light-cone wave functions.  
Firstly, these functions are universal, a property
which is essential for the whole
approach, similarly as the universality of the
vacuum condensates in the case of short-distance sum rules.
Secondly, the asymptotic form of the wave functions is 
given by conformal symmetry and asymptotic freedom of 
QCD \cite{CZ,BF2}, while the nonasymptotic
features incorporate long-distance quark-gluon interactions.
Thirdly, the factorization of  
collinear logarithms generated by gluon radiation 
and the renormalization procedure
induce scale dependences. Usually, 
a common scale $\mu$ is chosen for simplicity.

In leading-order approximation (LO), the twist 2 wave function
$\varphi_\pi(u,\mu)$ obeys
the Brodsky-Lepage evolution equation \cite{exclusive}:
\begin{equation}
\frac{d\varphi_\pi (u,\mu)}{d\ln \mu} = \int\limits_{0}^{1} dw
V(u,w)\varphi_\pi(w,\mu)~,
\label{BLL}
\end{equation}
where 
\begin{eqnarray}
V(u,w)  = \frac{\alpha_s(\mu) C_{F}}{\pi} \left[ \frac{1-u}{1-w}
\left( 1 + \frac{1}{u-w} \right) \Theta(u-w)
+ \frac{u}{w} \left( 1 + \frac{1}{w-u} \right)
 \Theta(w-u) \right]_{+}
\label{BL}
\end{eqnarray}
and the $+$ regularization is defined by
\begin{eqnarray}
 R(u,w)_{+} & = & R(u,w) - \delta(u-w) \int\limits_{0}^{1} R(v,w) dv ~.
\label{plus}
\end{eqnarray}
The evolution equation effectively sums up the leading logarithms
to all orders. 
It can be solved by expanding $\varphi_\pi(u,\mu)$ in
terms of Gegenbauer polynomials 
\begin{equation}
\varphi_\pi(u,\mu) = 6 u(1-u)\Big[1 + 
  \sum_i a_{2i}(\mu)C^{3/2}_{2i}(2u-1)\Big]\,,
\label{expansion}
\end{equation}
in which case the coefficients $a_{2i}(\mu)$
are multiplicatively renormalizable \cite{exclusive}.
The scale dependence is given by
\be 
a_n(\mu)=a_n(\mu_0)\left( 
\frac{\alpha_s(\mu)}{\alpha_s(\mu_0)}\right)^{\gamma_n/b}~,
\label{anom}
\ee 
where $b=11- 2n_f/3$ is the 1-loop coefficient of the QCD beta function,
$n_f$ being the number of active flavours, and
\be
\gamma_{n}=C_F\left[-3 -\frac{2}{(n+1)(n+2)}+4\left(\sum_{k=1}^{n+1}
\frac1k\right)\right]
\label{gamman}
\ee
are the anomalous dimensions \cite{CZ}.
We see that $a_n(\mu)$, $n \geq 2$ vanishes for
$\mu \rightarrow \infty $. Therefore,
these terms describe the nonasymptotic effects in the
wave function (\ref{expansion}).
The $\mu$-independent term is the asymptotic wave function, 
$\varphi_\pi(u,\mu = \infty) = 6u(1-u)$. As already pointed out,
the  normalization is such that
\be
\int^1_0 du \varphi_\pi(u,\mu)=1 ~.
\ee

The initial values of the nonasymptotic coefficients 
can be estimated from two-point sum rules \cite{CZ} for the moments
$\int u^n \varphi_\pi(u,\mu) du$ at low $n$. 
The nonperturbative information encoded 
in the quark and gluon condensates is 
thereby transmuted into the long-distance properties of the wave function.
Alternatively, one can determine the coefficients directly
from light-cone sum rules for known hadronic quantities
such as the $\pi NN$ and $\omega\rho\pi$ couplings.
Here, 
the estimates at $\mu_0=0.5$ GeV given in \cite{BF} are adopted:
\be
a_2(\mu_0)=\frac23, ~a_4(\mu_0)=0.43~.
\label{aa}
\ee 
The corresponding Gegenbauer polynomials are given by
\bq
C_2^{3/2}(2u-1)&=&\frac{3}2[5(2u-1)^2-1]~,
\nonumber\\
C_4^{3/2}(2u-1)&=&\frac{15}8[21(2u-1)^4-14(2u-1)^2+1]~.
\label{GP}
\eq
On the basis of the approximate conformal symmetry 
of QCD it has been shown \cite{BF2} that the expansion (\ref{expansion})
converges sufficiently fast so that the terms with $n>4$ are
negligible. 

For completeness, the twist 3 and 4 two-particle light-cone wave functions
of the pion are listed below \cite{CZ,BF}:
\begin{eqnarray}
\varphi_p(u,\mu)&=&1+B_2(\mu)\frac12(3(u-\bar{u})^2-1)
+B_4(\mu)\frac18(35(u-\bar{u})^4
\nonumber\\
&&{}-30(u-\bar{u})^2+3) ~,
\label{bbb}
\\
\varphi_\sigma (u,\mu)&=&6u\bar{u}\Big[ 1+C_2(\mu)\frac{3}2(5(u-\bar{u})^2-1)
\nonumber\\
&&{}+C_4(\mu)\frac{15}8(21(u-\bar{u})^4-14(u-\bar{u})^2+1)\Big] ~,
\label{tw3}
\\
g_1(u,\mu)&=&\frac{5}2\delta^2(\mu)\bar{u}^2u^2+\frac{1}{2}\varepsilon(\mu)
\delta^2(\mu)[\bar{u}u(2+13\bar{u}u)+10u^3\ln u(2-3u+\frac65u^2)
\nonumber\\
&&{}+10\bar{u}^3\ln \bar{u}(2-3\bar{u}+\frac65\bar{u}^2)] ~,
\label{g1}
\\
g_2(u,\mu)&=&\frac{10}3\delta^2(\mu)\bar{u}u(u-\bar{u}) ~, 
\label{g2}
\\
\int\limits_{0}^{u} dv g_2(v,\mu) &=&\frac{5}3\delta^2(\mu)\bar{u}^2u^2 
\label{G2}
\end{eqnarray}
with $\bar{u}=1-u$ and
\begin{eqnarray}
B_2=30\frac{f_{3\pi}}{\mu_\pi f_\pi}\,, ~~
B_4=\frac32\frac{f_{3\pi}}{\mu_\pi f_\pi}
(4\omega_{2,0}-\omega_{1,1}-2\omega_{1,0})\,,
\nonumber\\
C_2=\frac{f_{3\pi}}{\mu_\pi f_\pi}(5-\frac12\omega_{1,0})\,,~~
C_4=\frac1{10}\frac{f_{3\pi}}{\mu_\pi f_\pi}(4\omega_{2,0}-\omega_{1,1})~.
\label{ccc}
\end{eqnarray}
The additional parameters appearing in the above are numerically given by
$$ 
f_{3\pi}(1\mbox{GeV}) 
= 0.0035~\mbox{\rm GeV}^2~, 
$$
$$
\omega_{1,0}(1\mbox{GeV}) = -2.88\,
,~~ \omega_{2,0}(1\mbox{GeV})= 10.5\,,~~ 
\omega_{1,1}(1\mbox{GeV}) = 0\,,
$$
\be
\delta^2(1 \mbox{GeV})=0.2 ~ \mbox{GeV}^2\,,~~ 
\varepsilon(1 \mbox{GeV}) =0.5~. 
\label{varepsil}
\ee
For the quark-antiquark-gluon wave functions the reader is referred to
\cite{BB} and \cite{review}.

\subsection{Sum rules for heavy-to-light form factors}

Now we are ready to proceed with the discussion of the $B\to \pi$ 
transition form factors.
In order to determine $f^{\pm}(p^2)$ from the
correlation function $F_{\mu}(p,q)$ derived in (\ref{1a4}), 
we apply QCD sum rule methods 
to the $B$-meson channel following essentially the same
steps as in the derivation of  the sum rule for $f_B$ in sect. 2.2.
The hadronic representation of (\ref{1a4}) is obtained by
inserting a complete set
of intermediate states with $B$-meson quantum numbers
between the currents on the r.h.s. of (\ref{1a4}).
Using the matrix elements (\ref{fB2}) and (\ref{form3}),
and representing the sum over
excited and continuum states by a dispersion integral
with the spectral density $\rho^h_\mu = \rho^h q_\mu + \tilde{\rho}^hp_\mu$,
one obtains the following relations for the invariant amplitudes
$F$ and $\tilde{F}$:
\be
F(p^2,(p+q)^2)=
\frac{2m_B^2f_Bf^+(p^2)}{m_b(m_B^2-(p+q)^2)}
+\int\limits_{s_0^h}^{\infty}
\frac{\rho^h (p^2,s)ds}{s-(p+q)^2}~,
\label{dispa}
\ee
\be
\widetilde{F}(p^2,(p+q)^2)=
\frac{m_B^2 f_B(f^+(p^2)+f^-(p^2))}{m_b(m_B^2-(p+q)^2)}
+\int\limits_{s_0^h}^{\infty} ds \frac{\tilde{\rho}^h(p^2,s)}{s-(p+q)^2}~.
\label{hadr4}
\ee
Similarly as in (\ref{Borel12}),
the integrals over 
$\rho^h$ and $\tilde{\rho}^h$ are approximated by corresponding
integrals over the spectral densities 
calculated from the light-cone 
expansion of (\ref{1a4}) using the substitution
\be
\rho^h(p^2,s) \Theta(s - s_0^h)
 = \frac1\pi \mbox{Im} F(p^2, s) \Theta( s-s_0^B) ~,
\label{rhoh5}
\ee
and the analogous relation for $\tilde{\rho}_h$ and 
$\mbox{Im}\tilde{F}$.
The calculation of the imaginary parts of
$F$ and $\tilde{F}$ from the functions 
given in (\ref{F}) and (\ref{Ftilde}), respectively,
is outlined, e.g., in \cite{review}.
With the above duality approximation 
it is again rather straightforward to subtract
the continuum contributions from 
the l.h.s. of (\ref{dispa}) and (\ref{hadr4}).
After performing 
the Borel transformation in $(p+q)^2$, one arrives at the   
sum rules \cite{BKR,BBKR}
$$
f_Bf^+( p^2)=\frac{f_{\pi} m_b^2}{2m_B^2}
\exp \left( \frac{m_B^2}{M^2}\right)
\Bigg\{
\int\limits_{\Delta}^{1}\frac{du}{u}
exp\left[-\frac{m_b^2-p^2(1-u)}{uM^2} \right]
$$
$$
\times\Bigg( \varphi_\pi(u)
+ \frac{\mu_\pi}{m_b}
\Bigg[u\varphi_{p}(u) + \frac{
\varphi_{\sigma }(u)}{3}\left(1 + \frac{m_b^2+p^2}{2uM^2}\right) \Bigg]
$$
$$
- \frac{4m_b^2 g_1(u)}{ u^2M^4 }
+\frac{2}{uM^2} \int\limits_{0}^{u} g_2(v)dv
\left(1+ \frac{m_b^2+p^2}{uM^2} \right ) \Bigg)+ t^+(s_0^B,p^2,M^2)
$$
\be
+f^+_G(p^2,M^2)
+ \frac{\alpha_sC_F}{4\pi}\delta^+(p^2,M^2) \Bigg\} ~,
\label{fplus}
\ee
and \cite{KRW} 
$$
f_B(f^+(p^2) + f^-(p^2))=\frac{f_\pi \mu_\pi m_b}{m_B^2 }\exp\left(
\frac{m_B^2}{M^2}\right)\Bigg\{
\int\limits_{\Delta}^{1}
\frac{du}{u}
\exp \left[ - \frac{m_b^2-p^2(1-u)}{u M^2}\right]
$$
\be
\times\Bigg[
\varphi_{p}(u)
+ \frac{\varphi_\sigma(u)}{6u}
\left(1-
\frac{m_b^2-p^2}{uM^2}\right)
+ \frac{2m_bg_2(u)}{\mu_\pi uM^2}\Bigg]+ t^{\pm}(s_0^B,p^2,M^2)\Bigg\} ~.
\label{fplusminus}
\ee
The sum rule for the scalar form factor $f^0$ follows from the relation 
(\ref{f0}).
In the above, $t^+$ and $t^{\pm}$ are numerically small corrections
left from continuum subtraction, and  
$f^+_G$ denotes the contribution from the 
three-particle wave functions.
The explicit expressions of these terms are given in \cite{BBKR,KRW}.
Also indicated is the perturbative QCD correction $\delta^+$ to the 
leading twist 2 term in $f^+$ which has been calculated only recently 
\cite{KRWY,Bagan}. Other
perturbative corrections are still unknown.

Concerning the $O(\alpha_s)$ correction to the leading twist 2 result,
a few comments are in order.
Hard gluon exchanges give rise to radiative effects
in the correlation function (\ref{1a4}) 
which can be calculated perturbatively.
The first-order diagrams are shown in Fig. 4. 
The expression (\ref{Fzeroth4}) of the invariant
function $F$ can be more generally written  
as a convolution of a hard scattering 
amplitude $T$ with the twist 2 wave function:
\begin{eqnarray}\label{represent}
F(p^2,(p+q)^2)=f_\pi\int^1_0 du \varphi_\pi (u,\mu) T(p^2,(p+q)^2,u,\mu)~,
\end{eqnarray}
where $T$ is determined by a perturbative series
\begin{eqnarray}
T(p^2,(p+q)^2,u,\mu)=T_0(p^2,(p+q)^2,u)+\frac{\alpha_s C_F}{4\pi}
T_1(p^2,(p+q)^2,u,\mu)
+ O(\alpha_s^2)~.
\label{T}
\end{eqnarray}

\begin{figure}[htb]
\centerline{
\epsfig{file=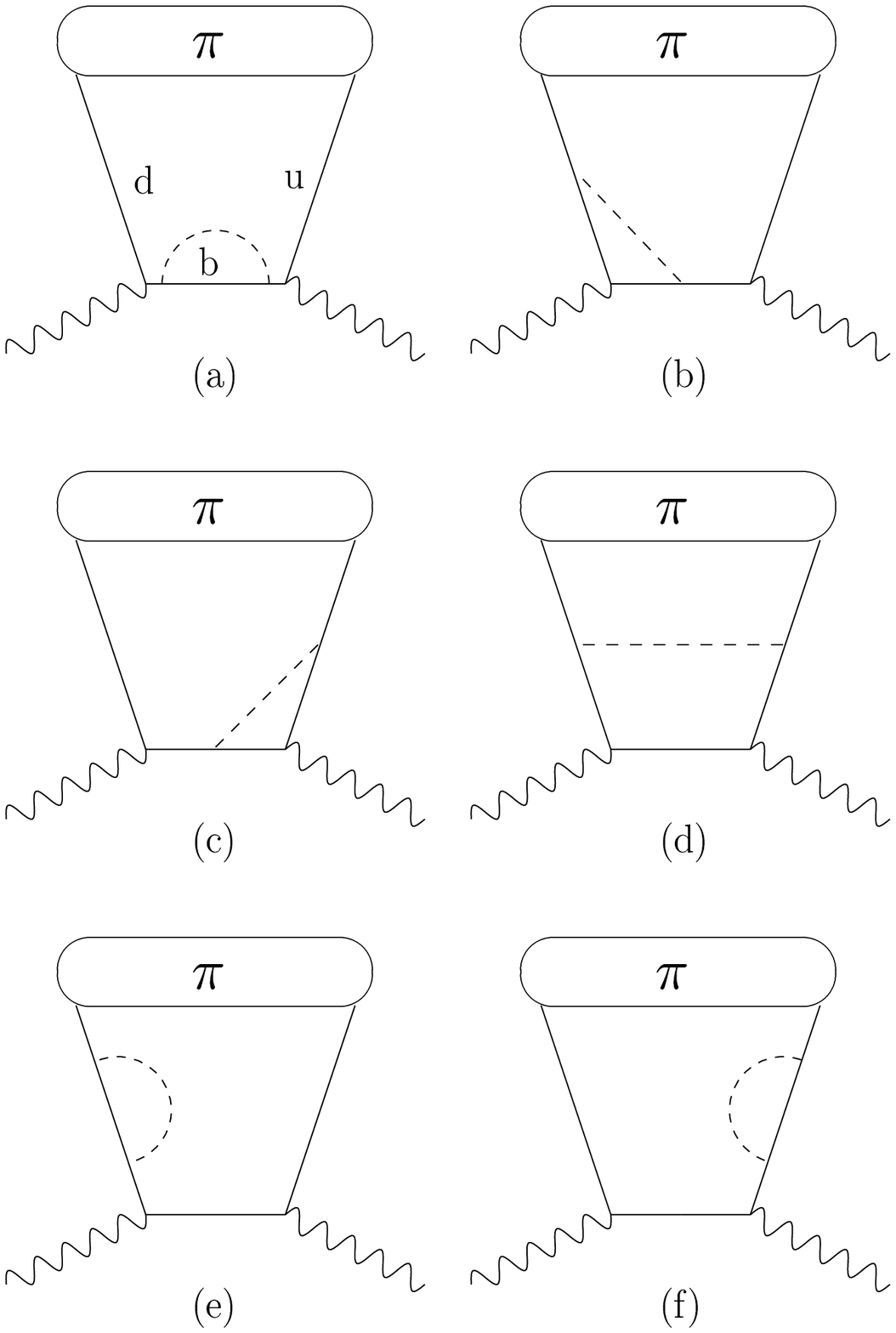,width=12cm,bbllx=0pt,bblly=100pt,bburx=600pt,%
bbury=720pt,clip=}
}
\caption{\it Diagrams of the $O(\alpha_s)$ corrections to the correlation
function (\ref{1a4}).}
\end{figure}

The LO approximation is given by the zero-order amplitude  
\begin{equation}
 T_0(p^2,(p+q)^2,u)=\frac{m_b}{m_b^2-p^2(1-u)-(p+q)^2u}
\label{Born}
\end{equation}
convoluted with the wave function obeying 
the LO Brodsky-Lepage evolution equation (\ref{BLL}).
In next-to-leading order (NLO),
the hard amplitude (\ref{T}) 
includes the first-order term $T_1$, while 
the wave function $\varphi_\pi(u,\mu)$ solves the NLO evolution 
equation \cite{kad}. For 
calculational details and analytic results one may consult  
\cite{KRWY, Bagan}.  
To date, the NLO program is completed
only for the twist 2 contribution to the invariant amplitude $F$. 

It is important to note that the correlation function (\ref{1a4})
actually leads to sum rules for products of decay constants and 
form factors as made explicit in (\ref{fplus}) and (\ref{fplusminus}).
When dividing out the decay constants, consistency requires
to take both the sum rule and $f_P$ in LO or in NLO. 
This appears obvious, but 
has not always been respected in the literature.
Some of the discrepancies in the calculations originate just from  
mixing orders. In fact,
as will be quantified later, the 
remaining radiative corrections to $f^+(p^2)$ turn
out to be rather small because of a cancellation of the corrections
to the sum rules (\ref{fplus}) for  $f_Bf^+(p^2)$ and (\ref{fB1}) for $f_B$.

As a final remark, the light-cone sum rules 
provide a unique possibility
to investigate the heavy-mass dependence of the
heavy-to-light form factors.
To this end, one employs the following scaling
relations for mass parameters and decay constants:
\be
m_P = m_Q+\bar{\Lambda}~,~~~ s_0^P = m_Q^2 + 2m_Q\omega_0 ~,
~~~M^2= 2m_Q\tau~,
\label{hqet}
\ee
\be
f_p = \hat{f}_P/\sqrt{m_Q},
\label{fBhat}
\ee
where in the heavy quark limit 
$\bar{\Lambda}$, $\omega_0$, $\tau$, $\hat{f}_P$ are
$m_Q$-independent quantities.
With these substitutions it is straightforward to expand
the sum rules (\ref{fplus}) and (\ref{fplusminus}) in $m_Q$. 
In both cases, the light-cone expansion
in terms of wave functions with increasing twist is 
consistent with the heavy mass expansion, that is
the higher-twist contributions either 
scale with the same power of $m_Q$ as the leading-twist term,
or they are suppressed by extra powers of $m_Q$.

The asymptotic scaling behaviour
of the form factors turns out to
differ sharply at small
\footnote{see also ref. \cite{CZ1}} and large momentum transfer
\cite{KRW}.
At $p^2 = 0$  
\be
f^+(0)= f^0(0)\sim m_Q^{-3/2}~,
\label{0limit}
\ee
\be
f^+(0) +f^-(0)\sim m_Q^{-3/2}~,
\label{0limit1}
\ee
whereas at $p^2 = m_Q^2-2m_Q\chi$,
$\chi$ being independent of $m_Q$,
\bq
f^+(p^2) \sim m_Q^{1/2}~,
\label{scal1}
\\
f^+(p^2) + f^-(p^2) \simeq f^0(p^2) \sim m_Q^{-1/2}~.
\label{scal2}
\eq
The sum rules thus nicely reproduce
the asymptotic dependence of the form factors on the heavy quark mass 
derived in \cite{IW,Vol}
for small pion momentum in the rest frame of the $P$ meson.
In addition, the sum rules allow to investigate the opposite
region of large pion momentum where neither HQET nor the single-pole 
model can be trusted. Clearly, from the sum rule
point of view it is expected
that the excited and continuum states become
more and more important as $p^2 \rightarrow 0$. This is reflected 
in the change of the asymptotic mass dependence.
Claims in the literature which
differ from (\ref{0limit}) to (\ref{scal2}) are 
often based on the pole model and therefore not reliable. A similar
analysis has been carried out in \cite{BallBraun} for $B \ra \rho$
form factors with
essentially the same conclusions.

\subsection{Numerical analysis}

For most of the following illustrations, it suffices 
to work in LO. Strictly speaking, since the NLO order corrections
to $f_B(f^+ + f^-)$ are unknown, it is also more consistent
to work in LO when combining or comparing $f^+$ and $f^+ + f^-$.
Therefore, if not stated otherwise, the values of 
$m_b$, $s_0^B$, and $f_B(\alpha_s = 0)$  
specified in (\ref{bmass}), (\ref{s0B}) and (\ref{fB0}), respectively,
are used.
The pion wave functions are as given in sect. 3.3 with the coefficients
scaled to 
\be
\mu_b =\sqrt{m_B^2-m_b^2} \simeq 2.4\, \mbox{\rm GeV}~.
\label{bscale}
\ee  
With this input we have checked that for 
$M^2 = 10 \pm 2$ GeV$^2$ the twist 4 corrections are very small,
and that the continuum contributions
do not exceed $30\%$. Up to $p^2 \simeq 18$ GeV$^2$, the sum rules are also
quite stable under 
variation of $M^2$. At larger momentum transfer 
the sum rules become unstable, and the twist 4 contribution grows rapidly.
This is not surprising because the light-cone expansion and the sum rule method
are expected to break down as $p^2$ approaches $m_b^2$. 

\begin{figure}[htb]
\centerline{
\epsfig{bbllx=100pt,bblly=209pt,bburx=507pt,%
bbury=490pt,file=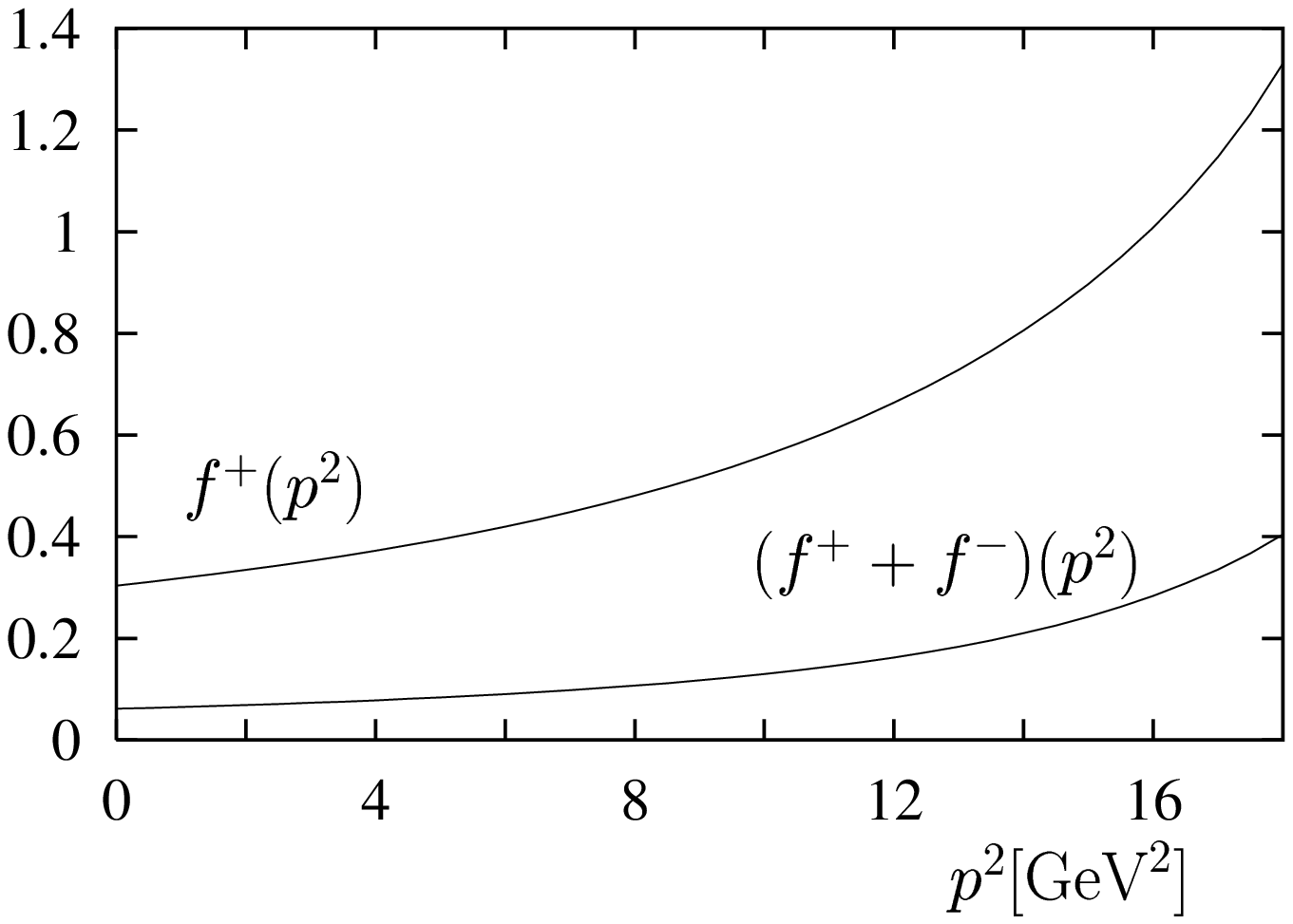,scale=0.9,%
clip=}
}
\caption{\it $B \to \pi$ form factors
obtained from light-cone sum rules in LO.}
\end{figure}

The momentum dependence of the form factors 
$f^+$ and $f^+ + f^-$ can be seen in Fig. 5.
Specifically, at zero momentum transfer we predict
\bq
f^+(0)=0.30,
\nonumber
\\
f^+(0) + f^-(0)= 0.06.
\label{fpl0}
\eq
If the known $O(\alpha_s)$ corrections are included
in the sum rules for $f_Bf^+(p^2)$ and $f_B$, one gets instead
\be 
f^+(0)=0.27.
\label{Bfpnlo}
\ee

\begin{figure}[htb]
\centerline{
\epsfig{file=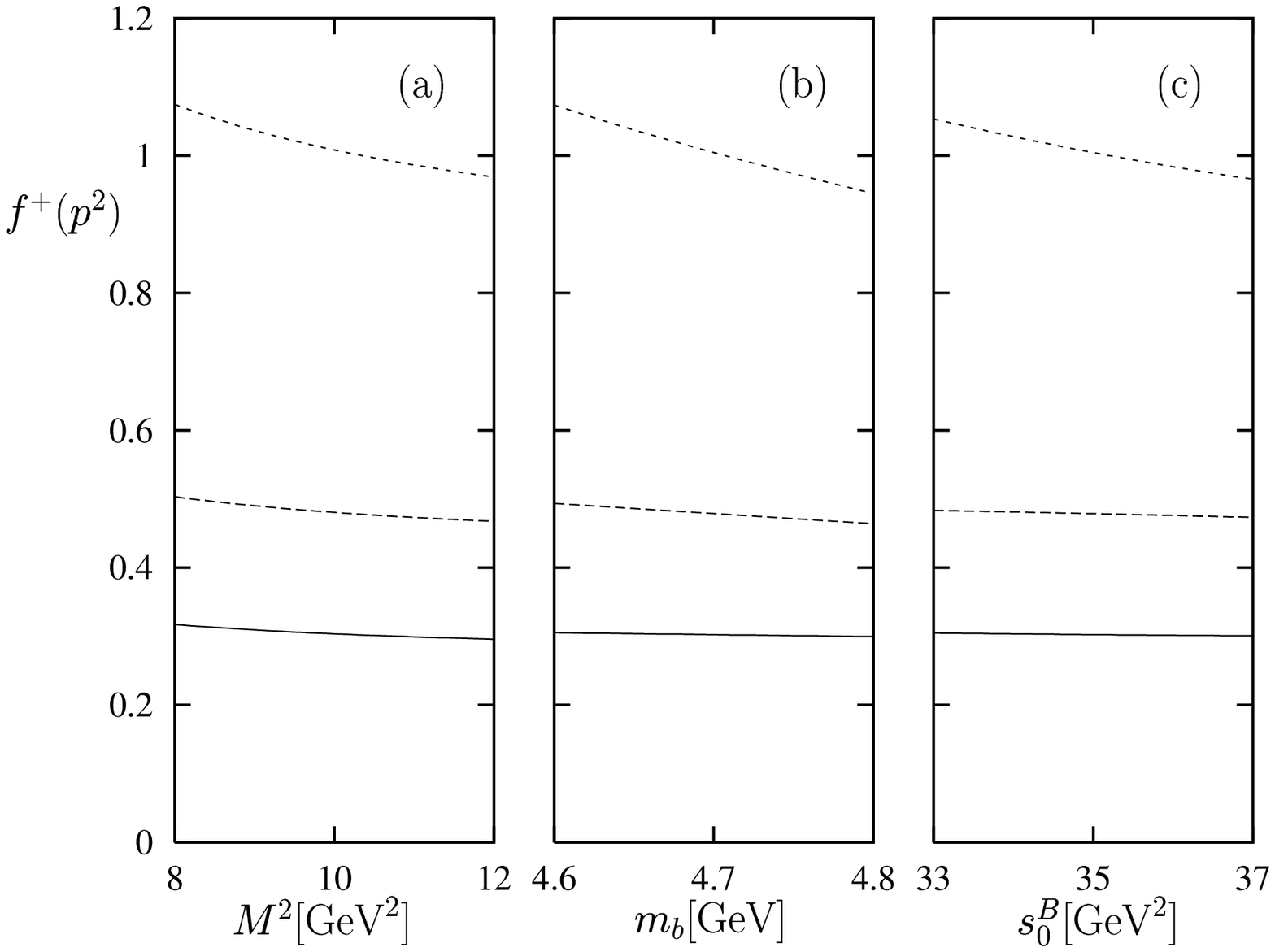,width=\textwidth,bbllx=38pt,bblly=207pt,bburx=580pt,%
bbury=660pt,clip=}
}
\caption{\it Variation
of the sum rule prediction on $f^+(p^2)$ 
with the Borel parameter (a), the $b$-quark mass (b),
and the continuum threshold $s_0^B$ (c).
Considered are three typical values of momentum transfer:
$p^2=0$ (solid),
$p^2=8$ GeV$^2$ (long-dashed), and $p^2=16$ GeV$^2$ (short-dashed).}
\end{figure}

Allowing the parameters to vary within the ranges specified
one can estimate the uncertainties in the sum rule predictions.
The main sources of uncertainty are discussed below:

(a) Borel mass parameter

\noindent The dependence of $f^+$ on $M^2$ is illustrated in Fig. 6a.
In the allowed interval of $M^2$, $f^+$ deviates by only
$\pm$(3 to 5)~\% from the nominal value, depending slightly on $p^2$.

(b) $b$-quark mass and continuum threshold 

\noindent Fig. 6b shows the variation of $f^+$ with $m_b$
keeping all other parameters except $f_B$ fixed. 
The value of $f_B$ is taken from the sum rule ({\ref{fB1})  
after dropping the $O(\alpha_s)$ corrections. 
The analogous test for $s_0^B$ is performed in Fig. 6c.  
If $s_0^B$ and $m_b$ are varied simultaneously such that 
one achieves maximum stability of the sum rule for $f_B$, 
the change in $f^+$ is negligible at 
small $p^2$ and only about $\pm$ 3~\% at large $p^2$.

\begin{figure}[p]
\centerline{
\epsfig{file=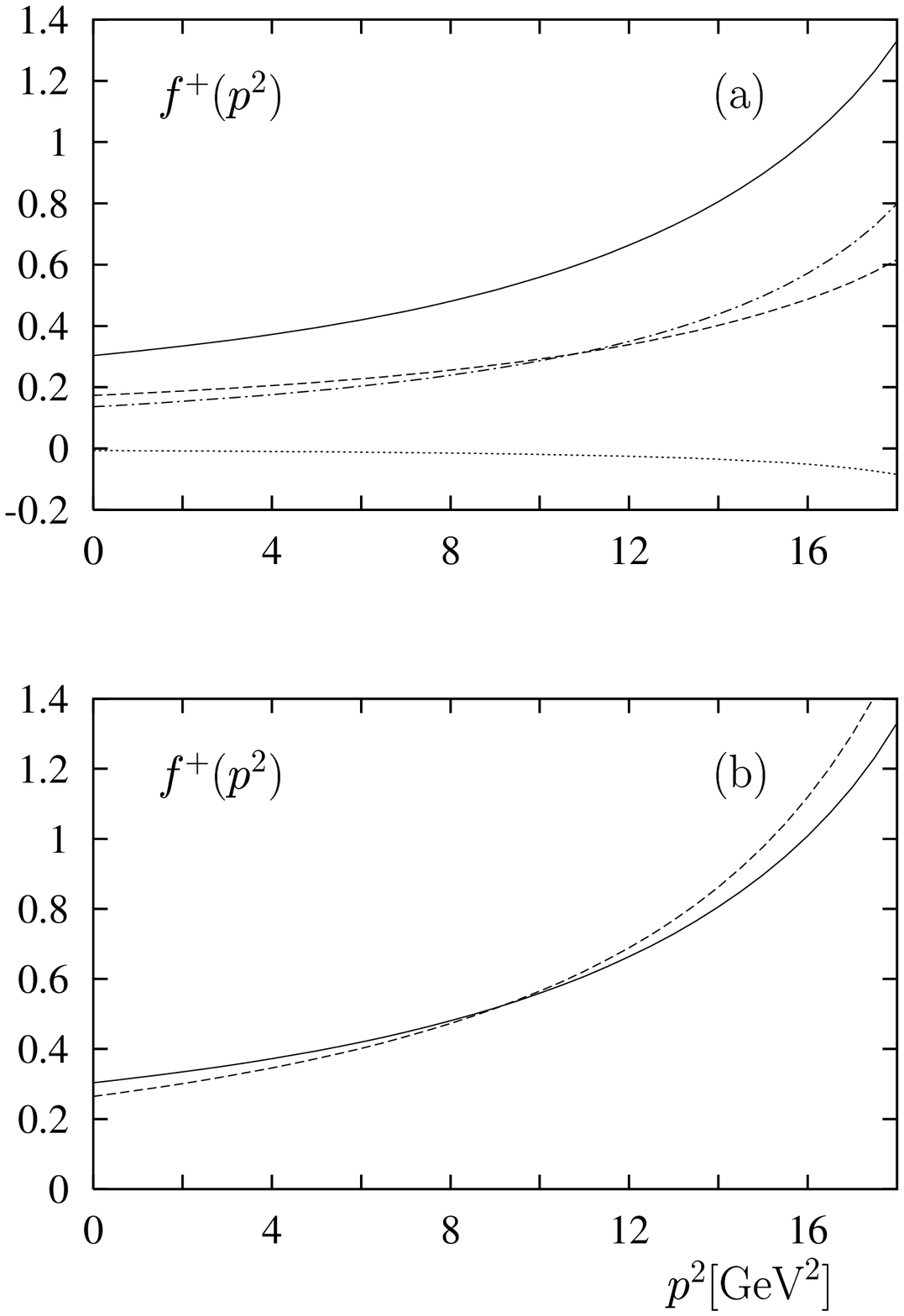,scale =0.6,width=\textwidth,bbllx=0pt,bblly=80pt,
bburx=600pt,%
bbury=800pt,clip=}
}
\caption{\it  $B \to \pi$ form factor $f^+$: (a) individual contributions
of twist 2 (dashed), 3 (dash-dotted), and 4 (dotted), 
and the total sum (solid); 
(b) wave functions with (solid) and without (dashed) 
nonasymptotic corrections.}
\end{figure}

(c) higher-twist contributions

\noindent No reliable estimates exist for wave functions beyond twist 4.
Therefore, we use the magnitude of the twist 4 contribution to $f^+$
as an indicator for the uncertainty due to the neglect of
higher-twist terms. 
From Fig. 7a we see that the impact of the twist 4 components is 
comfortably small, less than 2~\% at low
$p^2$ and about 5~\% at large $p^2$.
This suggests a conservative estimate
of $\pm$ 5~\% uncertainty due to unknown higher-twist.

(d) light-cone wave functions

\noindent
In order to clarify the sensitivity of $f^+$ to the nonasymptotic effects
in the wave functions, the corresponding coefficients are put to zero 
and the result is compared, in Fig. 7b, to the 
nominal prediction. The change amounts to about -10~\% 
at small $p^2$ and +10~\% at large $p^2$, while
the intermediate region around $p^2\simeq 10$ GeV$^2$ is very little affected.
Since this exercise is rather extreme, the real uncertainty from the
errors in the nonasymptotic coefficients is considerably smaller.

\begin{figure}[htb]
\centerline{
\epsfig{bbllx=100pt,bblly=209pt,bburx=507pt,%
bbury=490pt,file=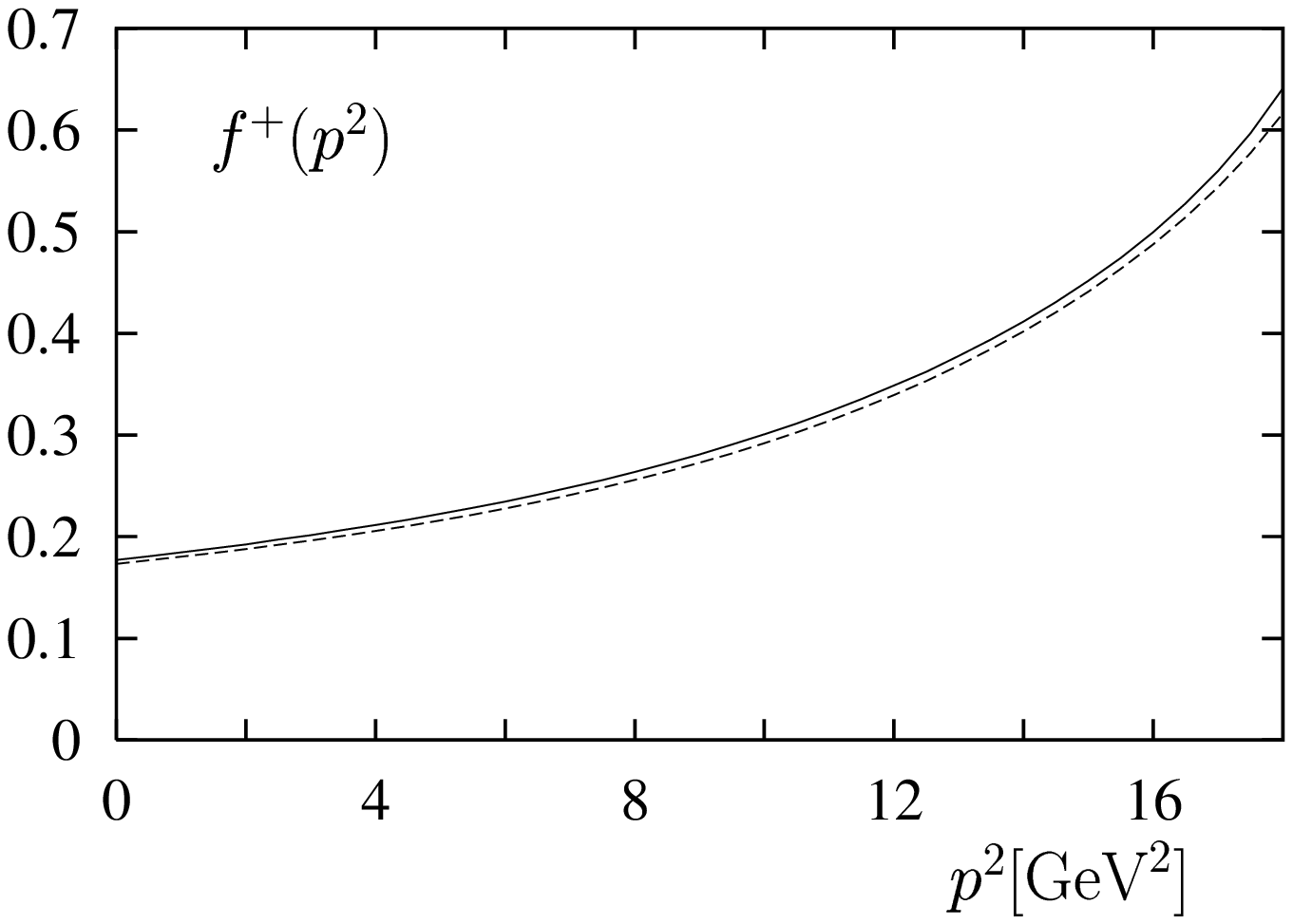,scale=0.9,%
clip=}
}
\caption{\it $B\to \pi$ form factor $f^+(p^2)$ in leading twist 2
approximation: LO (dashed) in comparison to NLO (solid).} 
\end{figure}

(e) perturbative corrections

\noindent The $O(\alpha_s)$ corrections
to the leading twist 2 term in the sum rule (\ref{fplus}) for $f_Bf^+$ 
and to the sum rule (\ref{fB1}) for 
$f_B$, both being about 30~\%, cancel in the ratio.
The net effect is the surprisingly small correction to $f^+$
shown in Fig. 8. 
This result \cite{KRWY,Bagan} eliminates one main 
uncertainty. Unfortunately, the perturbative corrections to the higher-twist
terms are not known. Considering that
the twist 3 terms contribute about 50~\% to the sum rule for $f_B f^+$, 
and assuming again radiative corrections of about 30~\% but no 
cancellation, the remaining uncertainty is 15~\%.

In summary, the present total uncertainty in the 
normalization of $f^+(p^2)$ is estimated to
be about 25~\%, if the individual contributions (a) to (e) with the 
exception of (d) are added up linearly, 
and about 17~\% if they are added in quadrature
as is often done in the literature. 
Ultimately, this may be reduced to about 15 to 10~\%. 
In addition, there is a 
shape-dependent uncertainty from (d) of less than -10~\% at low and 
+10~\% at high $p^2$.
Moreover, in the integrated width the latter
uncertainty averages out almost completely.
The uncertainties from (a) to (d) on $f^+ + f^-$ are of comparable size,
while the effect of radiative corrections is still unknown.

Finally, it is interesting to compare 
the predictions on the $B \rightarrow \pi$ form factors
from the light-cone sum rules with the results of other approaches. 
Such a comparison is presented in Fig. 9 and 10. Although there is rough
agreement on the normalization at low momentum transfer, the light-cone
sum rules predict the fastest increase of $f^+$ with $p^2$.

\begin{figure}[htb]
\centerline{
\epsfig{bbllx=100pt,bblly=209pt,bburx=507pt,%
bbury=490pt,file=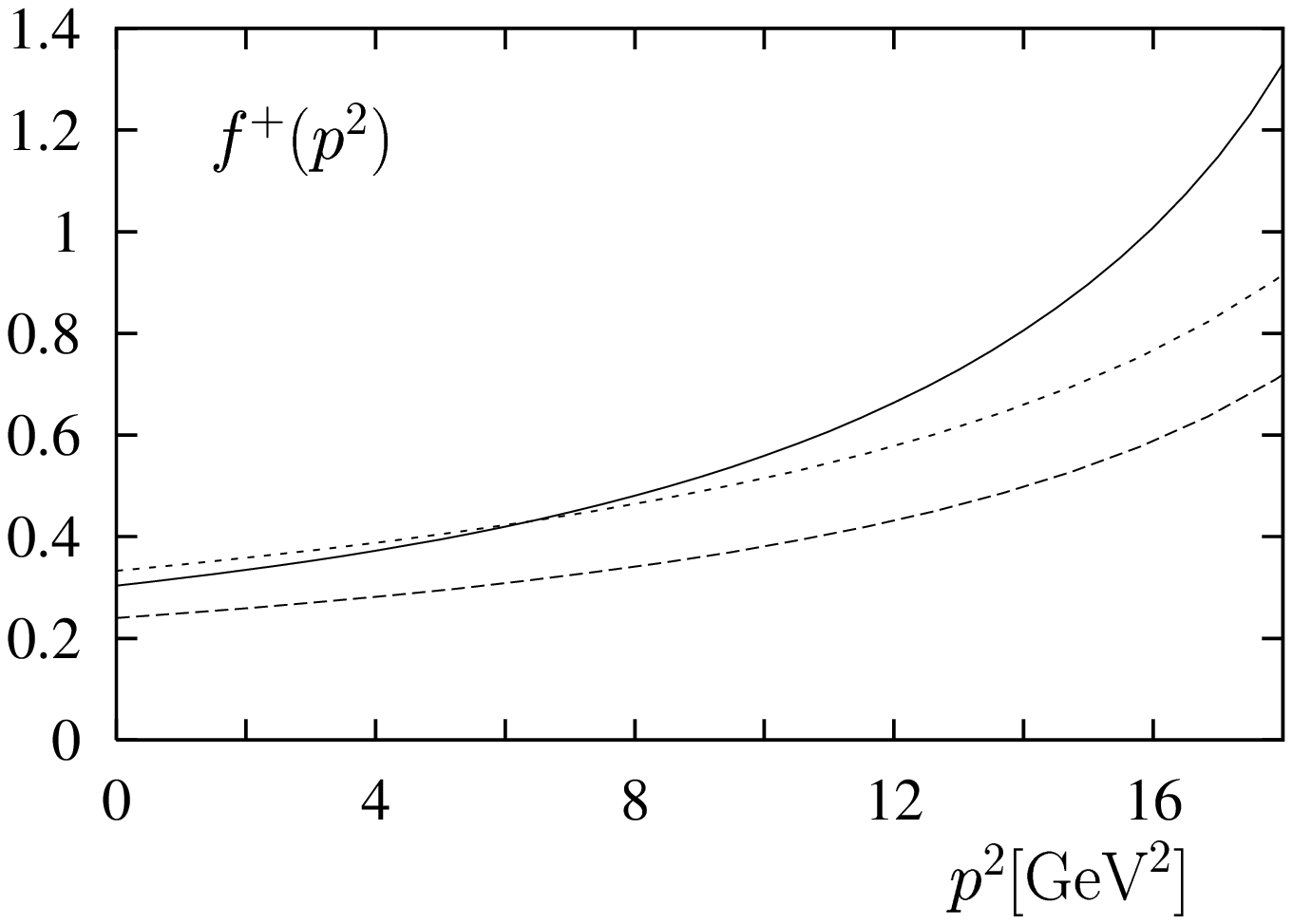,scale=0.9,%
clip=}
}
\caption{\it $B \to \pi$ form factor $f^+$ 
calculated in different approaches: 
three-point sum rule \cite{BBD2} (dashed),
quark model \cite{BSW} (dotted), and
light-cone sum rule \cite{BKR,KRW} (solid).}
\end{figure}

\begin{figure}[htb]
\centerline{
\epsfig{bbllx=100pt,bblly=209pt,bburx=507pt,%
bbury=490pt,file=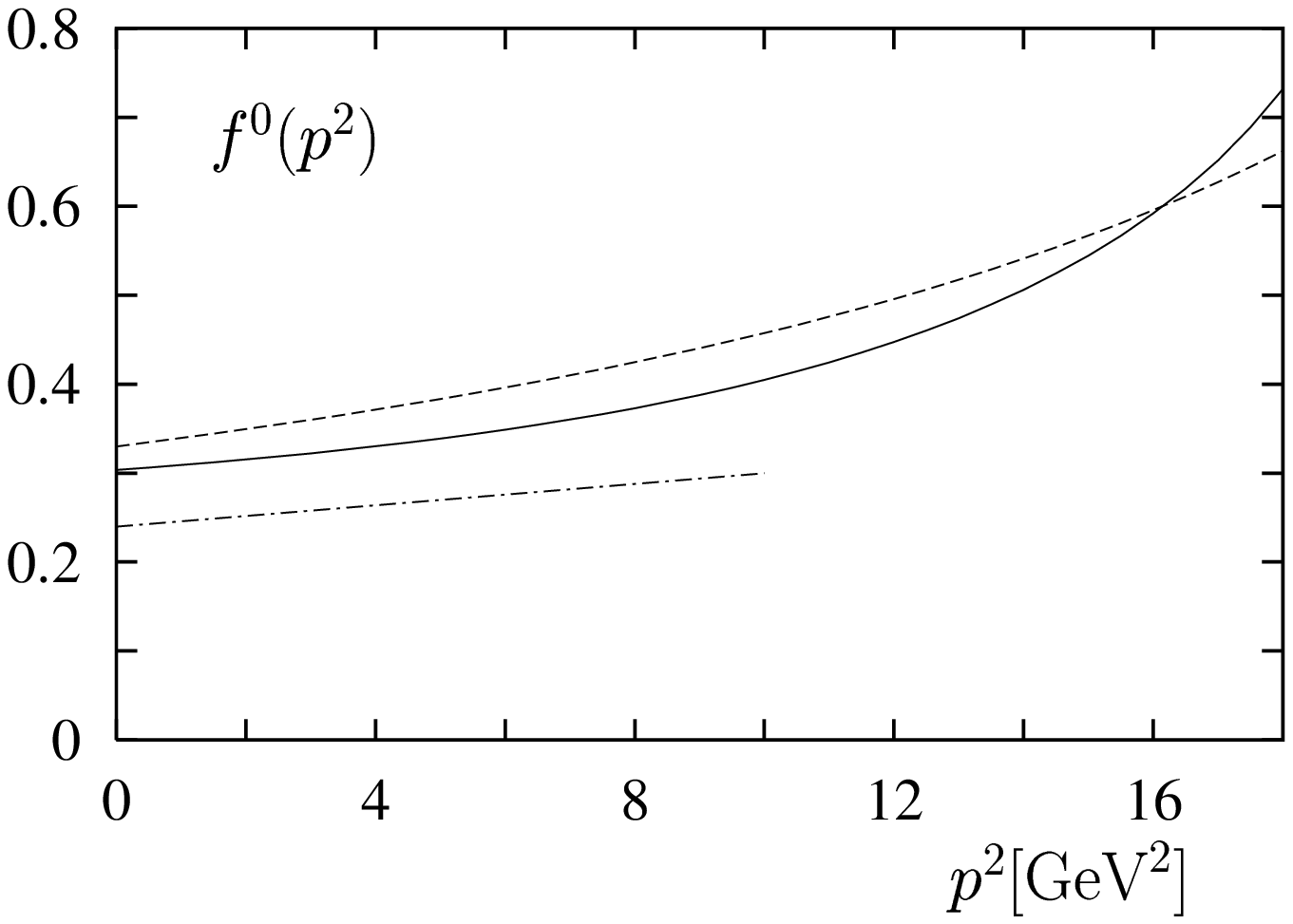,scale=0.9,%
clip=}
}
\caption{\it 
The $B \rightarrow \pi$   form factor $f^0$:  
light-cone sum rule prediction (solid) in comparison to  
the quark model result from
\cite{BSW}  (dashed)  and the sum rule estimate from
\cite{ColSan} (dash-dotted).}
\end{figure}

The sum rules  (\ref{fplus}) and
(\ref{fplusminus}) for the $ B\rightarrow \pi$ form factors
are formally converted into sum rules for
the $D\rightarrow \pi$  form factors by replacing
$b$ with $c$ and $B$ with $D$.
The input parameters are taken over from the calculation of
$f_D$ in sect. 2.2.
In addition, one has to rescale the wave functions
from $\mu_b \simeq 2.4$ GeV to $\mu_c \simeq 1.3$ GeV.  
The allowed Borel mass window is 
$ 3 ~\mbox{GeV}^2 < M^2 < 5 ~\mbox{GeV}^2$. With this choice, the $D \to \pi$
form factor at $p^2 = 0$ is predicted to be \cite{BBKR,KRW} 
\bq f^{+}(0) = 0.68.
\label{Dfplo}
\eq
$O(\alpha_s)$ corrections are not included here. 
The momentum dependence of $f^+$ in the range
$p^2 \le m_c^2 - O(1 ~\mbox{GeV}^2)$ is shown in the next
section, together with an extrapolation to higher $p^2$.

Other important applications of light-cone sum rules
include the estimate of the $B\to K$ form factor \cite{BKR}
which determines the factorizable part of the $B\to J/\psi K$ 
amplitude, the prediction  
of the matrix element of the electromagnetic penguin operator for 
$B \to K^* \gamma$ \cite{ABS}, and the
calculation of the $B\to \rho$ form factors \cite{BallBraun}. 
In the latter work, essentially 
the same procedure is applied as outlined above. However, the relevant weak
currents and light-cone wave functions are very different. 

\subsection{Single pole approximation and the $B^*B\pi$ and 
            $D^*D\pi$ couplings}

As repeatedly stressed, the light-cone sum rules derived 
in sect. 3.2 and 3.4
are only valid in the momentum range $p^2 \leq m_Q^2 - 2 m_Q \chi$.
In order to calculate the integrated semileptonic width
$\Gamma(P \rightarrow h l \bar{\nu}_l)$ one must therefore find 
another way to estimate the form factors in the high momentum range
$m_Q^2 - 2 m_Q \chi \leq p^2 \leq (m_P - m_h)^2$. 
Near the pole $p^2 = m_V^2$ of the vector ground state $V$ it is
justifiable to use the
single pole approximation
\be
f^+(p^2)= \frac{f_V g_{VP\pi}}{2m_V(1-p^2/m_V^2)}~.
\label{singlepole}
\ee
Here, $f_V$ is the decay constant of the vector state defined by
the matrix element
\bq
\langle 0 \mid \bar{q} \gamma_\mu Q \mid V \rangle = m_V f_V 
\epsilon_\mu ~,
\label{fBstar2}
\eq
and $g_{VP\pi}$ is the strong coupling between the heavy vector  
and pseudoscalar mesons and the pion defined by 
\begin{equation}
\langle V\pi\mid P\rangle = -g_{VP\pi}(q \cdot\epsilon )~,
\label{16}
\end{equation}
$\epsilon_\mu$ being the $V$ polarization vector.

The $B^*B\pi$ and $D^*D\pi$ couplings have
been studied with different variants of sum rules and in a variety of 
quark models.
In \cite{BBKR}, a light-cone sum rule for
$g_{B^*B\pi}$ has been suggested which is derived from
the same correlation function (\ref{1a4}) as the
$B\to \pi$ form factors. Hence, the nonperturbative input
is exactly the same in both calculations .
The key idea is to write a double dispersion integral for the
invariant function $F(p^2, (p+q)^2)$.
Inserting in (\ref{1a4})
complete sets of intermediate hadronic states
carrying $B$ and $B^*$ quantum numbers, respectively, and using 
the matrix elements (\ref{fB2}), (\ref{fBstar2}) and (\ref{16})
one obtains
\bq
F(p^2,(p+q)^2)=\frac{m_B^2m_{B^*}f_Bf_{B^*}g_{B^*B\pi}}{m_b
(p^2-m_{B^*}^2)((p+q)^2-m_B^2)}
+\int_{\Sigma}\frac{\rho^h(s_1,s_2)ds_1ds_2}{(s_1-p^2)(s_2-(p+q)^2)}
\label{2216}
\eq
$$
+ ~(subtractions)~.
$$
The first term is obviously the ground-state contribution
containing the $B^*B\pi$ coupling,
while the spectral function $\rho^h(s_1,s_2)$
represents higher resonances and continuum states in the
$B^*$ and $B$ channels. The integration region in the $(s_1,s_2)$ -- plane
is denoted by $\Sigma$. The subtraction terms are polynomials in
$p^2$ and/or $(p+q)^2$ which vanish by Borel transformation of  
(\ref{2216}) with respect to
both variables $p^2$ and $(p+q)^2$. The resulting
hadronic representation of $F$ is thus given by
\bq
F(M_1^2,M_2^2)&\equiv&
{\cal B}_{M_1^2}{\cal B}_{M_2^2}F(p^2,(p+q)^2)=
\frac{m_B^2m_{B^*}f_Bf_{B^*}g_{B^*B\pi} }{m_b}
\exp\left[-\frac{m_{B^*}^2}{M_1^2}-\frac{m_{B}^2}{M_2^2}\right]
\nonumber
\\
&+&\int_{\Sigma_{12}} \exp\left[-\frac{s_1}{M_1^2}-\frac{s_2}{M_2^2}\right]
\rho^h(s_1,s_2)ds_1ds_2~,
\label{2226}
\eq
where  $M_1^2$ and $M_2^2$  are the Borel parameters associated with
$p^2$ and $(p+q)^2$, respectively.
The same transformation is also applied to $F(p^2,(p+q)^2)$ given in
(\ref{F}):
\be
F(M_1^2, M_2^2)
= m_bf_\pi\varphi_\pi(u_0)M^2 e^{-\frac{m_b^2}{M^2}} + ~(higher ~twist)~
\label{form6}
\ee
with 
$
u_0= M_1^2/(M_1^2+M_2^2)
$
and 
$
M^2= M_1^2M_2^2/(M_1^2+M_2^2)
$.
The derivation of the complete result and, in particular, the subtraction
of the continuum is quite tricky. Further explanations 
can be found in \cite{review}. It turns out that
the subtraction greatly simplifies if $M_1^2=M_2^2$,
i.e., $u_0= 1/2$. With this choice,
the following sum rule is obtained from (\ref{2226}) and (\ref{form6}) 
after continuum subtraction:
$$
f_Bf_{B^*}g_{B^*B\pi}=\frac{m_b^2f_\pi}{m_B^2m_{B^*}}
e^{\frac{m_{B}^2+m_{B^*}^2}{2M^2}}
\Bigg\{M^2 \Big[e^{-\frac{m_b^2}{M^2}} - e^{-\frac{s_0^B}{M^2}}\Big]
$$
$$
\times
~\Big[\varphi_\pi(u_0)
+ \frac{\mu_\pi}{m_b} \left( u_0\varphi_p(u_0)
+\frac13\varphi_\sigma (u_0)+\frac16u_0\frac{d\varphi_\sigma}{du}(u_0)\right)
+ \frac{2f_{3\pi}}{m_bf_\pi}I^G_3(u_0)\Big]
$$
\be
+e^{-\frac{m_b^2}{M^2}}
\Big[\frac{\mu_\pi m_b}3 \vp_\sigma (u_0)
+2u_0g_2(u_0)-\frac{4m_b^2}{M^2}
(g_1(u_0)+G_2(u_0)) + I^G_4(u_0)\Big]\Bigg\}_{u_0=1/2}~.
\label{fin}
\ee
Here, $I^G_3(u_0)$ and $I^G_4(u_0)$ 
are contributions from the three-particle amplitude shown
in Fig. 3b and given formally in \cite{BBKR,review}.
Since $G$-parity implies
$g_2(u_0)= \frac{d\varphi_\sigma}{du}(u_0)=0$ at $u_0 = 1/2$, 
these terms can be dropped in the above sum rule due to the choice
$M_1^2=M_2^2$.

One of the numerically crucial parameters is the leading twist 2
wave function $\varphi_{\pi}$ at the symmetry point $u_0=1/2$. 
The same parameter
also enters the sum rules for other
hadronic couplings involving the pion. For the following predictions the
value
\be
\varphi_\pi(u=\frac12) = 1.2\pm0.2
\label{phi12}
\ee
obtained from the light-cone sum rule for the
pion-nucleon coupling \cite{BF} has been used. This value 
is consistent with the coefficients $a_{2,4}$ given in (\ref{aa}).
The values of the remaining parameters
are as in the calculation of the form factor
$f^+$ in sect. 3.5. With this input the sum rule (\ref{fin})
yields
\be
f_Bf_{B^{*}}g_{B^*B\pi }=  0.64 \pm 0.06 \,\mbox{\rm GeV}^2 ~.
\label{combinB}
\ee
Furthermore, using the sum rule estimate (\ref{fB0}) for $f_B$, 
and 
\be
\bar{f}_{B^*} = f_{B^*} (\alpha_s = 0) = 160 \pm 30 ~\mbox{MeV}
\label{fBstar0}
\ee
derived from a similar two-point sum rule for $f_{B^*}$, 
one gets
\bq
g_{B^*B\pi}=29\pm 3~.
\label{41}
\eq
Note that the couplings of the different charge states 
are related by isospin symmetry:
\bq
g_{B^*B\pi} \equiv g_{\bar{B}^{*0}B^-\pi^+}=
-\sqrt{2}g_{\bar{B}^{*0}\bar{B}^{0}\pi^0}=\sqrt{2}g_{B^{*-}B^-\pi^0}
m=-g_{B^{*-}\bar{B}^0\pi^-}~.
\label{gd}
\eq
 
The uncertainties quoted above reflect the variation of the results
with the Borel mass in the allowed window
$6 \le M^2 \le 12$ GeV$^2$. In this window
the excited and continuum states contribute
less than 30~\% and the twist 4 corrections do not exceed 10~\%.
There are other sources of uncertainties. In particular,
the above predictions can and should be improved by calculating
radiative gluon corrections.

The sum rule (\ref{fin}) for $g_{B^*B\pi}$ is translated into  
a sum rule for $g_{D^*D\pi}$
by formally changing $b$ to $c$, $\bar{B}$ to $D$, and $\bar B^*$ to $D^*$.
With (\ref{phi12}) and the input parameters from the
calculation of the $D\to \pi$ form factor,
one finds
\be
f_Df_{D^{*}}g_{D^*D\pi }~= ~0.51 \pm 0.05~ GeV^2  ~.
\label{combinD}
\ee
Use of (\ref{fDres}) for $f_D$ and of the analogous sum rule estimate 
\be
\bar{f}_{D^*} = f_{D^*} (\alpha_s = 0) = 240 \pm 20 ~\mbox{MeV}
\label{fDstar0}
\ee
yields
\bq
g_{D^*D\pi}= 12.5\pm 1.0~.
\label{constD*Dpi}
\eq
The variation with 
$M^2$ in the allowed interval, here, $2 \le M^2 \le 4$ ~GeV$^2$,
is again quoted as an uncertainty.

%
{\footnotesize
\begin{table}
\caption{\it Theoretical estimates of the strong
$B^*B\pi$ and $D^*D\pi$ couplings (from \cite{BBKR}).}
\bigskip
\begin{center}
\begin{tabular}{lllll}
\hline
\hline
\\
Reference & $\hat{g}$ &
$g_{B^*B\pi}$ & $g_{D^*D\pi}$ &$\Gamma(D^{*+}\ra D^0\pi^+)$ (keV)\\
\\
\hline
\\
\cite{BBKR} & 0.32 $\pm$ 0.02 & 29 $\pm$ 3 & 12.5 $\pm$ 1.0 & 32
 $\pm$ 5\\
\\
\cite{BBKR}$^a$ & -- & 28 $\pm$ 6& 11 $\pm$ 2 & --\\
\\
\cite{Ovch89}$^a$ &--& 32 $\pm$ 6& -- & --\\
\\
\cite{GY}$^a$ &0.2 $\div$ 0.7 &-- & -- & --\\
\\
\cite{Colang}$^a$ &0.39 $\pm$ 0.16& 20 $\pm$ 4&9 $\pm$ 1& --\\
\\
\cite{Colang}$^{a *}$ &0.21 $\pm$ 0.06& 15 $\pm$ 4 &7 $\pm$ 1& 10
 $\pm$ 3\\
\\
\cite{NW}$^b$ & 0.7&--& --& --\\
\\
\cite{IW}$^{b}$ &--& 64&--&--\\
\\
\cite{Yan}$^b$ & 0.75 $\div$ 1.0&--& --& 100 $\div$ 180\\
\\
\cite{CG}$^c$ & 0.6 $\div $ 0.7 & -- & -- & 61 $\div$ 78\\
\\
\cite{Ametal}$^c$ & 0.4 $\div$ 0.7 & -- & -- & --\\
\\
\cite{BardeenHill}$^d$ & 0.3 & -- & -- & --\\
\\
\cite{Eichtetal}$^e$ & --& -- & 16.2&53.4\\
\\
\cite{DoXu}$^f$ & --& -- &19.5 $\pm$ 1.0&76 $\pm$ 7\\
\\
\cite{Miller}$^g$ & --& -- & 16.2 $\pm$ 0.3&53.3 $\pm$ 2.0\\
\\
\cite{Kam}$^h$ & --& -- &8.9&16\\
\\
\cite{KN}$^i$ & --& -- &8.2& 13.8\\
\\
\cite{PDG}$^k$&--& -- &$<$ 21 & $<$ 89\\
\\
\hline
\hline
\end{tabular}
\end{center}
\hspace*{2.4cm}
$^a$ QCD sum rules in external axial field or soft pion limit. \\
\hspace*{2.4cm}
$^*$ including perturbative correction to the heavy meson decay constants.  \\
\hspace*{2.4cm}
$^b$ Quark model + chiral HQET.  \\
\hspace*{2.4cm}
$^c$ Chiral HQET with experimental constraints on $D^*$ decays. \\
\hspace*{2.4cm}
$^d$ Extended NJL model + chiral HQET . \\
\hspace*{2.4cm}
$^e$ Quark Model + scaling relation. \\
\hspace*{2.4cm}
$^f$ Relativistic quark model. \\
\hspace*{2.4cm}
$^g$ Bag model. \\
\hspace*{2.4cm}
$^h$ SU(4) symmetry. \\
\hspace*{2.4cm}
$^i$ Reggeon quark-gluon string model.  \\
\hspace*{2.4cm}
$^k$ Experimental limits\\
\end{table} }

From (\ref{constD*Dpi}) one can calculate the width for the decay
$D^* \to D \pi$. The prediction,
\be
\Gamma( D^{*+} \rightarrow D^0 \pi^+)~ =
~ \frac{g_{D^*D\pi}^2}{24\pi m_{D^*}^2}|~\vec q ~|^3
~ = ~32 \pm 5 \,\mbox{\rm keV}~,  
\label{Gamma6}
\ee
lies well below the current experimental upper limit,
\be
\Gamma( D^{*+} \rightarrow D^0 \pi^+)~ < ~ 89 ~\mbox{\rm keV}~.
\label{exp}
\ee
The latter is derived from the upper limit
$\Gamma_{tot}(D^{*+}) < 131$ keV and from the branching ratio 
$BR(D^{*+} \to D^0 \pi^+) = (68.3 \pm 1.4)~\%$ \cite{PDG}.
Predictions for other charge combinations are
readily obtained from (\ref{Gamma6}) and isospin relations 
analogous to (\ref{gd}). Accounting also for the
differences in phase space, one expects
\be
\Gamma( D^{*+} \rightarrow D^0 \pi^+)~ =
~2.2\,\Gamma (D^{*+} \rightarrow D^+ \pi^0)=~
1.44\,\Gamma (D^{*0} \rightarrow D^{0} \pi^0)~.
\label{widths}
\ee
In contrast to $g_{D^*D\pi}$, the coupling constant $g_{B^*B\pi}$
cannot be measured directly,
since the corresponding decay $B^* \rightarrow B \pi $ is kinematically
forbidden.

\begin{figure}[htb]
\centerline{
\epsfig{bbllx=100pt,bblly=209pt,bburx=507pt,%
bbury=490pt,file=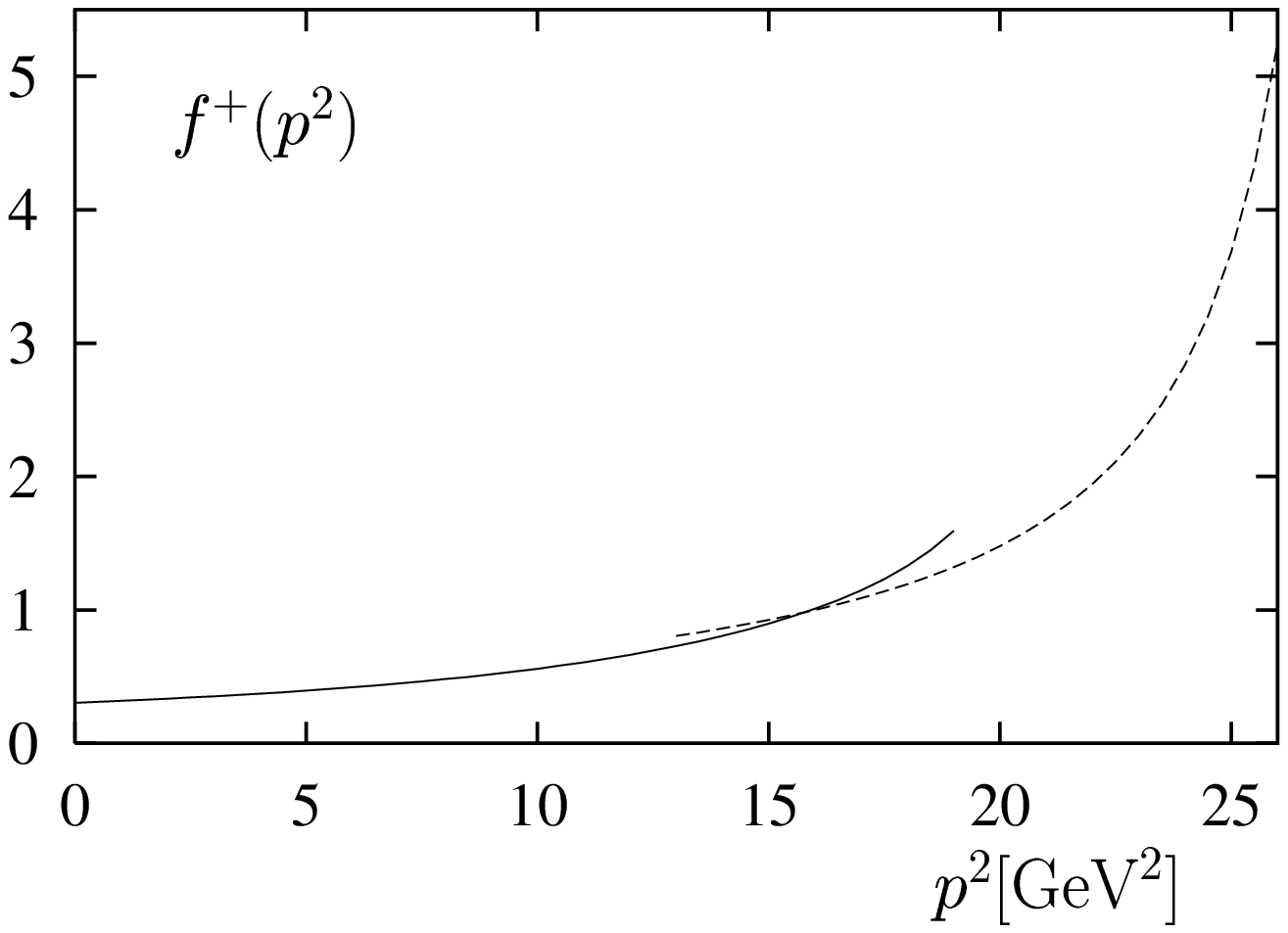,scale=0.9,%
clip=}
}
\caption{\it The $B\to\pi$ form factor $f^+$: direct 
sum rule prediction (solid), and 
single-pole approximation normalized by 
the sum rule estimate for $g_{B^*B\pi}$ (dashed).}
\end{figure}

\begin{figure}[htb]
\centerline{
\epsfig{bbllx=100pt,bblly=209pt,bburx=507pt,%
bbury=490pt,file=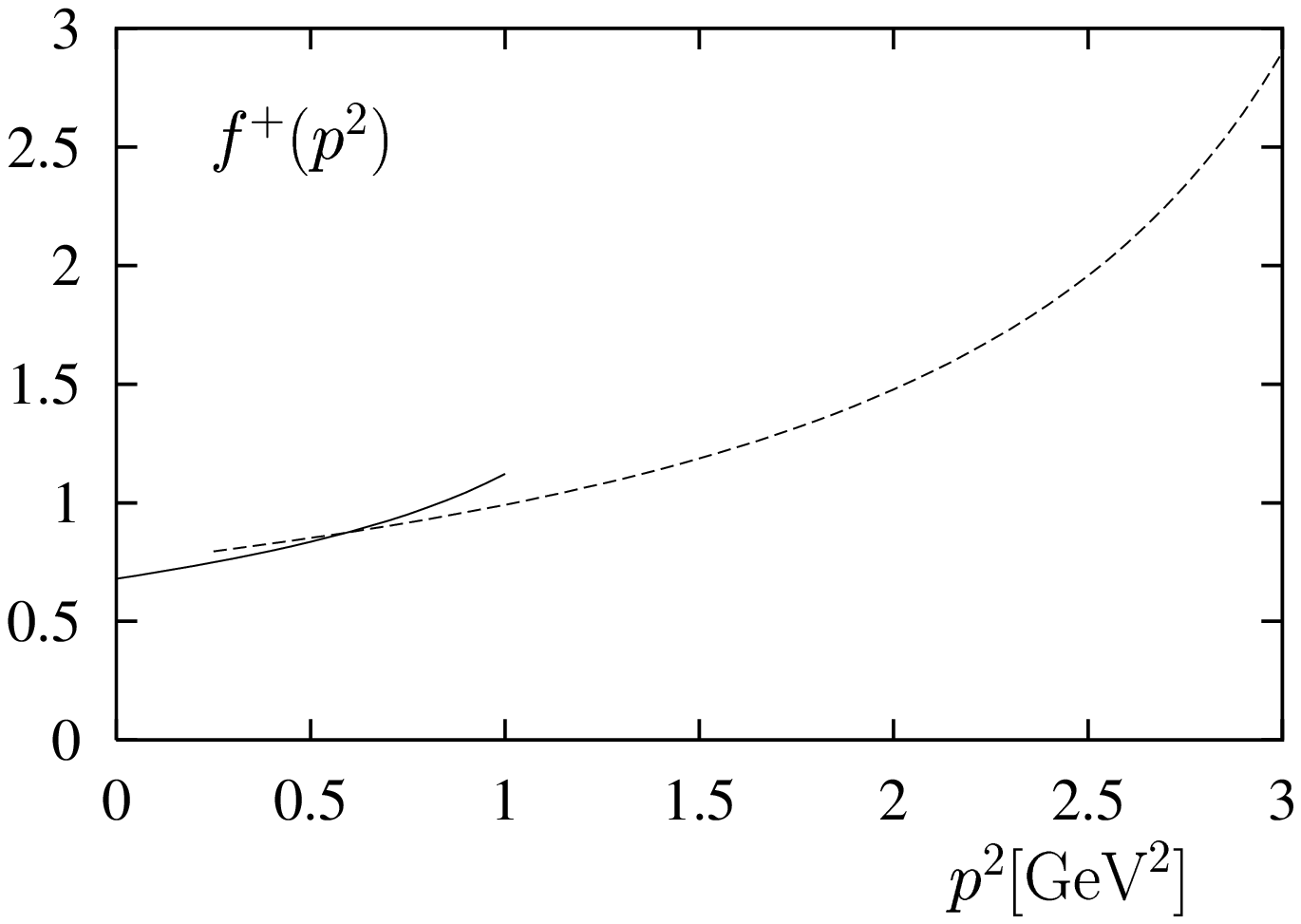,scale=0.9,%
clip=}
}
\caption{\it The $D\to\pi$ form factor $f^+$: direct 
sum rule prediction (solid), and 
single-pole approximation normalized by the sum rule estimate 
for $g_{D^*D\pi}$ (dashed).}
\end{figure}

\begin{figure}[htb]
\centerline{
\epsfig{bbllx=100pt,bblly=209pt,bburx=507pt,%
bbury=490pt,file=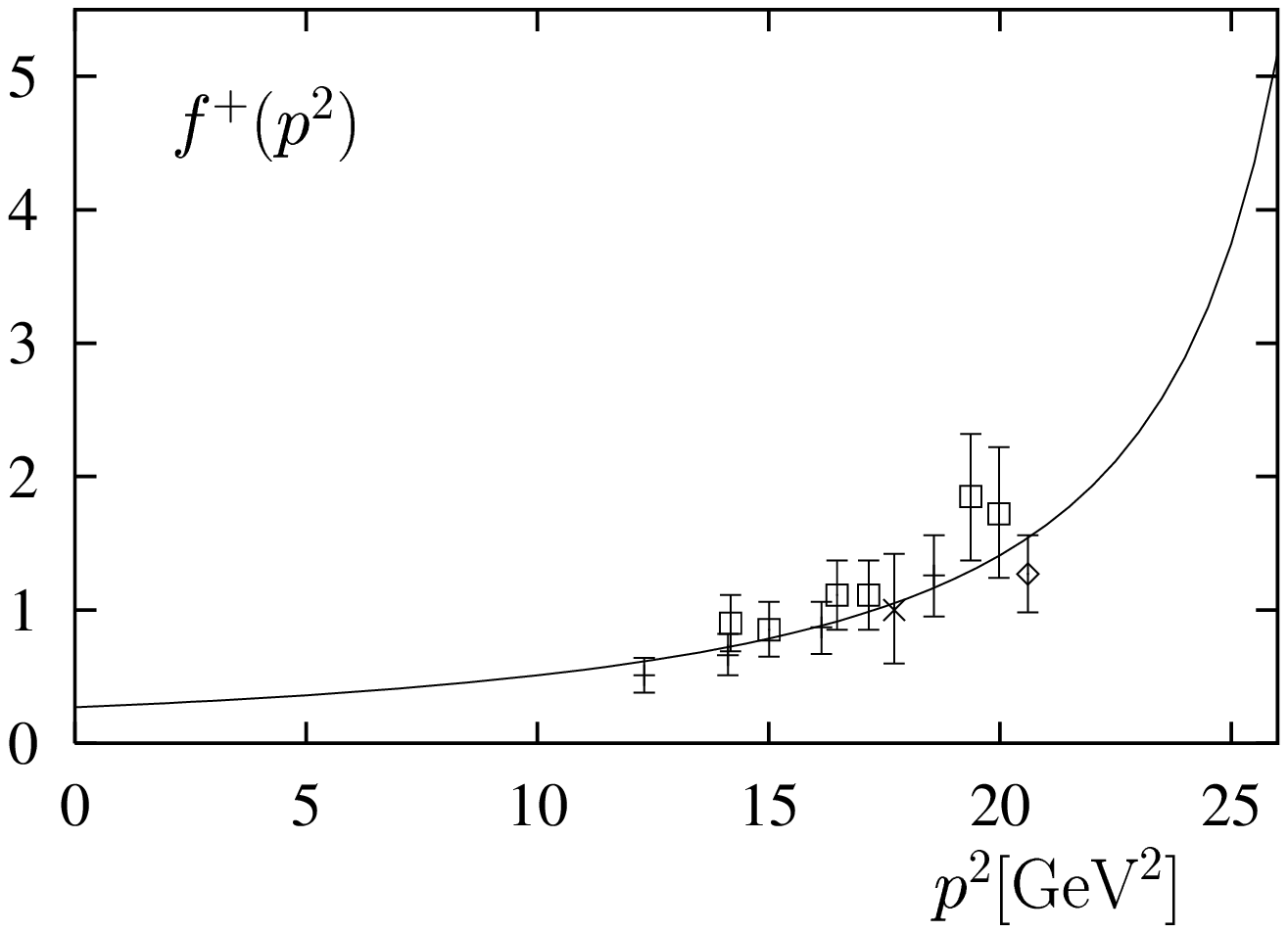,scale=0.9,%
,clip=}
}
\caption{\it The sum rule prediction for the $B\to\pi$ form factor $f^+$
in comparison to lattice results \cite{Flynn}.}
\end{figure}

In Tab. 4, the numerical results discussed above are summarized
and compared with other estimates. One observes significant
differences. Some of the predictions 
are rather close to the experimental upper limit, some already 
violate the bound.
It would be very interesting to have more precise data.

As far as the asymptotic dependence of the $VP\pi$ coupling 
on the heavy quark mass is concerned, 
the sum rule (\ref{fin}) suggests
\be
g_{VP\pi} \sim m_Q~,
\label{scalg}
\ee
in agreement with the expectation from HQET \cite{IW,NW,Wise}.
The sum rules can also be used to
investigate the $1/m_Q$ corrections. A simple quantitative 
estimate of the latter is obtained
by fitting the numerical predictions (\ref{41}) and (\ref{constD*Dpi})
to the expression 
\be\label{1/m}
g_{B^*B\pi} = \frac{2 m_B}{f_\pi}\cdot \hat g
\Bigg(1+\frac{\Delta}{m_B}\Bigg)
\ee
and the analogous one for $g_{D^*D\pi}$.
The result is \cite{BBKR}
\be
\hat g =0.32\pm 0.02~,~~
\Delta =(0.7 \pm 0.1)~ \mbox{GeV} ~.
\label{fit}
\ee
In Tab. 4 , the above  
value of the reduced
coupling constant $\hat{g}$ is 
compared to the estimates from other approaches.
The $1/m_Q$ correction being
about 15\% for $g_{B^*B\pi}$ and  40\% for $g_{D^*D\pi}$
is non-negligible. 
Furthermore, in the heavy quark limit the ratio
\be
r=\frac{g_{B^*B \pi}f_{B^*}\sqrt{m_{D}}}
{g_{D^*D \pi}f_{D^*}\sqrt{m_{B}}}
\label{ratio3}
\ee
is expected to approach unity. Moreover, it has been shown 
\cite{BLN} that $r$ is subject to $1/m_Q$ corrections 
only in next-to-leading order. Indeed,
the ratio $r=0.92$ 
derived from the light-cone sum rules deviates surprisingly
little from unity, in qualitative agreement with the HQET expectation.

Finally, let us return to the original problem of 
calculating the heavy-to-light form factors
at large momentum transfer.
As pointed out in the beginning of this section,
the $VP\pi$ on-shell vertex is of great
importance for the understanding of the
form factor $f^+$ at large momentum transfer, since
near the kinematic limit the $V$ pole is expected
to dominated the behaviour of $f^+$. For $B\to \pi$,
the single-pole approximation,
\be
f^+(p^2)= \frac{f_{B^*}g_{B^*B\pi}}{2m_{B^*}(1-p^2/m_{B^*}^2)}
\label{onepole}
\ee
is illustrated in Fig. 11 taking $g_{B^*B\pi}$ from (\ref{41})
and $f_{B^*}$ from (\ref{fBstar0}). The
extrapolation of the single-pole model to smaller $p^2$ matches
quite well with the direct estimate from
the light-cone sum rule (\ref{fplus})
at intermediate momentum transfer
$p^2=15$ to $20$ GeV$^2$. One thus has a consistent and 
complete theoretical description of $f^+$. 
The extrapolation for the $D\to \pi$ form factor using
the analogous  single-pole formula
with $g_{D^*D\pi}$ from (\ref{constD*Dpi}) and $f_{D^*}$ from 
(\ref{fDstar0}) is shown in Fig. 12. Also in this case
we find the direct sum rule result and the pole
model to match nicely at $p^2 \simeq 0.7$ GeV$^2$. However, here
the matching point is a little bit too far away from the
$D^*$ resonance for the extrapolation to be completely trustworthy.

\subsection{Decay spectra, integrated widths, and $V_{ub}$}

With the $f^+$ form factors for $B$ and $D$ mesons at hand, 
we are now in the position to
discuss the differential distributions and integrated widths 
for the exclusive semileptonic decays $B\ra \pi \bar{l} \nu_l$
and $D\ra \pi \bar{l} \nu_l$ with $l=e$ or $\mu$. 
In these decays, the form factor $f^0$ being suppressed 
by $m_l^2$ plays a negligible role. 

\begin{figure}[htb]
\centerline{
\epsfig{bbllx=100pt,bblly=209pt,bburx=507pt,%
bbury=490pt,file=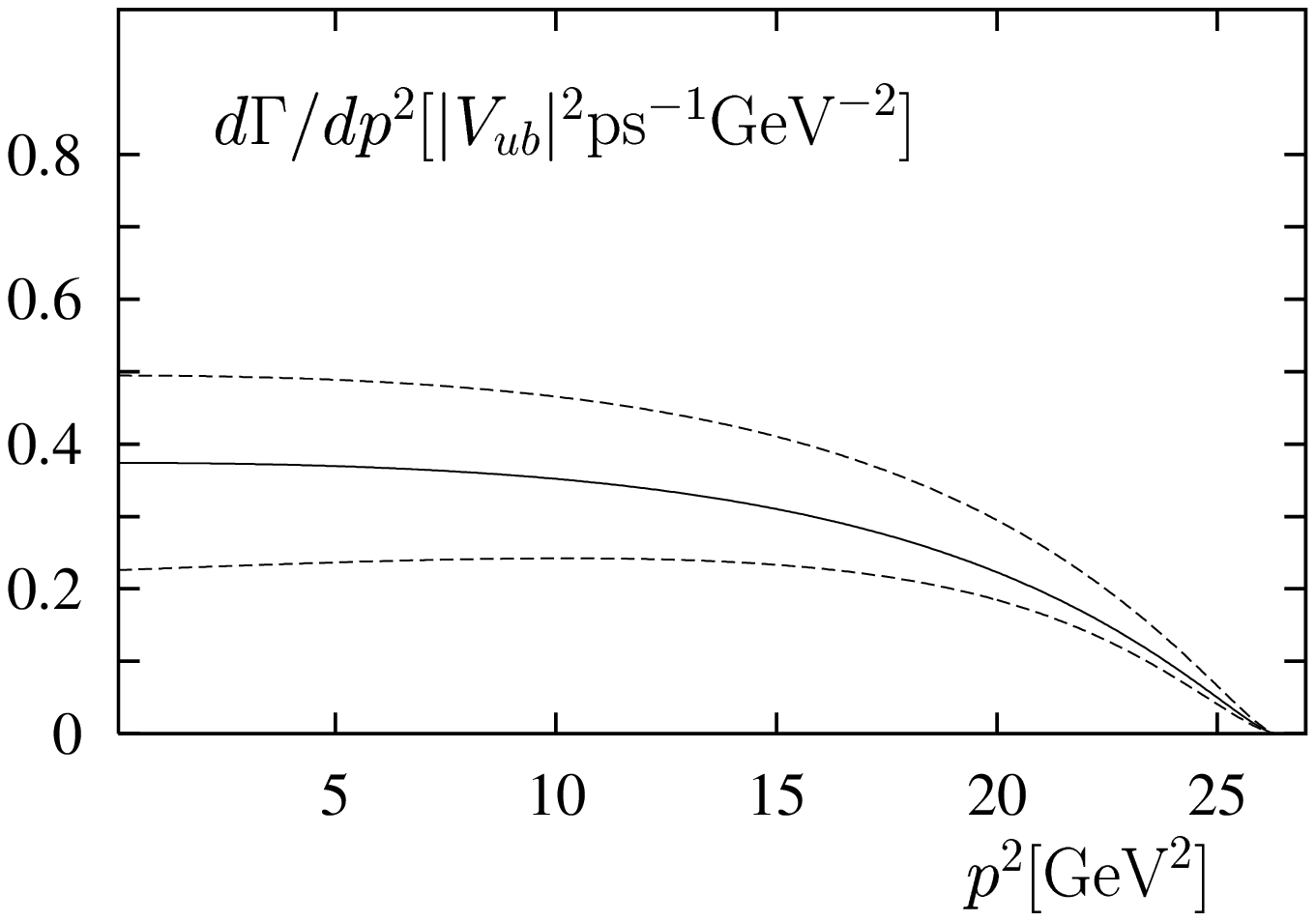,scale=0.9,%
,clip=}
}
\caption{\it Distribution of the momentum transfer squared in
$B \to \pi \bar l \nu_l$ ($ l=e, \mu$). 
The dashed curves indicate the theoretical uncertainty
discussed in detail in the text.}
\end{figure}

\begin{figure}[htb]
\centerline{
\epsfig{bbllx=110pt,bblly=209pt,bburx=507pt,%
bbury=490pt,file=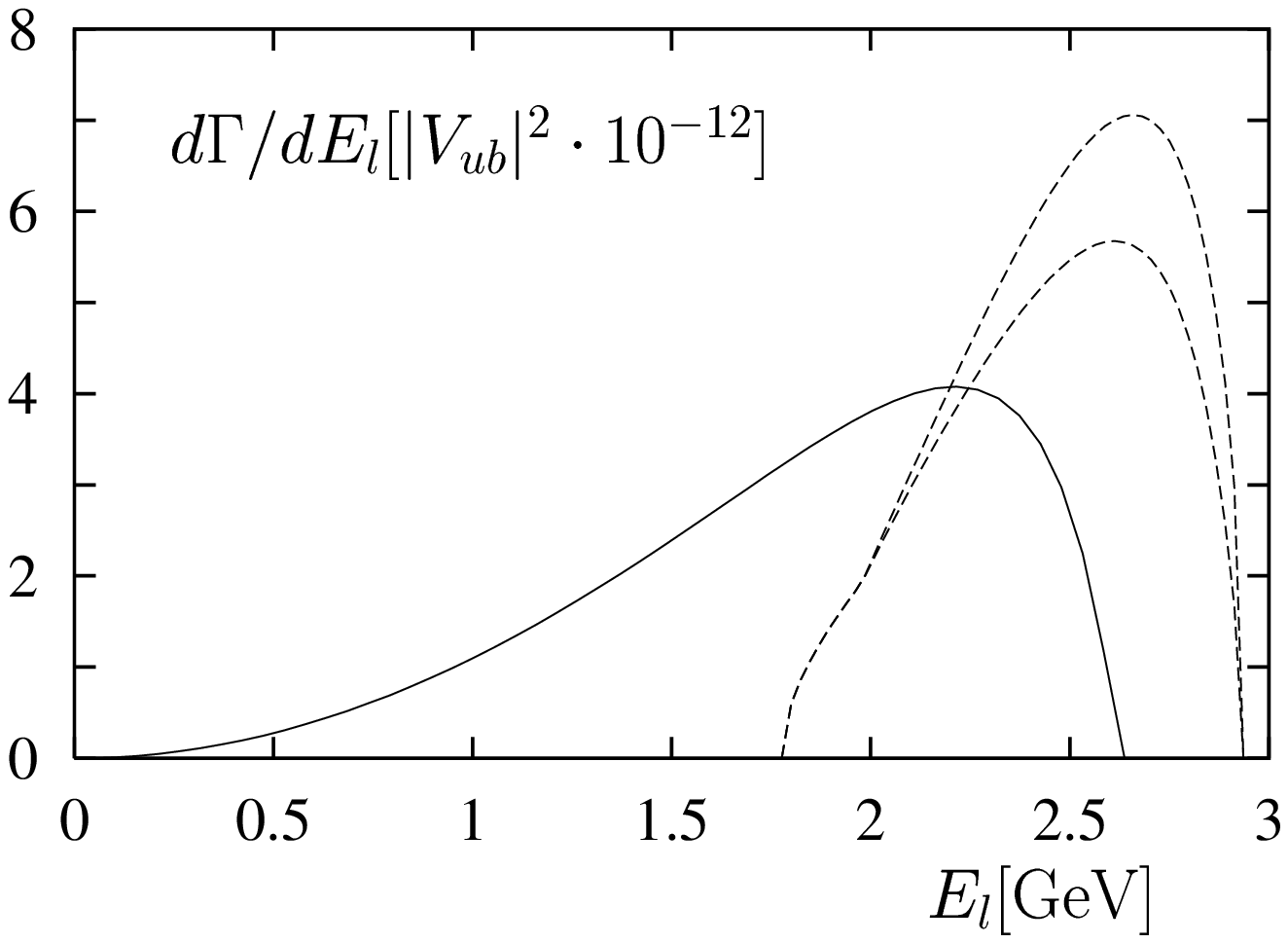,scale=0.9,%
clip=}
}
\caption{\it Distribution of the charged lepton energy in
$B \to \pi \bar{l} \nu_l$ for $l=e,\mu$ (solid) 
and $l=\tau$ (dashed). In the latter case 
the two curves correspond to the two extrapolations 
of $f^0$ shown in Fig. 16.}
\end{figure}

For convenience, one may fit a parametrization of the form
\be
f^+(p^2) = \frac{f^+(0)}{1-ap^2/m_P^2+b p^4/m_P^4}
\label{param}
\ee
to the theoretical curves plotted in Fig. 11 and 12.
For the $B \to \pi$ form factor this fit yields
\be
a = 1.50,~ b = 0.52~.
\label{param1}
\ee
Here, the known NLO corrections are included and, 
correspondingly, $f^+(0)$ has been fixed to the NLO value $f^+(0) = 0.27$ 
given in (\ref{Bfpnlo}).  
Fig. 13 shows the interpolation (\ref{param}) in comparison 
with recent lattice results
\footnote{For a comprehensive review see J.M. Flynn
and C.T. Sachraida, ref. \cite{flynnsachr}.}.
The agreement is very encouraging. 
The analogous fit of (\ref{param}) for the $D \to \pi$ 
form factor results in
\be
a = 1.16, ~b = 0.32~.
\label{param1D}
\ee  
In this case, the NLO effects are not yet included whence the LO prediction
$f^+(0) = 0.68$ given in (\ref{Dfplo}) is used.

The distribution in momentum transfer squared 
for $B \to \pi \bar{l} \nu_l$ is shown in Fig. 14. The band indicates
the present theoretical uncertainty as discussed in detail 
in sect 3.5. To be precise, the uncertainties from the different
sources are added in quadrature.
Integrating this distribution one obtains the partial width
\be
\Gamma(B\rightarrow\pi \bar{l} \nu_l) =
 (7.5\pm 2.5) ~|V_{ub}|^2\ \mbox{ps}^{-1}~. 
\label{elmu}
\ee
The corresponding distribution in the charged lepton energy is displayed
in Fig. 15.

Contrary to the semileptonic decays into $e$ and $\mu$,  
the decay $B \rightarrow \pi \bar{\tau} \nu_\tau$
is quite sensitive to the scalar form factor $f^0$.
As well-known \cite{BLN}, the single-pole approximation used 
to extrapolate the sum rule result on $f^+$
to maximum $p^2$ cannot be applied to $f^0$. 
One way to argue is that the scalar $B$ ground state  
is about 500 MeV heavier than the pseudoscalar $B$ and, therefore,
lies too far above the kinematical endpoint 
of the $B\to \pi$ transition in order
to dominate the form factor.
Nearby excited resonances and nonresonant states are expected to give 
comparable contributions.   
Thus, in order to demonstrate the sensitivity of the $\tau$-mode to $f^0$, 
the light-cone prediction shown in Fig. 10 is 
linearly extrapolated from the value at $ p^2 \simeq 15$ GeV$^2$ at which 
the sum rules (\ref{fplus}) and (\ref{fplusminus}) still hold  
to the value at $p^2 \simeq m_B^2$ dictated by the
soft pion limit \cite{Vol,BLN}:
\be
\lim_{p^2\to m^2_B} f^0(p^2) = \frac{f_B}{f_\pi}= 1.1 ~\mbox{to}~ 1.6~.
\label{CT}
\ee
Here, we have used the conservative range of estimates
$f_B = 150$ to 210 MeV.
This rough extrapolation of $f^0$ is illustrated in Fig. 16 \cite{KRW}.
Also shown are lattice data. They are systematically lower than the 
expectation. Note, however, that the sum rule estimate of $f^0$ 
still lacks NLO corrections.
Certainly, within the theoretical uncertainties there is no contradiction.

Fig. 15 shows the $\tau$-energy spectrum.
As anticipated, a measurement of $B \rightarrow \pi \bar{\tau} \nu_\tau$
has the potential 
to determine or at least constrain the scalar form factor.
For the integrated width one predicts
\be
\Gamma(B \to \pi \bar{\tau} \nu_\tau) = (6.1 \pm 0.4)
|V_{ub}|^2\ \mbox{ps}^{-1}~.
\label{tau}
\ee
Interesting is also the ratio of (\ref{tau}) and (\ref{elmu}):
\be
\frac{\Gamma(B\rightarrow\pi \bar{\tau} \nu_\tau)}
{\Gamma(B\rightarrow\pi \bar{e} \nu_e)}= 0.75 ~\mbox{to} ~0.85~.
\label{r}
\ee
This expectation is independent of 
$V_{ub}$, and less sensitive to uncertainties in the sum rule parameters
than the widths themselves. The range quoted above 
corresponds to the two extreme extrapolations of $f^0$
considered.

\begin{figure}[htb]
\centerline{
\epsfig{bbllx=100pt,bblly=209pt,bburx=507pt,%
bbury=490pt,file=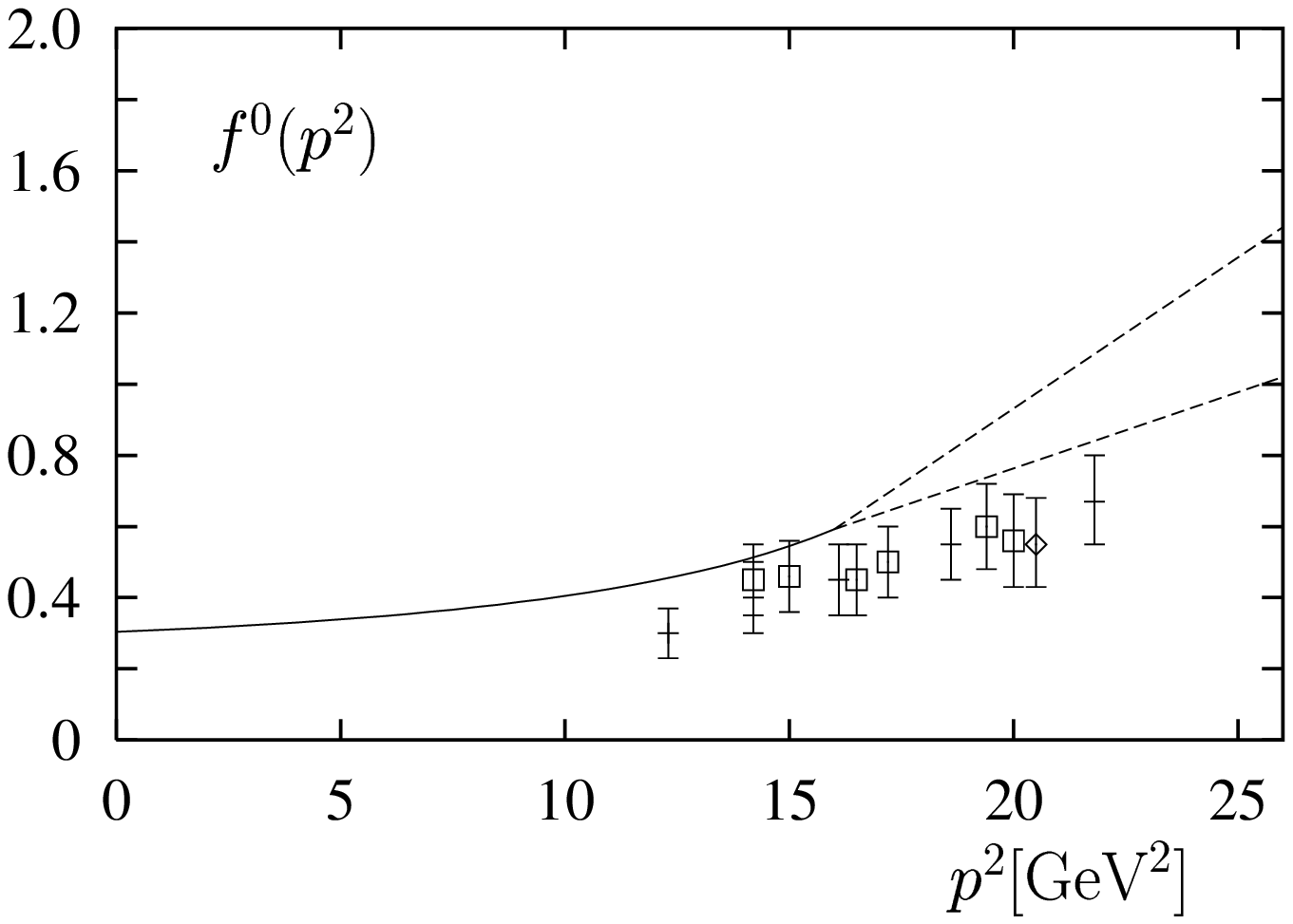,scale=0.9,%
,clip=}
}
\caption{\it The $B\to\pi$ form factor $f^0$: direct 
sum rule estimate (solid) and linear 
extrapolations to the limit (\ref{CT}) (dashed).  
The lattice results are from \cite{Flynn}.}
\end{figure}

Recently, the CLEO collaboration \cite{CLEO}
has reported the first observation of the semileptonic decays
$B\rightarrow \pi \bar{l}\nu$ and $B\rightarrow \rho \bar{l}\nu$
($l = e, \mu $). From the measured branching fraction,
$BR(B^0\to\pi^-l^+\nu_l) = (1.8 \pm 0.4\pm 0.3 \pm 0.2)\cdot 10 ^{-4}$,
and the world average of the $B^0$ lifetime
\cite{PDG},  $\tau_{B^0}=1.56 \pm 0.06 $ ps,
one derives
\be
\Gamma(B^0\rightarrow \pi ^- l^+ \nu_l) =
(1.15 \pm 0.35)\cdot10^{-4}~\mbox{ps}^{-1}~,
\label{CLEOpi7}
\ee
where the errors have been added in quadrature.
Comparison of (\ref{CLEOpi7}) with (\ref{elmu}) yields
\be
|V_{ub}| = 0.0039 \pm 0.0006 \pm 0.0006.
\label{vubpi}
\ee
Here, the first (second) error corresponds to the current
experimental (theoretical) uncertainty.

\begin{figure}[p]
\centerline{
\epsfig{file=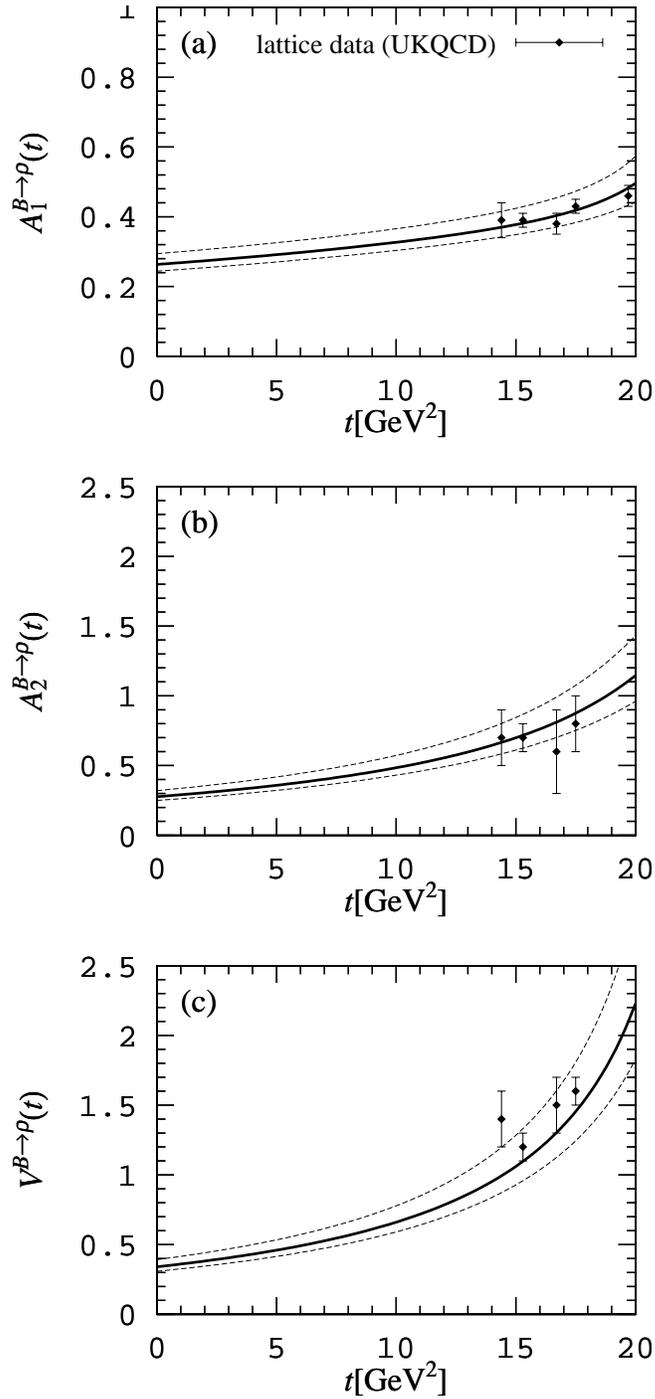,width=9cm,bbllx=50pt,bblly=50pt,bburx=300pt,%
bbury=570pt,clip=}
}
\caption{\it The $B \to \rho$ form factors: 
predictions from light-cone sum rules (solid)
in comparison to lattice results \cite{Flynn}.
The dashed curves indicate the uncertainties in the sum rule
results (from \cite{BallBraun}).}
\end{figure}

\begin{figure}[htb]
\centerline{
\epsfig{file=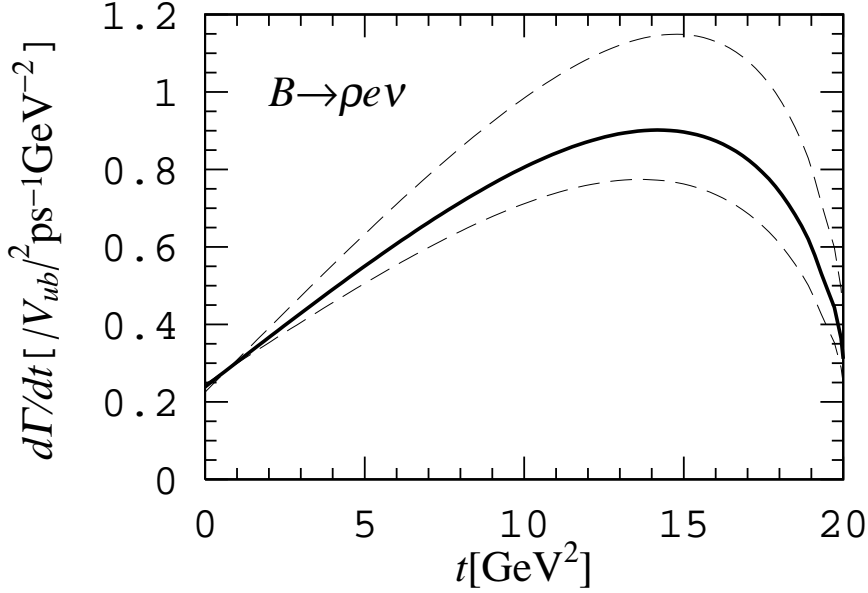,width=12cm,bbllx=50pt,bblly=50pt,bburx=300pt,%
bbury=220pt,clip=}
}
\caption{\it Distribution of the momentum transfer squared in
$B\to\rho \bar{e} \nu_e$.
The dashed curves indicate the theoretical uncertainty 
(from \cite{BallBraun}).}
\end{figure}

A similar analysis has been performed for 
$B \rightarrow \rho \bar{l}\nu_l$ \cite{BallBraun}. 
There are four independent $ B\to \rho$ form factors:
\begin{eqnarray}
\langle \rho(\lambda) | \bar{u}\gamma_\mu(1-\gamma_5)b| B \rangle & = &
-i (m_B + m_\rho) A_1(t) \epsilon(\lambda)_\mu +
\frac{iA_2(t)}{m_B
+ m_\rho} (\epsilon(\lambda) p_B) (p_B+p_\rho)_\mu \nonumber\\
& & + \frac{iA_3(t)}{m_B + m_\rho} (\epsilon(\lambda) p_B)
(p_B-p_\rho)_\mu \\
& & + \frac{2V(t)}{m_B + m_\rho}
\epsilon_\mu^{\phantom{\mu}\alpha\beta\gamma}\epsilon(\lambda)_\alpha
p_{B\beta} p_{\rho\gamma} \nonumber
\label{Brho}
\end{eqnarray}
with $t=(p_B-p_\rho)^2$.  Only $V(t)$, $A_1(t)$ and $A_2(t)$
contribute to the semileptonic decay into $l=e$ or $\mu$.
The sum rule predictions on these form factors are plotted in Fig. 17, 
while Fig. 18 shows the resulting
distribution in momentum transfer squared.

From the integrated width \cite{BallBraun},
\be
\Gamma(B\rightarrow \rho \bar{l} \nu_l)= (13.5\pm 4)
~|V_{ub}|^2~\mbox{ps}^{-1}~,
\label{rhoenu}
\ee
and the CLEO result,
\be
\Gamma(B^0\rightarrow \rho ^- l^+ \nu_l) =
(1.60 \pm 0.6)\cdot10^{-4}~\mbox{ps}^{-1}~,
\label{CLEOrho7}
\ee
derived from the observed branching ratio
$BR(B^0\ra\rho^-l^+\nu_l) = (2.5 \pm 0.4^{+ 0.5}_{-0.7}\pm 0.5)\cdot
10 ^{-4}$
and the $B^0$ lifetime used in (\ref{CLEOpi7}), one obtains
\be
|V_{ub}| = 0.0034 \pm 0.0006 \pm 0.0005.
\label{vubrho}
\ee
Within errors, the values of $|V_{ub}|$ extracted from the
two exclusive measurements (\ref{CLEOpi7}) and (\ref{CLEOrho7}) are nicely 
consistent with each other, and also coincide with the inclusive 
determination of $V_{ub}$ \cite{uraltsev,vubincl}.

It is worth mentioning that the ratio 
\be
\frac {\Gamma (B^0 \rightarrow \rho^-l^+\nu_l)}
{\Gamma (B^0 \rightarrow \pi^-l^+\nu_l)} = 1.8 \pm 0.7
\label{neuGamma}
\ee
following from (\ref{rhoenu}) and (\ref{elmu})
is also consistent with the ratio
\be
\frac {BR (B^0 \rightarrow \rho^-l^+\nu_e)}
{BR (B^0 \rightarrow \pi^-l^+\nu_e)} = 1.4^{+0.6}_{-0.4}\pm 0.3 \pm 0.4 
\label{neuBR}
\ee
observed by CLEO. However, the uncertainties on both sides
are still too big to draw any firm conclusion.

The final application of the sum rule results on form factors which I
want to consider here is the Cabibbo-suppressed decay 
$D\rightarrow\pi \bar{e} \nu_e$.
The decay distribution (\ref{dG})
calculated from the parametrization (\ref{param}) with (\ref{param1D})
for $f^+$ is plotted in Fig. 19. Since the electron mass can be neglected,
the form factor $f^0$ plays no role.
The integrated width is predicted to be
\be
\Gamma(D\rightarrow\pi \bar{l} \nu_l) =
0.16 ~|V_{cd}|^2 ~\mbox{ps}^{-1} =
8.0 \cdot 10^{-3} ~\mbox{ps}^{-1}~,
\label{Gamma}
\ee
where $|V_{cd}|= 0.224 \pm 0.016 $ \cite{PDG} has been used.
This prediction should be compared with
the experimental result,
\be
\Gamma(D^0\rightarrow\pi^- e^+ \nu_e) =
(9.2^{+2.9}_{-2.4})\cdot10^{-3} ~\mbox{ps}^{-1}~,
\label{brD}
\ee
derived from
the branching ratio
$BR(D^0\rightarrow\pi^- e^+ \nu) =
(3.8^{+1.2}_{-1.0})\cdot 10^{-3}$
and the lifetime
$\tau_{D^0}= 0.415 \pm 0.004  ~\mbox{ps}$ ~\cite{PDG}.
The theoretical uncertainty is estimated to be 
of the order of the experimental error. 
Despite of the preliminary agreement, one cannot be satisfied
with the present status.
In order to really test the sum rule approach both
more accurate calculations and more precise data are needed.

\begin{figure}[ht]
\centerline{
\epsfig{bbllx=89pt,bblly=209pt,bburx=507pt,%
bbury=490pt,file=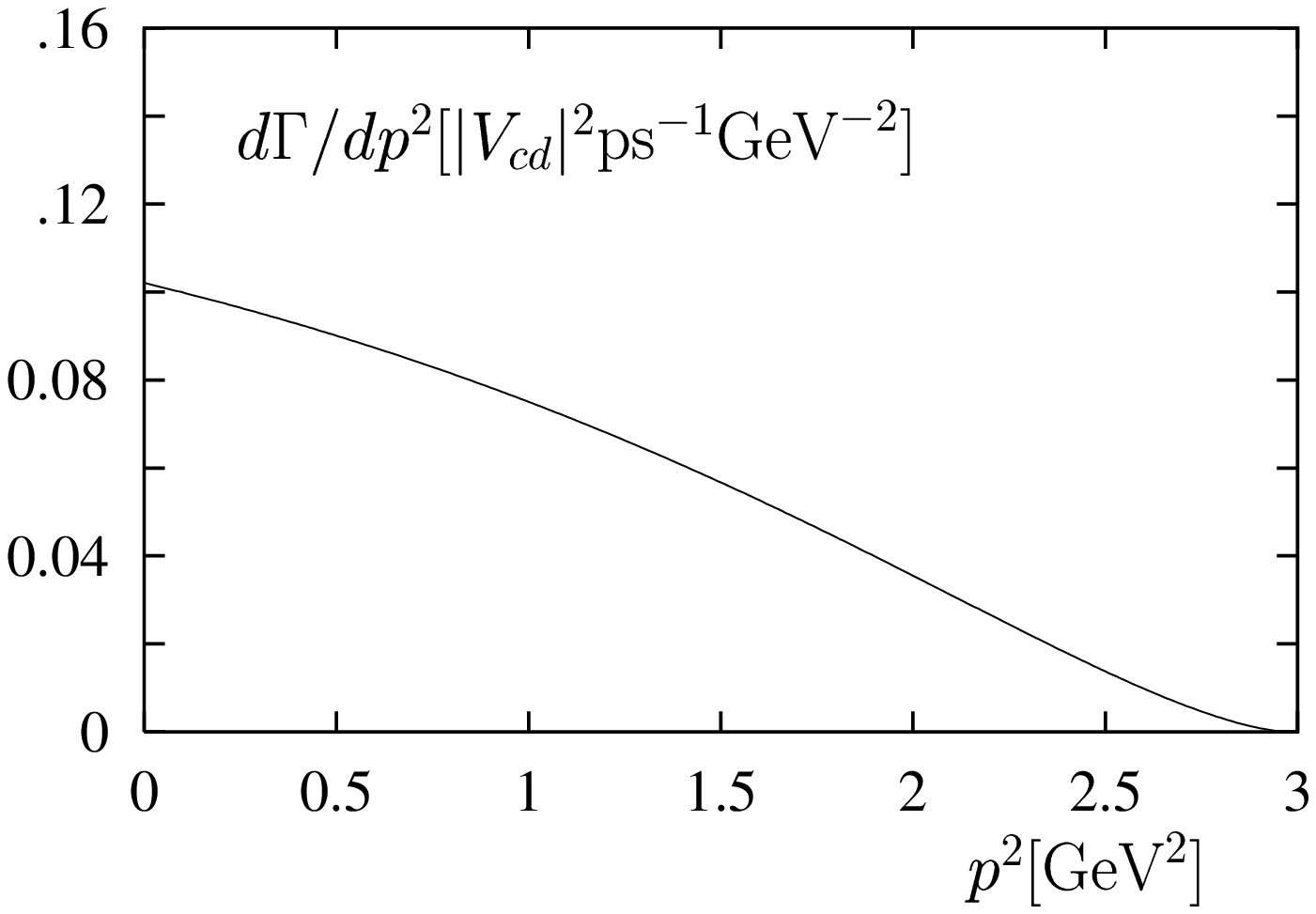,scale=0.9,%
,clip=}
}
\caption{\it Distributions 
of the momentum transfer squared in 
$D \rightarrow \pi \bar e \nu_e$ .}
\end{figure}

%
\section{Nonleptonic two-body decays}

\subsection{Effective hamiltonian}

As illustrated in Tab. 2, exclusive nonleptonic decays are the
theoretically most complicated processes
discussed in these lectures. This class of decays
is strongly influenced by
hard and
soft interactions of quarks and gluons, by initial bound state and
nonspectator effects, as well as by hadronization and 
final state interactions.
Although
over the years one has developed a qualitative understanding and even an
amazingly consistent phenomenological description of two-body decays 
\footnote{For a recent comprehensive review see
M. Neubert and B. Stech, ref. \cite{NS}.},
some features are still lacking a quantitative theoretical explanation. 

Up to now, only the hard-gluon effects can be systematically taken
into account in the framework of renormalization group improved 
QCD perturbation theory. As already pointed out,
the essential tool is OPE
which allows to separate the dynamics
at short and long distances. 
The result is an effective weak hamiltonian 
given by a sum of renormalized local operators 
with scale-dependent Wilson coefficients as sketched in (\ref{HW}).
Specifically for nonleptonic $B$ decays, one has
\be
H_{NL}= \frac{G_F}{\sqrt{2}}\sum_{q=c,u}V_{qb}
\left[c_1(\mu) O_{1q}(\mu) + c_2(\mu) O_{2q}(\mu)\right] ~+~h.c. ~,  
\label{Hb}
\ee
where 
\be
O_{1q}=(\bar{d}'\Gamma^\rho u + \bar{s}'\Gamma^\rho c)
(\bar{q}\Gamma_\rho b),\
O_{2q}=(\bar{q}\Gamma^\rho u)(\bar{d}'\Gamma_\rho b) +
(\bar{q}\Gamma^\rho c)(\bar{s}'\Gamma_\rho b)
\label{ob}
\ee
with $\Gamma_\rho = \gamma_\rho(1-\gamma_5)$. 
Similarly, the effective hamiltonian relevant for nonleptonic 
$D$ decays is given by
\be
H_{NL}= \frac{G_F}{\sqrt{2}}\sum_{q=s,d}V_{cq}
\left[c_1(\mu) O_{1q}(\mu) + c_2(\mu) O_{2q}(\mu)\right] ~+~h.c. ~,  
\label{Hc}
\ee
with 
\be
O_{1q}=(\bar{d}'\Gamma^\rho u)(\bar{c} \Gamma_\rho q),\
O_{2q}=(\bar{c} \Gamma^\rho u)(\bar{d}'\Gamma_\rho q) \,.
\label{oc}
\ee
In the above, $d' = V_{ud}d + V_{us}s$ and $s' = V_{cd}d + V_{cs}s$
are weak eigenstates. In addition to the four-quark operators
$O_1$ and $O_2$, the operator-product expansion includes
so-called penguin operators \cite{penguin}. These are mainly  
relevant for rare nonleptonic decays such as 
$B \ra K\pi$ or radiative decays such as $B \ra K^*\gamma$, but they 
play no role in the following discussion.

The Wilson coefficients $c_{i}(\mu)$ contain the effects
from hard gluon and quark exchanges with virtualities larger than
$\mu$. In perturbation theory, these interactions generate 
large logarithmic terms, $\alpha_s^m ln^n \frac{m_W}{\mu}$,
which must be summed up to all orders. This is achieved  
by solving the renormalization group equation
\be
\left(\frac{d}{dln\mu} - \gamma_{\pm}\right) c_{\pm}(\mu) = 0 ~,
\label{rg}
\ee
where 
\be
\gamma_{\pm} = \frac{\alpha_s}{4\pi}\gamma_{\pm}^{(1)}
+ \left(\frac{\alpha_s}{4\pi}\right)^2\gamma_{\pm}^{(2)} + .....
\label{anomdim}
\ee 
are the anomalous dimensions of the operators 
$O_{\pm} = \frac{1}{2}(O_1 \pm O_2)$, and 
$c_{1,2} = \frac{1}{2}(c_+ \pm c_-)$.
With the boundary conditions 
$c_{\pm}(m_W) = 1$ following from the unrenormalized weak hamiltonian
indicated in (\ref{HF}) the solution is given by
\be
c_{\pm}(\mu) = \left(\frac{\alpha_s(\mu)}{\alpha_s(m_W)}\right)^
\frac{\gamma_{\pm}^{(1)}}{2b}\left(1 + 
R_{\pm}\frac{\alpha_s(\mu)-\alpha_s(m_W)}{4\pi}\right) ~.
\label{solution}
\ee
The first factor is the LO result \cite{LOcoeff}, while the
second bracket incorporates the NLO corrections \cite{NLOcoeff}.
The LO coefficients are determined by the one-loop 
anomalous dimensions, 
\be
\gamma_+^{(1)} = +8\,, ~\gamma_-^{(1)} = -4 \,,
\label{anomal1}
\ee
and by the one-loop coefficient of the QCD $\beta$-function,
\be
b = 11 - \frac{2}{3}n_f ~. 
\label{beta1}
\ee
The NLO term $R_{\pm}$ is given in \cite{NLOcoeff}.

The relevant physical scale is expected to be of the order of
the heavy quark mass.
For $\mu=m_b=4.8$ GeV one obtains \cite{NS}, numerically,
\be
c_1(m_b)=1.108, \,\, c_2(m_b)=-0.249 \,\, (\mbox{LO})\,,
\label{cbLO}
\ee
\be
c_1(m_b)=1.132, \,\, c_2(m_b)=-0.286 \,\, (\mbox{NLO})\,,
\label{cbNLO}
\ee
while
for $\mu=m_c=1.4$ GeV one gets
\be
c_1(m_c)=1.263, \,\, c_2(m_c)=-0.513 \,\, (\mbox{LO})\,,
\label{ccLO}
\ee
\be
c_1(m_c)=1.351, \,\, c_2(m_b)=-0.631 \,\, (\mbox{NLO})~.
\label{ccNLO}
\ee
Here, the running coupling is normalized to $\alpha_s(m_Z)=0.118$.
As anticipated, the effects are smaller in $b$ decays than in $c$ decays.
Furthermore, the NLO effects re-enforce the LO trend.

\subsection{Decay amplitude and factorization}

Given the effective hamiltonian, the amplitude for a
two-body decay is calculated from 
the corresponding hadronic matrix elements of the local four-quark operators: 
\begin{eqnarray} 
A(P \ra h_1 h_2)  &=& \langle h_1 h_2 \mid H_{NL} \mid P \rangle
\nonumber\\
{}&=& \frac{G_F}{\sqrt{2}} \sum _q  V_{qb} \sum_i c_i(\mu)
 \langle h_1 h_2 \mid O_{iq}(\mu) \mid P \rangle ~.    
\label{weakampl}
\end{eqnarray}
Since these matrix elements contain soft QCD interactions
characterized by virtualities smaller than $\mu$,
the problem of calculating them is extremely
difficult and still far from a satisfactory solution. 
It may not be completely needless to stress that
the scale dependence of the matrix elements
must cancel in (\ref{weakampl}) against the  
scale dependence of the Wilson coefficients since 
the decay amplitude is an observable. 
In practice, there is always some residual scale dependence 
because of the truncation of the perturbative series.
In terms of the weak amplitude (\ref{weakampl}) 
the partial width for two-body decays is given by
\be
\Gamma(P \ra h_1 h_2) = \frac{1}{16\pi m_P^3} 
\sqrt{(m_P^2+m_1^2-m_2^2)^2-4m_P^2m_1^2} ~|A|^2 ~.
\label{gammanl}
\ee 

The present knowledge and understanding
of nonleptonic $B$ decays has recently been 
reviewed in a very comprehensive article 
by M. Neubert and B. Stech \cite{NS}. 
For $D$ decays one may consult the classic papers
by M. Bauer, B. Stech and M. Wirbel \cite{BSW}.
Here, I will restrict the 
discussion to a particular example which brings
the main theoretical difficulties to light, 
namely $B \rightarrow J/\psi K$.
This decay mode also plays an outstanding role 
in $B$-physics at hadron colliders and
future $B$-factories. 

After Fierz-transformation of the operator $O_{1c}$ in (\ref{Hb})
the piece of the
effective hamiltonian relevant for $B \rightarrow J/\psi K$
can be written in the form 
\be
H_{NL}= \frac{G_F}{\sqrt{2}}V_{cb}V^*_{cs}\{(c_2
+\frac{c_1}3) O_2+2c_1\tilde{O}_2\}  
\label{H}
\ee
with 
\be
O_2=(\bar{c}\Gamma^\rho c)(\bar{s}\Gamma_\rho b),\
\tilde{O}_2=(\bar{c}\Gamma^\rho \frac{\lambda^a}2c)(\bar{s}\Gamma_\rho
\frac{\lambda^a}2 b) ~,
\label{o}
\ee
$\lambda^a$ being the usual $SU(3)$-colour matrices.
The colour indices and the $\mu$-dependence 
are suppressed for simplicity. 

In a radical first approximation, one may factorize
the matrix elements of the operators $O_2$ and $\tilde{O}_2$
into products of 
matrix elements of the currents that compose these operators.
Long distance strong interactions  
between quarks belonging to different 
currents are thereby completely neglected. 
Furthermore, the matrix element of the octet operator
$\tilde{O}_2$ vanishes because of colour conservation, while
the factorized matrix element of the singlet operator $O_2$
is approximated by
\be
\langle J/\psi K\mid O_2(\mu)\mid B\rangle
= \langle J/\psi\mid \bar{c}\Gamma^\rho c  \mid 0 \rangle
\langle K  \mid \bar{s}\Gamma_\rho b \mid B\rangle 
= 2f_\psi f_{B \ra K}^+m_\psi(\epsilon^\psi  \cdot q) ~.
\label{factoriz}
\ee
Obviously, $f_{\psi}$ is the $J/\psi$ decay constant,
$f_{B \ra K}^+$ is the $B\ra K$ form factor at the
momentum transfer $p^2=m_\psi^2$, 
$\epsilon^\psi$ denotes the $J/\psi$ polarization vector,
and $q$ the $K$ four-momentum.  
The leptonic width 
$\Gamma( J/\psi$\-$ \ra l^+l^-) = 5.26 \pm 0.37$ keV 
implies
\be
f_{\psi} = 405~ \mbox{MeV}~,
\label{fpsi}
\ee
while a sum rule estimate \cite{BKR} similar 
to the one for the $B \ra \pi$ form factor in sect. 3.5 yields
\be
f_{B \ra K}^+ =  0.55 \pm 0.05 ~.
\label{fBK}
\ee

At this point one encounters a first principal problem:
since the matrix elements of the quark currents in (\ref{factoriz})
are scale-independent, the $\mu$-dependence of the 
short-distance coefficients cannot be compensated.
Hence, the approximation
\be
\langle J/\psi K\mid H_{NL}\mid B\rangle 
= \sqrt{2} G_F V_{cb}V^*_{cs} \left(c_2(\mu)
+\frac{c_1(\mu)}3 \right) 
f_\psi f_{B \ra K}^+m_\psi(\epsilon^\psi  \cdot q)
\label{factO}
\ee
can at best be 
valid at a particular scale $\mu$ which could be called
the factorization scale. The conventional assumption
is $\mu = O(m_b)$. 

Using the next-to-leading order coefficients $c_{1,2}(\mu)$ 
in the HV scheme with 
$\Lambda^{(5)}_{\overline{MS}}=225$ MeV from 
\cite{Buras2} and
taking $\mu=m_b \simeq 5$ GeV, 
one has 
\be
c_2(\mu) + \frac{c_1(\mu)}{3} = 0.156~.
\label{c2c1}
\ee
Together with  
(\ref{fpsi}) and (\ref{fBK}), this yields the branching ratio
\be
BR( B\rightarrow J/\psi K) = 0.025\%~, 
\label{BRfact} 
\ee
a value which is considerably 
smaller than the measured branching ratios
\cite{PDG,CLEOnl}
\be
BR( B^- \rightarrow J/\psi K^- )= (0.101 \pm 0.014 )\%~,
\label{CLEOpsi}
\ee
\be
BR( B^0 \rightarrow J/\psi \bar K^0 )= (0.075 \pm 0.021) \% ~.
\label{CLEO0}
\ee
The discrepancy between experiment and expectation is worsened
by a factor 3 if the coefficients (\ref{cbNLO}) are used yielding
$c_2 + c_1/3 = 0.091$ instead of (\ref{c2c1}).
This provides a good example for the scale uncertainty
due to factorization. 

\subsection{Nonfactorizable contributions and effective coefficients}

The quantitative failure and the scale problem pointed out above
imply that naive factorization of matrix elements does not work.
Clearly, factorization has to be accompanied by a 
reinterpretation  of the Wilson coefficients. For  
$B \ra J/\psi K$, the short-distance coefficient
$c_2(\mu) + c_1(\mu) / 3 $
is substituted by an effective scale-independent coefficient $a_2$.
Phenomenologically \cite{BSW}, 
$a_2$ is treated as 
a free parameter to be determined from experiment. From 
(\ref{factO})
and (\ref{CLEOpsi}) one finds
\be
|a_2^{B\psi K}|= 0.31 \pm 0.02~, 
\label{a2cleo}
\ee
where the quoted error is purely experimental.
The sign of $a_2^{B\psi K}$ remains undetermined.
It is an interesting observation that 
the above value is similar to the coefficient $a_2$ found from 
so-called class III decays 
\cite{NS}.
This suggests, but does not prove a universal nature of $a_2$.

The drastic difference between the short-distance 
coefficient (\ref{c2c1}) and the effective coefficient (\ref{a2cleo})
points at the existence of sizeable nonfactorizable contributions.
The latter are also needed to
cancel or at least soften the strong 
$\mu$-dependence of the factorized amplitude (\ref{factO}).
Further investigation shows that the dominant nonfactorizable effects
should arise from the matrix element
of the operator  $\tilde{O}_2$. Writing the latter in the 
convenient parametrization 
\be
\langle J/\psi K\mid \tilde{O}_2(\mu) \mid B\rangle = 
2 f_\psi \tilde{f}_{B \psi K}(\mu) m_\psi (\epsilon^\psi \cdot q)~,
\label{nf}
\ee
and adding it to (\ref{factoriz}), one formally reproduces (\ref{factO}),
\be
\langle J/\psi K\mid H_{NL}\mid B\rangle 
= \sqrt{2}G~V_{cb}V^*_{cs}
a_2^{B\psi K}f_\psi f_{B \ra K}^+m_\psi(\epsilon^\psi\cdot q)~, 
\label{ampl}
\ee
with the effective coefficient \cite{KR} 
\be
a_2^{B\psi K}
=c_2(\mu)+\frac{c_1(\mu)}3 + 
2c_1(\mu)\frac{\tilde{f}_{B\psi K}(\mu)}{f_{B\ra K}^+}~.
\label{a2}
\ee
In Fig. 20, the partial width for $B\to J/\psi K$ is shown
as a function of the parameter $\tilde{f}_{B\psi K}$
associated with $\langle \tilde{O}_2 \rangle$. Note that 
$\tilde{f}_{B\psi K}= 0$ corresponds to naive factorization, while 
the fitted value of $a_2^{B\psi K}$ given in (\ref{a2cleo}) 
implies
\be
\tilde{f}_{B\psi K}(\mu=m_b)= +0.04~~\mbox{or}~ -0.12~.
\label{tildemb}
\ee
For a lower scale the value of $\tilde{f}_{B\psi K}$
is shifted to slightly more positive values, e.g.,
\be
\tilde{f}_{B\psi K}(\mu = \frac{1}{2} m_b)= +0.06 ~~\mbox{or}~ -0.09~.
\label{tildemu}
\ee
We see that a nonfactorizable amplitude of 10 to 20~\% of the size
of the factorizable one is sufficient to reconcile expectation 
with experiment. 

\begin{figure}[htb]
\centerline{
\epsfig{bbllx=100pt,bblly=209pt,bburx=507pt,%
bbury=530pt,file=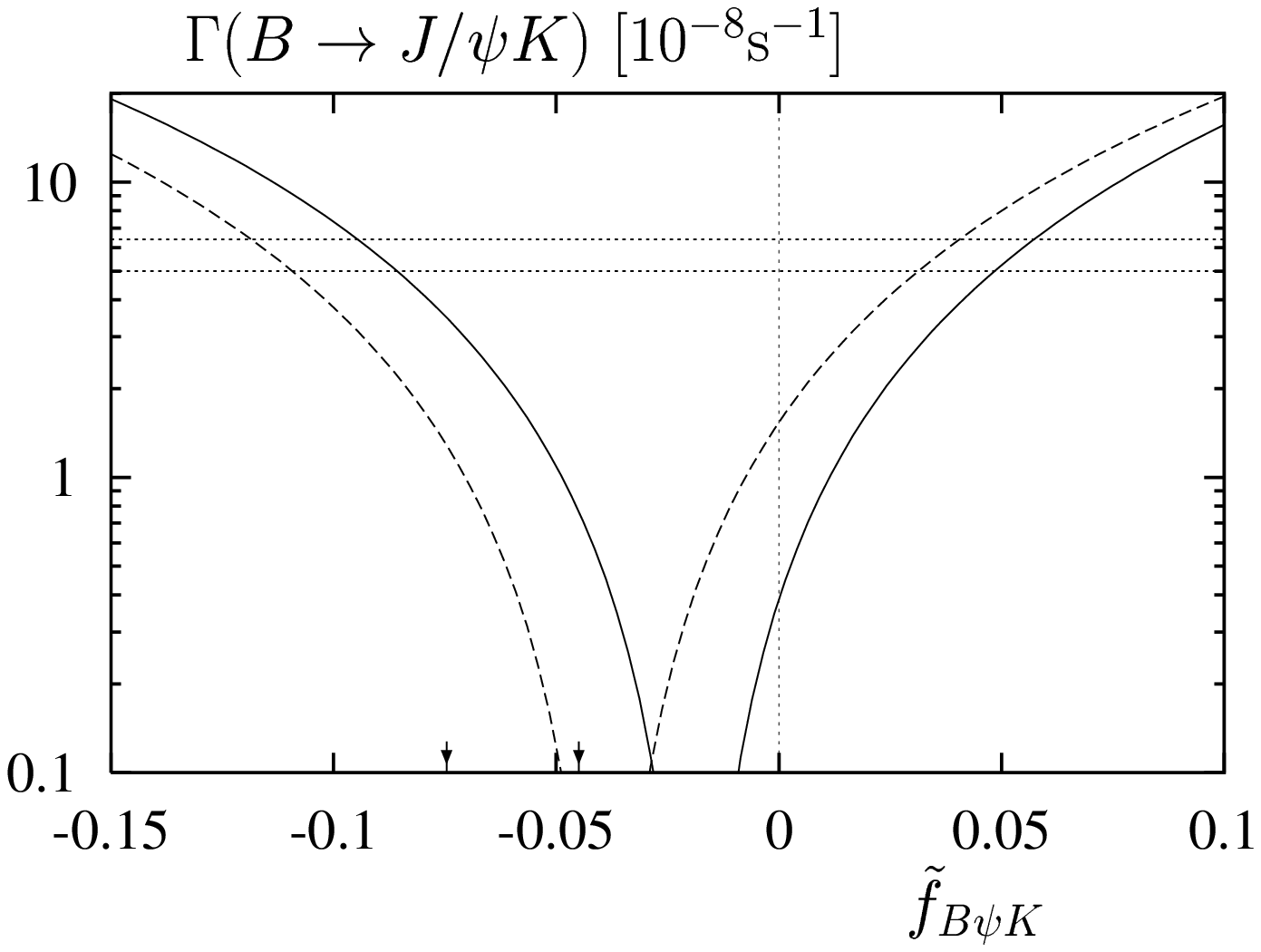,scale=0.9,%
,clip=}
}
\caption{\it The partial width for $B \rightarrow J/ \psi K $ 
as a function of $\tilde{f}_{B\psi K}$ parameterizing
the nonfactorizable contribution to the decay amplitude.
The horizontal dotted lines represent
the average of the experimental widths
given in the text. The solid (dashed) curves show the theoretical
expectation for $\mu=m_b/2$ ($\mu=m_b$).
The arrows indicate the QCD sum rule estimate. 
}
\end{figure}

\subsection{Sum rule estimate of the nonfactorizable amplitude}

The calculation of $a_2^{B\psi K}$ and of the analogous 
coefficients for other two-body decays \cite{BSW} 
from first principles is one
of the outstanding problems in weak decays.
As a first step in this direction, 
we have undertaken a rough
estimate of $\tilde{f}_{B \psi K}$ using again QCD sum rule methods
\footnote{This 
work was done in collaboration with B. Lampe and A. Khodjamirian. 
Results have been  
reported, e.g., in \cite{KR}.}.
Following the idea worked out for charm decays in \cite{BS}, 
we consider the four-point correlation function 
\be
\widetilde{\Pi}_{\mu\nu}(p,q)=\int~ d^4x~d^4y~d^4z~e^{iqx+ipy}
\langle 0 \mid T\{j_{\mu5}^K(x)j_\nu^\psi(y)\tilde{O_2}(z)j^B_5(0)\}\mid 0
\rangle~, 
\label{corr}
\ee
where $j^B_5= \bar{b}i\gamma_5 u$, $ j_\nu^\psi= \bar{c}\gamma_\nu c $, and
$j_{\mu5}^K= \bar{u}\gamma_\mu \gamma_5s$ 
are the generating currents of the mesons 
participating in the decay $B \to J/\psi K$, and 
$p+q$, $p$ and $q$ are the respective four-momenta.
$\tilde{O}_2$ is the octet operator given in (\ref{o})
which is suspected to give rise to the dominant effect.

Similarly as in the case of the two-point correlation function 
(\ref{corr2}) studied in sect. 2.2,
one writes a dispersion relation for (\ref{corr}) in terms of 
intermediate hadronic states in the $B$, $J/\psi$, and $K$ channel.
The ground state contribution to (\ref{corr}),
\be
\widetilde{\Pi}_{\mu\nu}(p,q)= 
i\frac{\langle
0\mid j_{\mu 5}^K\mid K \rangle
\langle
0\mid j_\nu^\psi \mid J/\psi\rangle
\langle J/\psi K \mid\tilde{O}_2 \mid B\rangle
\langle B\mid j_5^B \mid 0\rangle}
{(m_{K}^2-q^2)(m_{\psi}^2-p^2)(m_{B}^2-(p+q)^2)}+~.....~, 
\label{res}
\ee 
contains the desired matrix element (\ref{nf}).
In addition, one has contributions from  
excited resonances and continuum states denoted above by ellipses 
which lead to a very complicated singularity structure. 
Furthermore, in the euclidean momentum   
region $ Q^2 =-q^2 > O(1~\mbox{GeV}^2)$, $p^2 \leq 0$, 
$(p+q)^2 \leq 0$ 
one can expand the correlation function (\ref{corr})
in terms of local operators:
\be
\widetilde{\Pi}^{\mu\nu}(p,q)
= \sum_d \widetilde{C}^{\mu\nu}_d(p,q,\mu)
\langle  \Omega_d(\mu) \rangle~.
\label{ope9}
\ee
For the operators with dimension $d\leq6$ given in (\ref{oper2}),
one can calculate the necessary 
coefficients $\tilde{C}^{\mu\nu}_d(p,q,\mu)$ 
from the diagrams shown in Fig. 21.
From (\ref{res}) and (\ref{ope9})
it is possible, at least in principle, 
to derive a sum rule for the matrix element   
$\langle J/\psi K \mid\tilde{O}_2 \mid B\rangle$.

However, there are two complications that are 
not present in the two point sum rules discussed in sect. 2.2. 
One problem is the presence of a light continuum in the 
$B$ channel below the pole of the ground state $B$-meson. 
This contribution to (\ref{res}) can be associated with processes 
of the type $B \to$ ``$D^*D_s$'' $\to J/\psi K$, 
where an intermediate state carrying 
$D^*D_s$ quantum numbers rescatters into the final
$J/\psi K$ state. Formally, in (\ref{corr}) this intermediate
state is created from
the vacuum by the combined 
action of the operator product $\tilde{O}_2 j_5^B$. 
As a reasonable solution we suggest to 
cancel this unwanted contribution against the terms in 
(\ref{ope9}) which have the
quark content $c\bar{c}s\bar{q}$ and
which develop a nonzero
imaginary part at $(p+q)^2 \geq 4m_c^2$.
In the approximation considered this is the case for
the four-quark condensate contribution represented by  
Fig. 21c.  

The second problem concerns the continuum 
subtraction in the remainder.
A detailed examination of (\ref{corr})
indicates that the higher charmonium resonances
contribute with alternating signs. 
Therefore, the usual subtraction procedure  
in which the dispersion integral over the excited and continuum 
states is approximated by the perturbative counterpart
is not reliable here. In order to proceed we employ 
explicit, although rough
models for the hadronic spectral functions. 
Since the number of additional parameters has to be manageable, only 
the first excited resonances are included in each channel
besides the $B$, $J/\psi$ and $K$ ground states.
In total, in this approximation the correlator (\ref{res})  
contains three free parameters in addition to $\tilde{f}_{B\psi K}$.
Finally,
one takes the Borel transform of (\ref{res}) and (\ref{ope9})  
in the $B$-meson channel and moments in the charmonium channel.
In the $K$-meson channel, $q^2$ is kept
spacelike. 

A fit of the hadronic representation (\ref{res}) 
to the OPE result (\ref{ope9}) for various values of the
Borel mass $M$ and $q^2$, and for different moments yields  
\be
\tilde{f}_{B\psi K} = -(0.045~to~0.075).
\label{ftilde}
\ee
The implicit scale of this estimate is given by
$M \simeq \sqrt{m_B^2-m_b^2} \simeq \frac{1}{2}m_b \simeq 2.4$ GeV.
Substituting (\ref{ftilde}) in (\ref{a2}), and evaluating the 
short-distance coefficients $c_{1,2}(\mu)$ also at $\mu= M$, 
one gets
\be
a_2^{B \psi K}= -0.29 +0.38 -(0.19 ~to~ 0.31) =-(0.10 ~to~ 0.22) ~.
\label{a2number}
\ee
Here, the three terms in the first relation refer to the three terms 
in (\ref{a2}) in the same order. 
Interestingly, the sum rule approach seems to 
favour the negative solution for $\tilde{f}_{B \psi K}$. 
Although in comparison with (\ref{tildemu}) the above estimate  
falls somewhat short, it narrows the gap between theory 
and experiment considerably as can be seen from Fig. 20. 

\begin{figure}[htb]
\centerline{
\epsfig{bbllx=35pt,bblly=235pt,bburx=523pt,%
bbury=580pt,file=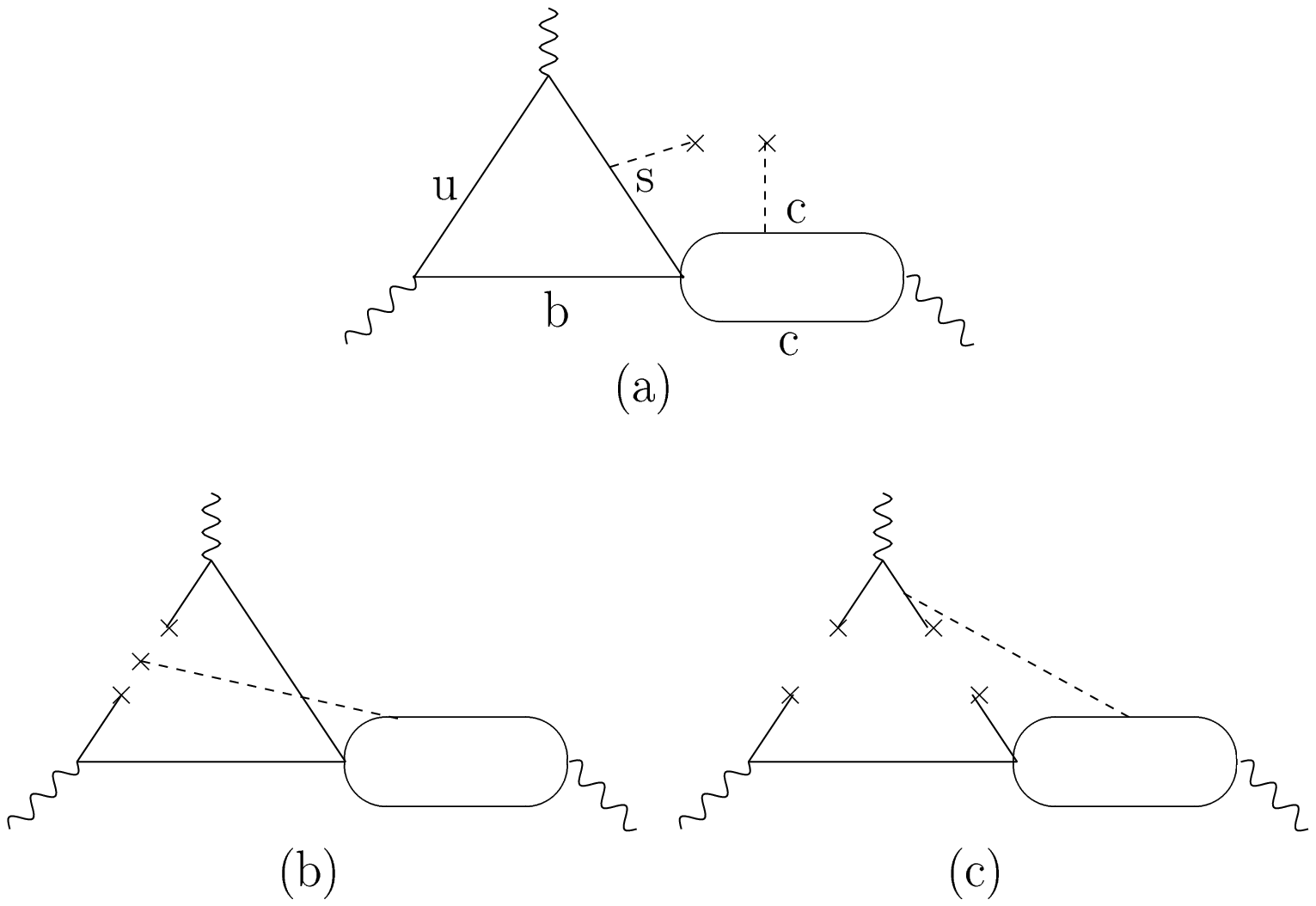,scale=0.9,%
,clip=}
}
\caption{\it Diagrams determining the Wilson coefficients of the OPE 
of the correlation function $\tilde{\Pi}_{\mu\nu}$.
The symbols are as in Fig. 1.}
\end{figure}

Several comments are in order. Firstly, the nonfactorizable matrix 
element (\ref{nf}) is rather small as compared to the factorizable one 
given in (\ref{factoriz}),
numerically, 
\be
|\tilde{f}_{B\psi K}/f^+_{B\ra K}|\simeq 0.1 ~.
\label{fftilde}
\ee 
Nevertheless, the impact on $a_2^{B\psi K}$ is strong 
because of the large coefficient as compared to the coefficient
of the factorized amplitude,
\be
|2c_1/(c_2+c_1/3)| \simeq 20 ~to~ 30 ~.
\label{ca2}
\ee 
Secondly, the factorizable amplitude proportional to $c_1/3$  
and the nonfactorizable one proportional to $\tilde{f}_{B\psi K}$
are opposite in sign and hence tend to cancel. 
In fact, if 
$|\tilde{f}_{B\psi K}|$ is taken at the upper end of the estimated 
range, the cancellation is almost complete resulting in a considerable 
enhancement of the branching ratio (\ref{BRfact}).
Note that both terms are nonleading in $1/N_c$.
This is exactly the scenario anticipated by the $1/N_c$ -- rule 
for $D$ decays \cite{BGR} which is
generally consistent with observation \cite{BSW} and has found theoretical
support by the sum rule analysis of \cite{BS}. 
In \cite{BS93,Halperin}, a similar trend was claimed for 
$B\to D \pi$.
Thirdly, our estimate 
yields a negative overall sign for $a_2^{B\psi K}$ in contradiction
to a description of the data on two-body $B$ decays based on 
factorized decay amplitudes  
with two universal coefficients $a_1$ and $a_2$ \cite{NS}.
It should be stressed, however, that in this fit 
the positive sign of $a_2$ actually results from 
the channels $B^- \ra D^0\pi^-, D^0 \rho^-$, $D^{*0} \pi^-$, 
and $D^{*0}\rho^-$,
and is then assigned also to the $J/\psi K$ channel.
This assignment may not be correct. Certainly, 
the sum rule approach described above 
provides no justification for such an assumption. 
On the contrary, diagrams of the kind shown
in Fig. 21 suggest some channel-dependence of the nonfactorizable
matrix elements. For class II processes 
the channel-dependence is enhanced by the large coefficient $2c_1$
in $a_2$ (see (\ref{a2})). The opposite is the case for class I processes 
involving 
\be
a_1 = c_1(\mu) + \frac{c_2(\mu)}{3} + 2c_2 \frac{\tilde{f}}{f^+}~.
\label{a1}
\ee
Here, the nonfactorizable contributions are damped
by the small coefficient $c_2$, while
the factorized term has a large coefficient.
Therefore, $a_1$ is indeed expected to be universal to a good 
approximation, whereas $a_2$ should exhibit some channel-dependence,
in particular when comparing decays with very different final states
such $D \pi$ and $J/\psi K$.

As a last remark, from the sum rule point of view
one does not expect a simple relation between $B$ and $D$ decays. 
At least, the OPE of the four-point correlation 
functions (such as (\ref{corr})) is significantly different, e.g.,  for   
the channels
$B \to D \pi$, $B \to J/\psi K$, and $D \to K \pi$  as can be
inferred from Fig. 21. Hence, the
arguments presented in \cite{NS}
in favour of a change of sign in $a_2$ when going from $D$ to $B$
may oversimplify the dynamics of nonleptonic decays.

%
\section{Conclusion}

These lectures have been devoted to the theory of exclusive decays of
$B$ and $D$ mesons. I have discussed examples of 
leptonic, semileptonic and
nonleptonic decays, focussing on the problem 
of calculating hadronic matrix elements of weak currents and 
four-quark operators. The aim has been to demonstrate the 
usefulness of QCD sum rule techniques. I have described
the derivation of sum rules for
decay constants, 
form factors, and two-body decay amplitudes.
In addition, I have outlined a
sum rule for couplings between heavy and light mesons.
Numerical predictions have been presented for the decay constants
$f_{D}$, $f_{D_s}$, $f_{B}$, and $f_{B_s}$, for the heavy-to-light
formfactors $f^+(p^2)$, $f^-(p^2)$, and $f^0(p^2)$, for the
strong couplings $g_{B^*B\pi}$ and $g_{D^*D\pi}$, and for the
nonfactorizable matrix element
$\langle J/\psi K\mid \tilde{O}_2(\mu) \mid B\rangle$. 

Using the sum rule results I have presented predictions 
on decay distributions and integrated
widths for $B \to \pi \bar{l} \nu_l$, $B \to \rho \bar{l} \nu_l$,
and $D \to \pi \bar{l} \nu_l$.
Comparison with the CLEO measurements \cite{CLEO}
of exclusive semileptonic $B$ decays 
yields values for $V_{ub}$ in good agreement with each other and
with the determination from inclusive data \cite{vubincl}.
Agreement between expectation and measurement is also found
for the Cabibbo-suppressed semileptonic $D^0$ decay \cite{PDG}.
Furthermore, a sum rule estimate is given
for $D^* \to D \pi$. The present experimental 
upper limit \cite{PDG} is still about three times larger than the 
expected width.
Finally, including the sum rule estimate of the 
nonfactorizable contribution to the amplitude for $B\ra J/\psi K$
a theoretical estimate of 
the effective coefficient $a_2$ is obtained which is rather close
to the value extracted from the experimental 
branching ratio \cite{PDG}. However, the sign of $a_2$ predicted
by the sum rule analysis is opposite to the sign determined from data,
if channel-independent, universal coefficients
$a_{1,2}$ are assumed \cite{NS}. I have explained why
in the sum rule approach universality can be expected for $a_1$, but
not for $a_2$. From the sum rule point of view
the apparent universality of 
$a_1$ and $a_2$ constitutes a major puzzle 
which needs further clarification.

Finally, I have addressed the present theoretical
uncertainties in the sum rule calculations, 
and the prospects for reducing the uncertainties.
On the theoretical side, one has to improve 
the light-cone wave functions of light mesons, 
and calculate the higher-order
perturbative corrections to the sum rules. On the experimental side,
new and more precise measurements at hadron facilities and at
future $B$ and tau-charm factories should allow to tightly
constrain the input parameters and to test the reliability
of the sum rule approach.  
It appears conceivable to 
ultimately decrease the uncertainties from presently
20 to 30 ~\% by a factor of 2.
Encouragement comes from the preliminary agreement of 
lattice and sum rule calculations on decay constants and form factors.
It would certainly be very fruitful to combine the flexibility 
of the sum rule method with the rigorous nature of the lattice
approach.

{\bf Acknowledgements}

I want to thank Ikarus Bigi and Luigi Moroni for having invitated me
to the School ``Enrico Fermi'', and the staff of the school for having
made my stay so enjoyable. I am particularly grateful to
Alexander Khodjamirian for his continuous collaboration and for his
great help with the present manuscript. Furthermore, I thank
V.M. Belyaev, V.M. Braun, B. Lampe, Ch. Winhart, S. Weinzierl, and 
O. Yakovlev for collaboration
on various subjects covered in these lectures and for useful discussions.
This work was supported by the Bundesministerium f\"ur Bildung,
Wissenschaft, Forschung und Technologie, Bonn, Germany,
Contract 05 7WZ91P (0).
%


\end{document}